\documentclass[journal,twoside]{IEEEtran}
\usepackage{cite}
\ifCLASSINFOpdf
  \usepackage[pdftex]{graphicx}
  \usepackage[position=bottom]{subfig}
  \DeclareGraphicsExtensions{.pdf,.jpeg,.png}
\else
\fi
\usepackage{svg}

\usepackage{amsmath}
\usepackage{amssymb}
\usepackage{mathtools}
\DeclareMathOperator{\supp}{supp}
\DeclareMathOperator*{\argmin}{arg\,min}

\usepackage{units}
\usepackage{algorithm}
\usepackage{algorithmic}
\newcommand\algorithmicinitialize{\textbf{Initialize:}}
\newcommand\INITIALIZE{\item[\algorithmicinitialize]}
\newcommand\algorithmicreturnbis{\textbf{Return:}}
\newcommand\RETURNBIS{\item[\algorithmicreturnbis]}
\usepackage{multirow}

\usepackage{hyperref}
\begin{document}
%
% paper title
% Titles are generally capitalized except for words such as a, an, and, as,
% at, but, by, for, in, nor, of, on, or, the, to and up, which are usually
% not capitalized unless they are the first or last word of the title.
% Linebreaks \\ can be used within to get better formatting as desired.
% Do not put math or special symbols in the title.
\title{Fast Approximate Construction of Best Complex Antipodal Spherical Codes}
%
%
% author names and IEEE memberships
% note positions of commas and nonbreaking spaces ( ~ ) LaTeX will not break
% a structure at a ~ so this keeps an author's name from being broken across
% two lines.
% use \thanks{} to gain access to the first footnote area
% a separate \thanks must be used for each paragraph as LaTeX2e's \thanks
% was not built to handle multiple paragraphs
%

\author{Miguel~{Heredia Conde},~\IEEEmembership{Member,~IEEE,}
        and~Otmar~Loffeld,~\IEEEmembership{Senior Member,~IEEE}% <-this % stops a space
\thanks{M. Heredia Conde and O. Loffeld are with the Center for Sensorsystems (ZESS), University of Siegen, Paul-Bonatz-Stra\ss e 9-11, 57076 Siegen, Germany. E-mail: \{heredia, loffeld\}@zess.uni-siegen.de.}% <-this % stops a space
%\thanks{Manuscript received MONTH DAY, 2017; revised MONTH DAY, 2017.}}
\thanks{A reduced version of this paper will be submitted to the IEEE Transactions on Signal Processing.}}

% note the % following the last \IEEEmembership and also \thanks - 
% these prevent an unwanted space from occurring between the last author name
% and the end of the author line. i.e., if you had this:
% 
% \author{....lastname \thanks{...} \thanks{...} }
%                     ^------------^------------^----Do not want these spaces!
%
% a space would be appended to the last name and could cause every name on that
% line to be shifted left slightly. This is one of those "LaTeX things". For
% instance, "\textbf{A} \textbf{B}" will typeset as "A B" not "AB". To get
% "AB" then you have to do: "\textbf{A}\textbf{B}"
% \thanks is no different in this regard, so shield the last } of each \thanks
% that ends a line with a % and do not let a space in before the next \thanks.
% Spaces after \IEEEmembership other than the last one are OK (and needed) as
% you are supposed to have spaces between the names. For what it is worth,
% this is a minor point as most people would not even notice if the said evil
% space somehow managed to creep in.

% The paper headers
%\markboth{IEEE Transactions on Signal Processing,~Vol.~00, No.~0, MONTH~DAY,~2017}%
\markboth{Arxiv, May~08,~2017}%
{{Heredia Conde} \MakeLowercase{and} Loffeld: Fast Approximate Construction of Best Complex Antipodal Spherical Codes}
%{{Heredia Conde} \MakeLowercase{\textit{et al.}}: Fast Approximate Construction of Best Complex Antipodal Spherical Codes}
% The only time the second header will appear is for the odd numbered pages
% after the title page when using the twoside option.
% 
% *** Note that you probably will NOT want to include the author's ***
% *** name in the headers of peer review papers.                   ***
% You can use \ifCLASSOPTIONpeerreview for conditional compilation here if
% you desire.

% If you want to put a publisher's ID mark on the page you can do it like
% this:
%\IEEEpubid{0000--0000/00\$00.00~\copyright~2015 IEEE}
% Remember, if you use this you must call \IEEEpubidadjcol in the second
% column for its text to clear the IEEEpubid mark.

% use for special paper notices
%\IEEEspecialpapernotice{(Invited Paper)}

% make the title area
\maketitle

% As a general rule, do not put math, special symbols or citations
% in the abstract or keywords.
\begin{abstract}

Compressive Sensing (CS) theory states that real-world signals can often be recovered from much fewer measurements than those suggested by the Shannon sampling theorem. Nevertheless, recoverability does not only depend on the signal, but also on the measurement scheme. The measurement matrix should behave as close as possible to an isometry for the signals of interest. Therefore the search for optimal CS measurement matrices of size $m\times n$ translates into the search for a set of $n$ $m$-dimensional vectors with minimal coherence. Best Complex Antipodal Spherical Codes (BCASCs) are known to be optimal in terms of coherence. An iterative algorithm for BCASC generation has been recently proposed that tightly approaches the theoretical lower bound on coherence. Unfortunately, the complexity of each iteration depends quadratically on $m$ and $n$.
In this work we propose a modification of the algorithm that allows reducing the quadratic complexity to linear on both $m$ and $n$. Numerical evaluation showed that the proposed approach does not worsen the coherence of the resulting BCASCs. On the contrary, an improvement was observed for large $n$.
The reduction of the computational complexity paves the way for using the BCASCs as CS measurement matrices in problems with large $n$. We evaluate the CS performance of the BCASCs for recovering sparse signals. The BCASCs are shown to outperform both complex random matrices and Fourier ensembles as CS measurement matrices, both in terms of coherence and sparse recovery performance, especially for low $m/n$, which is the case of interest in CS.

\end{abstract}

% Note that keywords are not normally used for peerreview papers.

% For peer review papers, you can put extra information on the cover
% page as needed:
% \ifCLASSOPTIONpeerreview
% \begin{center} \bfseries EDICS Category: 3-BBND \end{center}
% \fi
%
% For peerreview papers, this IEEEtran command inserts a page break and
% creates the second title. It will be ignored for other modes.
\IEEEpeerreviewmaketitle

\section{Introduction}
\label{int}
% The very first letter is a 2 line initial drop letter followed
% by the rest of the first word in caps.
% 
% form to use if the first word consists of a single letter:
% \IEEEPARstart{A}{demo} file is ....
% 
% form to use if you need the single drop letter followed by
% normal text (unknown if ever used by the IEEE):
% \IEEEPARstart{A}{}demo file is ....
% 
% Some journals put the first two words in caps:
% \IEEEPARstart{T}{his demo} file is ....
% 
% Here we have the typical use of a "T" for an initial drop letter
% and "HIS" in caps to complete the first word.
%\IEEEPARstart{T}{his} demo file is intended to serve as a ``starter file''
%for IEEE journal papers produced under \LaTeX\ using
%IEEEtran.cls version 1.8b and later.
%% You must have at least 2 lines in the paragraph with the drop letter
%% (should never be an issue)
%I wish you the best of success.
%
%\hfill mds
% 
%\hfill August 26, 2015
%
%\subsection{Subsection Heading Here}
%Subsection text here.
%
%% needed in second column of first page if using \IEEEpubid
%%\IEEEpubidadjcol
%
%\subsubsection{Subsubsection Heading Here}
%Subsubsection text here.

\IEEEPARstart{A} question that naturally arises by simply observing the nature is what is the best packing of $n$ points in an $m$-dimensional (hypo-)sphere, where $m$ is typically either two or three. \emph{Best} packing is to be understood as the arrangement that maximizes the minimum distance between points. Regular polygons, ubiquitously present in nature, are the best packing of $2$-dimensional points in the sphere of corresponding dimensionality, that is, the circle. Probably the most paradigmatic example is the hexagonal shape of honeycombs or paper wasp nests. Also the compound eyes of insects are packed hexagonally. Also some $3$-dimensional packing schemes seem to be present, e.\,g., in the arrangement of the places of exit on the surface of spherical pollen grains. It was the Dutch biologist Pieter M. L. Tammes \cite{Tammes30} who conjectured that, for a given size of the pollen grain and area required for each exit place, the arrangement was such that the number of exit places was maximized, i.\,e., it was the best packing of 3-dimensional points on the sphere.
In honor of Tammes, the problem of finding this optimal arrangement of points on a spherical surface or, equivalently, finding how many spherical caps of a given radius can be placed on a unit sphere without overlapping, is classically known as the \emph{Tammes' problem} or \emph{hard-spheres problem} \cite{Saff97}. This problem has been thoroughly analyzed and solutions exist for different numbers of points \cite{Mooers94,Fejes99}. A monograph on the pursuit of best packings with real life examples can be found in \cite{Weaire08}.

Note the close relationship between Tammes' problem and the so-called \emph{Thomson's problem}, namely, determining a stable distribution of electrons able to move freely on the surface of a $3$-dimensional sphere, motivated by the atom model first suggested about 1900 by Lord Kelvin and formally described later in \cite{Thomson04} by Sir J. J. Thomson. Several methods exist for accurately approaching solutions to Thomson's problem, e.\,g., based on random walk \cite{Weinrach90}, on steepest descent \cite{Erber91}, on constrained global optimization (CGO) \cite{Altschuler94}, using genetic algorithms \cite{Morris96}, a generalized simulated annealing algorithm \cite{Xiang97,Xiang00} (available as general-purpose software package \cite{Xiang13}) and Monte-Carlo-based simulations \cite{Wales06,Wales09}. 

At the light of the amount of research on this subject along the twentieth century, one is tempted to consider the problem of finding the best packing points on the sphere as solved. Nevertheless, such resolution is to be questioned when taking into consideration that this problem can be seen as a very specific instance of the more general problem of packing subspaces in Grassmann manifolds. A Grassmann manifold or \emph{Grassmannian} refers to a space parametrizing all linear subspaces of a vector space (which could be $\mathbb{R}^{m}$ or $\mathbb{C}^{m}$) of a given dimension. If the dimension of the subspaces is chosen to be one, the corresponding Grassmannian is the space of lines originating at the origin of the vector space. If the vector space is $\mathbb{R}^{m}$, this Grassmannian will be denoted as $G\left( \mathbb{R}^{m}, 1 \right)$ and finding the best packing of $n$ subspaces (lines) into it is equivalent to the problem of finding the optimal arrangement of $n$ $m$-dimensional points on the surface of the $m$-hypersphere. The general problem of subspace packing has been object of study since the 1960s \cite{Fejes65}, motivating a number of works which cannot be included here for space restrictions and the culmination of which can be seen in the detailed numerical study by J. H. Conway \cite{Conway96}.

Nowadays best subspace packings find multiple application areas, being of special interest in Multiple-Input Multiple-Output (MIMO) and Code Division Multiple Access (CDMA) wireless systems. In fact, communication strategies based on quantization via subspace packings have been incorporated into the IEEE wireless metropolitan area network (WirelessMAN\textsuperscript{TM}) standard \cite{WirelessMAN05}. %Importance for communications.
In general, it is clear that best packings are useful in any application where codes with low mutual coherence are required, since maximizing the minimal distance between codewords leads low mutual coherence. More specifically, codes with optimal or close-to-optimal mutual coherence have earned renewed importance thanks to the advent of the groundbreaking theory of Compressive Sensing (CS). According to this novel mathematical theory, many real-world signals can be exactly recovered from a number of measurements that is much lower than that suggested by the Shannon sampling theorem. CS takes profit of the fact that most real signals exhibit some structure and can be sparsely represented in an appropriate basis, in other words, they can be represented in a domain where their effective dimensionality is much lower than the original signal dimensionality. Nevertheless, some requirements are to be observed to attain successful signal reconstruction from few measurements, one of them being that the columns of the measurement matrix have to exhibit as low mutual coherence as possible. The measurement matrix models the linear transformation from the $n$-dimensional domain where the sparse signal lives to the $m$-dimensional measurement domain, with $m\ll n$. If intercolumn coherence is to be minimized, we face the problem of finding a set of $n$ $m$-dimensional codewords constituting an optimal code in terms of coherence.

In coding terminology, if the codewords lie on the surface of the unit sphere, they are known as \emph{Spherical Codes} (SC). We leave the details on the different classes of SCs and construction schemes for forthcoming sections. In general, an SC is called \emph{Best Spherical Code} (BSC) if the point arrangement it describes maximizes the minimum distances between points. At the time being, the sphere packings obtained by N. J. A. Sloane \cite{Sloane_spherical_codes} are often considered putative solutions to the problem, i.\,e., the definitive BSCs in real spaces. Nevertheless, as pointed out in \cite{Zoerlein15}, for a given problem dimensionality, the vector set of minimal coherence is given by a \emph{Best Complex Antipodal Spherical Code} (BCASC). In other words, the optimal CS measurement matrix in terms of coherence and supposing absence of \emph{a priori} information is given by a BCASC. Consequently, methods for generating BCASCs in an efficient manner are invaluably precious from a CS point of view.

In this work we propose an approximate implementation of the algorithm for constructing BCASCs proposed in \cite{Zoerlein15} which provides a considerable reduction of the complexity of the algorithm. This way, our approach can be used to generate BCASCs or, equivalently, CS measurement matrices, of very large dimensionality, thus paving the way to their implementation in real applications. We will also show that our approximation not only does not degrade the quality of the resulting BCASCs, but yields slightly better codes in terms of coherence for large $n$. The rest of the paper is structured as follows: Section~\ref{rel} reviews some of the related work, pointing out their main peculiarities. In Section~\ref{fun} we briefly provide the fundamentals we build our work upon, namely, Compressive Sensing, Spherical Codes and the Approximate Nearest Neighbor (ANN) search. Our approximate BCASC construction method is outlined and explained in Section~\ref{met}. The results of an experimental evaluation of the performance of our approach are presented in Section~\ref{exp}, where we compare to the original method in \cite{Zoerlein15}. The paper ends with the conclusions given in Section~\ref{con}.

\section{Related Work}
\label{rel}
The methods for construction BCASCs can be roughly divided in two categories: analytical approaches and numerical approaches. In principle, analytical approaches should be preferred over numerical methods, but in practice they can only provide solutions for specific combinations of number of codewords and codeword dimensionality. Additionally, analytical approaches may impose some further structure on the codewords, thus artificially reducing the degrees of freedom available to optimize the codes. If the aim is to generate BCASCs of arbitrary size, with special focus on large sizes, numerical methods are the only feasible way. In this work we build upon the recent BCASC construction algorithm proposed in \cite{Zoerlein15}. Nevertheless, there has been a number of previous works on numerical methods for approaching BCASCs. Note that many of the construction schemes that will be introduced in the following are not restricted to find BCASCs, but a MWBE (Maximum Welch Bound Equality) codebook or, in general, solutions to the subspace packing problem in Grassmann manifolds, which is the case in works related to MIMO communications. Nevertheless, for a subspace dimension of one, the latter general formulation boils down to the problem of packing Grassmanian lines, i.\,e., finding the best packing of points on the hypersphere.

In \cite{Hochwald00}, the authors draw from previous research on tight frame construction by R. Balan and I. Daubechies, who proposed constructing the tight frames by selecting the first $m$ components of the $n$-dimensional vectors of an $n\times n$ discrete Fourier transform (DFT). While this may be directly used for constructing $n$ $m$-dimensional complex codes for relatively low values of $m$, it is clear that when $m\to\infty$ the coherence between contiguous codewords tends to one. The novelty in \cite{Hochwald00} is that the selection of the $m$ DFT rows is not deterministic, but the result of an optimization process. The optimization consists in a random search that looks for the set of frequencies which yields minimal coherence between resulting codewords. In their simulations they consider $m=8$ and up to $n=2209$ codewords. 

The authors of \cite{Xia05} provide both an analytic construction of optimal codes approaching the Welch bound and a numerical search method for those cases for which analytical construction does not apply. The numerical method is based on Lloyd's algorithm. Lloyd's algorithm \cite{Lloyd82}, also known as Voronoi relaxation, is a well-known method for obtaining evenly spaced sets of points in subsets of Euclidean spaces. The original goal of the method was to attain an optimal finite quantization scheme, where optimality was meant in the sense of minimal average quantization error power. Given a probability density function, the quantization error power for each quantum partition takes the shape of a second order moment thereof, restricted to the domain covered by the partition. Consequently, if the quantum partitions are fixed, their optimal quantum values are their centers of mass, while if the set of quanta is fixed, the best partition for each quantum is given by those points of the domain for which the Euclidean distance with respect to the quantum value is lower than those obtained with respect to any other quantum. An iterative formulation with two alternative steps of center of mass calculation and construction of the partitions (Voronoi cells) converges to an optimal quantization scheme. For completeness, recall that well-known clustering approaches, such as $k$-means, are fully based on the Lloyd's algorithm, where the probabilistic distribution of data points is not given analytically, but as a set of particles. The K-SVD algorithm \cite{Aharon06} for dictionary learning is often regarded as an extension of $k$-means clustering and, consequently, also known as generalized Lloyd's algorithm (GLA). The idea in \cite{Xia05} was to use every quantizer codebook generated at each Voronoi iteration as candidate MWBE codebooks, supposing an uniform distribution of points on the unit hypersphere in $\mathbb{C}^{m}$. From all candidates, which could also come from many executions of Lloyd's algorithm, the complex codeword with smallest coherence was chosen as approximate MWBE codebook. In posterior work \cite{Xia06} the authors adopt the GLA for generating the codebooks and provide lower bounds on the rate-distortion performance. Another evolution of Lloyd's algorithm intended to generate optimal codebooks is proposed in \cite{Roh06}, where the objective is matrix quantization to minimize capacity loss.

Three different methods are provided in \cite{Gohary09} for generating evenly-distributed constellations of points on the complex unit hypersphere. The first is a greedy technique in which the constellation is constructed sequentially, while the second generates the entire constellation at once. The unaffordable computational cost of the latter alternative for large constellation sizes motivated a third approach, which consists of two sequential steps: generation of a low-size constellation using the latter approach and generation of the final constellation fusing several rotations of the initial one. The common denominator of the three approaches is that they all exploit
the smooth geometry of the Grassmann manifold and use derivative-based techniques to optimize a smooth cost function that approximates the original non-differentiable objective, in this case the pairwise chordal Frobenius norm, which is to be maximized. Similarly, in \cite{Agrawal01} the coherence, which is a non-smooth function to be minimized, is approached by a family of potential functions that are smooth. A smoothing approximation has also been used in \cite{Xavier08}, where the Euclidean distance is approximated by an hyperbolic function for solving the classical Tammes' problem.
Inspired by the objective-smoothing approach in \cite{Gohary09}, a construction method is proposed in \cite{Medra14} that uses a sequence of smooth approximations to the objective function. Since the smooth approximation functions have continuous first and second order derivatives, conventional smooth optimization techniques can be used to solve the problem. Several smooth approximations to the maximum distance (to minimize) and to the Fubini-Study distance are presented, highlighting the usual compromise between convergence rate and stability in the latter case. 

The construction method first introduced in \cite{Kammoun03} and further developed and analyzed in \cite{Kammoun07} exploits the fact that unitary transforms representing any point on the Grassmann manifold can be represented by the exponential of any element of the tangent space at the identity point. For clarity, in the rank one case we deal with in this paper this would mean an exponential of an $n$-dimensional vector with only $n-1$ nonzero components. This is due to the fact that we pack subspaces of dimension one (lines) on the Grassmannian, which yields a tangent space of dimensionality $n-1$. The core idea of the method is to design a coherent codebook in the tangent space, which yields an optimal non-coherent one on the Grassmannian. Further work exploiting the exponential parametrization of the Grassmannian can be found in \cite{Utkovski08}, where different rotated lattices are used as initial codes in the tangent space.

Provided that all rotations of a spherical code (or set of subspaces, in the case of non-unitary subspace dimensionality) around the origin are equivalent, in \cite{Dhillon08} each configuration of subspaces is associated with a block Gram matrix. Note that the off-diagonal elements of the Gram matrix are related to the distances between pairs of subspaces. The novelty in \cite{Dhillon08} is the use of an alternating projection scheme for approaching a solution to the best packing. The algorithm works directly on the Gram matrix and alternates two steps for enforcing both structural conditions and spectral conditions. In practice, the algorithm keeps two estimates of the Gramm matrix: one strictly satisfying the structural conditions and the other strictly satisfying the spectral conditions. The alternating scheme aims to minimizing the Frobenius norm of the difference, thus making both estimates recursively approach to each other. Clearly, the solution lies in the intersection of the two constrain sets where the temporal estimates live. The importance of \cite{Dhillon08} does not only lie in the novelty of the approach, but in the underlying connection to CS theory. One of the structural conditions is an upper bound on the norm of off-diagonal elements of the Gram matrix. In the specific case of line packing, i.\,e., subspaces of unit dimensionality, this is equivalent, in CS terms, to an upper bound on the coherence of the sensing matrix.

An expansion-compression algorithm (ECA) is proposed in \cite{Schober09} for finding packings in Grassmann manifolds. The ECA scheme seems to be motivated by the fact that using the chordal distance yields degenerated constellations if a simple max-min (maximization of the minimum distance between codewords) scheme is applied. To overcome this issue, an alternating scheme between a step of max-min and a subsequent step of min-max is proposed. The former step is called \emph{expansion} and the latter \emph{compression}. The authors observe that using the Fubini-Study distance as a metric degenerated constellations do not occur and one can do with a conventional max-min scheme, thus avoiding the compression step.

Building upon M. Elad's method for obtaining \emph{optimized projections} for CS \cite{Elad07}, the authors of \cite{Tsiligianni12} make use of frame theory concepts in order to construct tight frames that are the nearest to the obtained optimized projections. This results in significant mutual coherence reduction. In the sequel \cite{Tsiligianni14} the authors extended this work with an \emph{averaged projections} version of the algorithm instead of the initial \emph{alternating projections} formulation. The difference is that in the new version the temporal estimates of the Gram matrix obtained after applying each of the steps of the algorithm to the Gram matrix that resulted from the previous iteration are stored and the Gram matrix for the next iteration is obtained as the average of these stored \emph{projections}.
A further step into the very desirable overlap between optimal packings and CS was taken in \cite{Rusu13}. The author explicitly points out the equivalence between optimal unit-norm frame design and optimal packing of points on the unit sphere. Differently from the methods in \cite{Dhillon08} or \cite{Tsiligianni12}, which operate on the Gram matrix, a method called Iterative Decorrelation by Convex Optimization (IDCO) is proposed that works directly on the frame being optimized, i.\,e., on the code itself. The algorithm tries to find at each iteration a frame with lower coherence than the previous, while constraining the new frame to be close to the previous one. From an sphere packing point of view, the closeness constraint avoids that the spherical code being constructed rotates endlessly around the origin, precluding convergence, while the coherence minimization (a min-max scheme) forces the packing to be as uniform as possible. The performance of the algorithm was demonstrated creating frames of fairly large dimensionality.

%On Zörlein's work
Recently, a numerical method for computing BCASCs has been proposed in \cite{Zoerlein15}. The method can be regarded as an extension of the BASC construction in \cite{Lazic13} to the complex case, which, in turn, is based on Lazi\'{c}'s early works on the construction of BSCs \cite{Lazic86,Lazic88}. As it will be explained in Section~\ref{fun:SCs}, from a computational point of view, the novelty in \cite{Zoerlein15} is the introduction of an integral over all complex rotations of each temporal codeword that influence a specific codeword when computing the resulting force of the former over the latter. This is necessary to transfer the concept of antipodality to the complex case.

%Common problems
Some of the methods presented so far were either exclusively or, at least, initially designed to construct codes, packings, frames, sensing matrices, etc. in the real space. In some cases, complex extensions are available and, in fact, some of the previous methods were designed to work natively in complex space. Nevertheless, in such cases the computational cost is higher and the methods are only evaluated for relatively low numbers of codewords of low dimensionality. There is, indeed, a lack of work where constructions of large close-to-optimal complex codebooks are presented. To the best of our knowledge, the tables in \cite{Zoerlein15} provide the most complete comparative benchmark so far and are limited to codes with $m\leq 64$ codewords of dimensionality $n\leq 4$ (or only $m=16$ for $n=5$).

\section{Fundamentals}
\label{fun}
In the following we will provide the fundamentals of Compressive Sensing (CS) theory (Section~\ref{fun:CS}), a new sensing paradigm that aims to overcome the explosion in the volume of data acquired by high resolution sensors operating at the Nyquist rate. The exposition of the CS sensing model will immediately lead to the necessity of a matrix with low intercolumn coherence, that is, sets of vectors with low mutual coherence. The search of the set of vectors with lowest coherence brings the link to spherical codes, which will be presented and analyzed in detail in Section~\ref{fun:SCs}. As the methods for obtaining such set increase their ability to approach the theoretical lower bound on coherence, they also get more complex in computational terms and approximate methods become necessary. In Section~\ref{fun:ANN} we will give an overview of Approximate Nearest Neighbor (ANN) searches, the core idea for speeding up the construction of BCASCs.

\subsection{Compressive Sensing}
\label{fun:CS}
The Shannon sampling theorem is the keystone of classical sampling theory and states that a continuous signal is completely determined by a number of equidistant samples acquired at a rate that is twice the maximum frequency contained in the signal. While this is true and actually the basis of almost every signal acquisition system, it implies the generation of huge data volumes when the signals to sense are allowed to have large bandwidth. Such volumes of data are too large to be stored directly, often too large to be even transmitted, and thus a process of data compression right after sensing becomes necessary. In this scenario, \emph{compressive} (or \emph{compressed}) sensing (CS) \cite{Candes06a,Candes06b,Donoho06,Baraniuk07,Davenport10} arises as a a novel sensing paradigm, in which the compression is performed \emph{at sensing} and not immediately afterwards, thus obtaining fewer but more informative measurements.

While the Shannon sampling theorem requires the signal to be bandlimited, CS theory imposes the more general requirement of being \emph{sparse} in some known representation basis. More specifically, CS states that a finite-dimensional signal that admits a sparse or compressible representation in some known basis or tight frame can be exactly recovered from few non-adaptive measurements if certain conditions are satisfied. We restrict our attention to discrete signals with real or complex coefficients. Let $\vec{x}\in \mathbb{C}^{n}$ be the vector corresponding to a discrete signal. Then, the $l_0$ norm of $\vec{x}$ is defined as:

\begin{equation} \label{eq:l_0_norm}
\left\|\vec{x}\right\|_0\coloneqq \lim_{p\rightarrow 0} \left\|\vec{x}\right\|_{p}^{p}=|\supp{(\vec{x})}|
\end{equation}

\noindent that is, the cardinality of the support of $\vec{x}$, and $\vec{x}$ is called an $s$-sparse signal if:

\begin{equation} \label{eq:sparsity}
\left\|\vec{x}\right\|_0\leq s
\end{equation}

\noindent in other words, if $\vec{x}$ has, at maximum, $s$ non-zero elements. Provided that we know that the signal obeys a sparsity constraint, the challenge is to reconstruct it from a reduced number of linear measurements $m\ll n$. Thus, the classic CS measurement model is a severely underdetermined linear system of the form:

\begin{equation} \label{eq:CS_linear_model}
\vec{y} = \pmb{A} \vec{x}
\end{equation}

\noindent where $\pmb{A}\in \mathbb{C}^{m\times n}$ denotes the \emph{measurement matrix}, which explains how the vector of measurements $\vec{y}\in \mathbb{C}^{m}$ relates to the signal $\vec{x}$. This matrix is often the composition of a \emph{sensing matrix}, which models the real sensing process and a representation basis or \emph{dictionary}, which enables recovering signals that are not sparse, but admit a sparse representation. Clearly, provided that $m\ll n$, the challenge is to design $\pmb{A}$ in such a way that preserves the information in $\vec{x}$, despite the dimensionality reduction, in other words, that condenses a limited information content living in a high-dimensionality space into a space of lower dimensionality. %For this to be feasible,  ...

The linear system in Eq.~\ref{eq:CS_linear_model} is massively underdetermined and, therefore, cannot be solved directly. Solving Eq.~\ref{eq:CS_linear_model} via least squares would spread the power in the measurement vector over all coefficients of $\vec{x}$, yielding dense solution that does not fit our \emph{a priori} knowledge on the sparsity of $\vec{x}$. Therefore, we should rather look for the solution with minimal $l_0$ norm, between all the possible solutions satisfying Eq.~\ref{eq:CS_linear_model}. Unfortunately, it is well known that finding a solution to a constrained $l_0$ minimization is, in general, NP-hard \cite{Muthukrishnan03}. In the CS literature there are two main approaches to overcome this difficulty: either to relax the $l_0$ minimization into an $l_1$ minimization, or to use a \emph{greedy} algorithm to find the sparse support of the signal in an incremental fashion. Recovery guarantees are typically given under the assumption of solving Eq.~\ref{eq:CS_linear_model} by $l_1$ minimization, which yields:

\begin{equation} \label{eq:CS_l_1_min}
\hat{\vec{x}}=\argmin_{\vec{x}}\left\|\vec{x}\right\|_1 \ \text{subject to} \ \vec{y} = \pmb{A} \vec{x}
\end{equation}

Eq.~\ref{eq:CS_l_1_min} can be efficiently solved as a linear program \cite{Chen98}. A fundamental question in CS is under which conditions Eq.~\ref{eq:CS_l_1_min} is equivalent to the desired $l_0$ minimization \cite{Donoho06B}. The success of the signal recovery process, provided that the sparsity constraint is satisfied, depends on whether $\pmb{A}$ possesses some properties. The most intuitive property is the so-called \emph{Null Space Property} (NSP) \cite{Cohen09}, which is based on the analysis of the null space of $\pmb{A}$. There exist several formulations of the NSP, being the following one obtained in the common case of measuring the approximation errors in terms of $l_2$ norm: the matrix $\pmb{A}$ is said to satisfy the the NSP of order $k$ if: 

\begin{equation} \label{eq:NSP_l2-1}
\left\| \vec{x}_{\Omega_k} \right\|_{2}\leq \left\| \vec{x} \right\|_{2}\leq C_0 \frac{\sigma_k\left(\vec{x}\right)_{1}}{\sqrt{k}} = C_0 \frac{\left\| \vec{x}_{\bar{\Omega}_k} \right\|_{1}}{\sqrt{k}}, \ \vec{x}\in \mathcal{N}\left(\pmb{A}\right),
\end{equation}

\noindent where $\mathcal{N}\left(\pmb{A}\right)$ denotes the null space of $\pmb{A}$, $\Omega_k$ denotes any set of $k$ support indices, and $\bar{\Omega}_k$ the complement set of $\Omega_k$, so that $\vec{x}_{\bar{\Omega}_k}$ denotes the vector $\vec{x}$ with support restricted to $\bar{\Omega}_k$, with $\left|\Omega_k\right|=k$. $\sigma_k\left(\vec{x}\right)_{1}=\left\| \vec{x}_{\bar{\Omega}_k} \right\|_{1}$ is the so-called best $k$-term approximation error. The NSP prescribes that vectors in $\mathcal{N}\left(\pmb{A}\right)$ should not be too concentrated in a small subset of indices \cite{Eldar12}, i.\,e., should be as dense as possible. Satisfaction of the NSP of order $2s$ can be linked to the satisfaction of an upper bound on the recovery error of compressible signals and exact recovery of $s$-sparse signals.

A different condition for recoverability is given by the well-known \emph{Restricted Isometry Property} (RIP), also known as \emph{Uniform Uncertainty Principle} (UUP), first introduced in \cite{Candes05} and further analyzed in \cite{Candes08B}. A matrix $\pmb{A}$ is said to satisfy the RIP of order $k$ if there exists a constant $\delta_k\in (0,1)$ such that

\begin{equation} \label{eq:RIP}
(1-\delta_k) \left\|\vec{x}\right\|_2^2\leq \|\pmb{A}\vec{x}\|_2^2 \leq (1+\delta_k) \left\|\vec{x}\right\|_2^2, \ \forall \vec{x}\in\Sigma_k
\end{equation}

\noindent where $\Sigma_k$ is the subset of all $k$-sparse vectors and $\delta_k$ is known as the $k$-\emph{restricted isometry constant}. The RIP ensures that the sensing matrix is close to an isometry for $k$-sparse vectors, i.\,e., that the transformation preserves the $l_2$ distances between pairs of $k$-sparse vectors to some extent. If $\pmb{A}$ satisfies the RIP of order $2s$ with $\delta_{2s}$ low enough, e.\,g., $\delta_{2s}<\sqrt{2}-1$ from \cite{Candes08B} or $\delta_{2s}<\frac{3}{4+\sqrt{6}}$ from \cite{Foucart10B}, then successful recovery of the $s$-sparse vector $\vec{x}$ via $l_1$ minimization is guaranteed.
The RIP of order $k$ implies that any set of $k$ columns of $\pmb{A}$ chosen at random should behave approximately as an orthonormal system \cite{Candes05B}. In an orthonormal system, every basis vector is perfectly incoherent with all others, thus yielding null coherence. For completeness, we provide the following definition of coherence for the case of an arbitrary set of $n$ vectors of dimension $m$, which are to be identified with the columns of $\pmb{A}$ and in CS is often named \emph{matrix coherence} \cite{Candes11} of $\pmb{A}$:

\begin{equation} \label{eq:coherence_A}
\mu\left(\pmb{A}\right)=\max_{u<v\leq n}\frac{\left|\left\langle\vec{a}_u,\vec{a}_v \right\rangle\right|}{\|\vec{a}_u\|_2 \|\vec{a}_v\|_2}
\end{equation}

\noindent where $\left\langle\cdot,\cdot \right\rangle$ denotes (complex) scalar product. Clearly, in the typical case of normalized vectors, $\mu\left(\pmb{A}\right)$ is equivalent to the maximum scalar product between vector pairs and, consequently, only dependent on the angles between them. In other words, finding the optimal CS measurement matrix with normalized columns is equivalent to finding the arrangement of points on the unit sphere that maximizes the minimal angle between them when regarded as vectors. In terms of coherence, we face a minimization problem, which, in combination with Eq.~\ref{eq:coherence_A} yields the following min-max formulation:

\begin{equation} \label{eq:min_coherence_A}
\mu_{\mathrm{min}}\left(\pmb{A}\right)=\min_{\pmb{A}\in\mathbb{C}^{m\times n}}{\mu\left(\pmb{A}\right)}=\min_{\pmb{A}\in\mathbb{C}^{m\times n}}{\max_{u<v\leq n}\frac{\left|\left\langle\vec{a}_u,\vec{a}_v \right\rangle\right|}{\|\vec{a}_u\|_2 \|\vec{a}_v\|_2}}
\end{equation}

The remaining question is how lower coherence translates into a better RIP. As observed in \cite{Eldar12}, one can make use of the \emph{Ger\u{s}gorin circle theorem} (Theorem~2 of \cite{Gershgorin31}), which states that the eigenvalues of a matrix $\pmb{G}\in \mathbb{R}^{n\times n}$ lie in the union of $n$ discs, $\displaystyle\bigcup_{i=1}^{n} d_i\left(c_i,r_i\right)$, centered at $c_i=g_{i,i}$ and with radius $r_i=\sum_{j\neq i}\left|g_{i,j}\right|$. Bounding the eigenvalues of the partial $k\times k$ Gram matrix $\pmb{G}_{\Omega_k}=\pmb{A}_{\Omega_k}^{\ast} \pmb{A}_{\Omega_k}$ obtained from the columns of $\pmb{A}$ indexed by an arbitrary set of indices $\Omega_k$ with $\left|\Omega_k \right|=k$, one can directly obtain the restricted isometry constant for Eq.~\ref{eq:RIP}. More specifically, if the columns of $\pmb{A}$ are of unit norm, one can show that $\pmb{A}$ satisfies the RIP of order $k$ with $\delta_{k}$ given by:

\begin{equation} \label{eq:RIC_from_coherence}
\delta_{k}=(k-1)\mu\left(\pmb{A}\right), \ \forall k<1/\mu\left(\pmb{A}\right)
\end{equation}

\subsection{Spherical Codes and Optimal Coherence}
\label{fun:SCs}
\subsubsection{Minimum Coherence}
\label{fun:SCs:coh}
If the goal is to obtain a set of vectors with optimal, i.\,e., lowest, coherence, as defined above, then a fundamental question is what is the value of this minimum coherence. Several lower bounds on the coherence exist in the literature that depend only on the problem dimensionality, that is, on the number of vectors $n$ and their dimensionality $m$. Obviously, for $n\leq m$ one can select the $n$ vectors from an $m$-dimensional orthonormal basis, thus attaining zero coherence. Lower bounds are only meaningful for $n>m$. A well-known lower bound for the coherence of $\pmb{A} \in \mathbb{R}^{m\times n}$ is the Welch bound \cite{Welch74}, named after L. Welch, also known as \emph{simplex} bound, and given by

\begin{equation} \label{eq:Welch_bound}
\mu\left(\pmb{A}\right)\geq\sqrt{\frac{n-m}{m(n-1)}}
\end{equation}

The Welch bound can also be considered to have been implicitly given in \cite{Rankin56} for the case of real vectors, despite the bound in Theorem~5 of \cite{Rankin56} was originally an upper bound for mean square value of the sine of the angle between vectors in the set. When rewritten in terms of cosine it would yield a lower bound on the \emph{mean square} coherence, while the coherence defined in Eq.~\ref{eq:coherence_A} is rather a \emph{maximum} coherence. In principle one would like to fulfill Eq.~\ref{eq:Welch_bound} with equality or as close as possible to equality. Two different criteria exist to determine whether the Welch bound is satisfied with equality. One possibility considers the root mean square (RMS) value of the inner product between vectors in the set as coherence and names the set as Welch Bound Equality (WBE) sequence \cite{Massey93} if Eq.~\ref{eq:Welch_bound} holds with equality. Alternatively, if one adopts the more strict definition of coherence given by Eq.~\ref{eq:coherence_A}, sets of vectors satisfying Eq.~\ref{eq:Welch_bound} with equality receive the name of Maximum Welch Bound Equality (MWBE) sequences \cite{Sarwate99}. In this work and for coherence with CS literature we restrict our attention to the \emph{maximum} coherence defined in Eq.~\ref{eq:coherence_A} and, therefore, we focus on finding MWBE sequences. Provided that the coherence defined in Eq.~\ref{eq:coherence_A} has to be necessarily greater than or equal to the RMS counterpart, MWBE sequences are a subset of WBE sequences. The validity of the Welch bound is restricted to $n$ not too large with respect to $m$. More specifically, the following necessary conditions can be derived from the absolute bounds for $A$-sets given in \cite{Delsarte75} (second and third rows of the Table~II for the real and complex case, respectively):

\begin{equation} \label{eq:Welch_bound_nec_conds}
\begin{split}
n&\leq\frac{m(m+1)}{2}, \ \text{for} \ \pmb{A}\in\mathbb{R}^{m\times n} \\
n&\leq m^2, \ \text{for} \ \pmb{A}\in\mathbb{C}^{m\times n}
\end{split}
\end{equation}

If $n$ is greater than the upper bounds in Eq.~\ref{eq:Welch_bound_nec_conds}, the $n$ columns of $\pmb{A}$ cannot form an equiangular system \cite{Bodmann15} and Eq.~\ref{eq:Welch_bound} no longer applies. In such cases, one can adopt the \emph{orthoplex} bound (initially stated in \cite{Conway96} in terms of chordal distance for the case of real-valued vectors and extended later to the complex case \cite{Pitaval11}), given by:%or \emph{Rankin} bound?!

\begin{equation} \label{eq:orthoplex_bound}
\mu\left(\pmb{A}\right)\geq\frac{1}{\sqrt{m}}
\end{equation}

Similarly to the Welch bound, the orthoplex bound can only be achieved for some values of $n$ that depend on $m$, namely if the following bounds hold:

\begin{equation} \label{eq:orthoplex_bound_nec_conds}
\begin{split}
\frac{m(m+1)}{2}&<n\leq(m+1)(m+2), \ \text{for} \ \pmb{A}\in\mathbb{R}^{m\times n} \\
m^2&<n\leq2(m^2-1), \ \text{for} \ \pmb{A}\in\mathbb{C}^{m\times n}
\end{split}
\end{equation}

For cases of very large $n$, when Eq.~\ref{eq:orthoplex_bound_nec_conds} cannot be fulfilled, one can still adopt the lower bound developed by Kabatiansky and Levenshtein \cite{Kabatiansky78,Levenshtein83} for $n\to\infty$:

\begin{equation} \label{eq:Levenshtein_bounds}
\mu\left(\pmb{A}\right)\geq\begin{cases}
\sqrt{\frac{3n-m^2-2m}{(m+2)(n-m)}}, \ \text{for} \ \pmb{A}\in\mathbb{R}^{m\times n} \\
\sqrt{\frac{2n-m^2-m}{(m+1)(n-m)}}, \ \text{for} \ \pmb{A}\in\mathbb{C}^{m\times n}
\end{cases}
\end{equation}

Alternatively, for large $n$ one can also make use of the lower bound given in Eq.~\ref{eq:Mukkavilli_bound}, originally derived in \cite{Mukkavilli03} for the case of low-coherence beamformer codebooks and explicitly presented as a lower bound on the coherence in \cite{Xia05}.

\begin{equation} \label{eq:Mukkavilli_bound}
\mu\left(\pmb{A}\right)\geq1-2n^{-\frac{1}{m-1}}
\end{equation}

Taken all previous theoretical bounds into consideration, one can formulate a composite lower bound on the coherence \cite{Zoerlein15}, which for the complex case reads:

\begin{equation} \label{eq:composite_bound}
\begin{split}
\mu\left(\pmb{A}\right)&\geq\begin{cases}
\sqrt{\frac{n-m}{m(n-1)}}, \ \text{for} \ n\leq m^2 \\
\max{\left\{\frac{1}{\sqrt{m}},\sqrt{\frac{2n-m^2-m}{(m+1)(n-m)}}, 1-2n^{-\frac{1}{m-1}}\right\}}, \\ \text{for} \ m^2<n\leq2(m^2-1) \\
\max{\left\{\sqrt{\frac{2n-m^2-m}{(m+1)(n-m)}}, 1-2n^{-\frac{1}{m-1}}\right\}}, \\ \text{for} \ n>2(m^2-1)
\end{cases} \\
\forall\pmb{A}&\in\mathbb{C}^{m\times n}
\end{split}
\end{equation}

Note that the given bounds on coherence often derive from more general upper bounds on the minimum of some distance (be geodesic, chordal, etc.) between subspaces packed in some Grassmann manifold. The relatively simple expressions we provide for the bounds are for the case of packing subspaces of unit dimensionality (lines) and thus do not show dependency on this parameter.

\subsubsection{Best Spherical Codes}
\label{fun:SCs:BSC}
In this work we deal with a particular case of the Grassmannian subspace packing problem, namely, finding the best packing of $n$ one-dimensional subspaces in $\mathbb{R}^{m}$ or $\mathbb{C}^{m}$. This is known as the Grassmannian line packing problem, where the corresponding Grassmannians are denoted by $G\left(\mathbb{R}^{m}, 1 \right)$ and $G\left(\mathbb{C}^{m}, 1 \right)$, respectively, often simplified to $G\left(m, 1 \right)$ in the real case. It is known that in this case the metrics used to determine the best packing (e.\,g., geodesic, chordal) are equivalent \cite{Conway96} and thus its selection does not influence the result. Solving this problem is equivalent to solving the optimization in Eq.~\ref{eq:min_coherence_A} and yields the desired MWBE sequences.

Without loss of generality, we restrict our attention to codes with normalized codewords, in other words, to points on the surface of the unit sphere. Thanks to the normalization factors in the definition of coherence (Eq.~\ref{eq:coherence_A}), arbitrary scaling of the vectors has no effect on the coherence. Already in Section~\ref{int} it was introduced that a \emph{Spherical Code} (SC) refers to a code whose codewords lie on the surface of the unit sphere. We deal with the general case of codewords with complex entries and will drop the reference to the full space $\mathbb{C}^{m}$ out of the notation. Thus a SC of $n$ codewords in $\mathbb{C}^{m}$ is denoted by $C_{\mathrm{S}}\left(m,n\right)$. An SC is called \emph{Best Spherical Code} (BSC) and denoted $C_{\mathrm{BS}}\left(m,n\right)$ if the point arrangement it describes maximizes the minimum distances between points. A BSC is defined by its distribution of distances between codewords, rather than by the codewords themselves, and all rotations of the code around the origin are regarded as the same BSC.

Regarding the construction of the BSCs, a numerical method was proposed in \cite{Lazic88} for the real-valued case. In this method the points are regarded as charged particles confined on some spherical surface of unit radius. Each particle suffers the effect of the repelling fields generated by all other particles. Consequently, the particles will move until the system reaches some local minimum of the potential energy. The mutual interaction between particles is described by a generalized potential function \cite{Lazic86}. For some initial arrangement described by the SC $C_{\mathrm{S}}\left(m,n\right)$, this generalized potential function reads

\begin{equation} \label{eq:g_BSCs}
g_{\nu}\left(C_{\mathrm{S}}\left(m,n\right)\right)=\sum_{u<v}^{n}\frac{1}{\|\vec{c}_{u}-\vec{c}_{v}\|_{2}^{\nu-2}}
\end{equation}

\noindent where $\nu\in\mathbb{N}$, $\nu>2$, is a custom exponent and $\vec{c}_k$ denotes the $k^{\mathrm{th}}$-codeword (column when structured as a matrix). As already pointed out in \cite{Lazic86}, as $\nu$ increases, the difference between the maximum minimal distance between codewords and the minimal distance attained letting the system reach equilibrium (minimize $g_{\nu}\left(C_{\mathrm{S}}\left(m,n\right)\right)$) decreases. Consequently, as $\nu\to\infty$, the obtained arrangement tends exactly to a BSC. In order to minimize Eq.~\ref{eq:g_BSCs} constrained to $\|\vec{c}_k\|$ one can apply the method of Lagrange multipliers. We omit here further details for brevity, but the derivations can be found in \cite{Lazic86,Lazic88,Lazic13,Zoerlein15}, for instance. The solution can be expressed by two equivalent equilibrium formulas \cite{Lazic13}, which map the BSC into itself. The first and most intuitive form is the so-called Equilibrium of Rescaled Differences of (code) Words (ERDW), which expresses the equilibrium of mutual forces between codewords, given by \cite{Lazic88}:

\begin{equation} \label{eq:ERDW}
\left\{\underline{\vec{c}}_{u}=\underline{\sum_{v\neq u}\frac{\underline{\vec{c}}_{u}-\underline{\vec{c}}_{v}}{\left\|\underline{\vec{c}}_{u}-\underline{\vec{c}}_{v}\right\|_{2}^{\nu}}}%=\underline{\sum_{v\neq u}\vec{\delta}_{uv}}
\right\}_{u=1}^{n}
\end{equation}

\noindent where the underline denotes unit vectors, i.\,e., $\underline{\vec{u}}=\frac{\vec{u}}{\left\|\vec{u}\right\|_{2}}$. The second option is called Equilibrium of Rescaled code Words (ERW) and dispenses with the first term in the summation's numerator. This formulation exploits the fact that properly rescaled BSC codewords sum up to zero and reads

\begin{equation} \label{eq:ERW}
\left\{\underline{\vec{c}}_{u}=\underline{\sum_{v\neq u}\frac{-\underline{\vec{c}}_{v}}{\left\|\underline{\vec{c}}_{u}-\underline{\vec{c}}_{v}\right\|_{2}^{\nu}}}%=\underline{\sum_{v\neq u}\vec{\sigma}_{v}}
\right\}_{u=1}^{n}
\end{equation}

Constructing the BSC means finding the fixed point of either the ERDW or the ERW mapping. For coherence with \cite{Zoerlein15} but without loss of generality, we adopt the ERDW equilibrium formulation. During construction, the right hand side of Eq.~\ref{eq:ERDW} can be regarded as the set of aggregate forces $\underline{\vec{f}}_{u}$ acting on each of the codewords, $1\leq u\leq n$. Finding the fixed point of a mapping of the form

\begin{equation} \label{eq:BSC_mapping}
\pmb{F}\left[C_{\mathrm{S}}\left(m,n\right)\right]=\left\{\underline{\vec{f}}_{u}\left(C_{\mathrm{S}}\left(m,n\right)\right)\right\}_{u=1}^{n}
\end{equation}
 
\noindent turns out to be unfeasible, since, regardless of the initial SC used as starting point, the sequence of temporal solutions given by the mapping does not converge. This holds both for ERDW and ERW mappings and is a direct consequence of the fact that such formulations allow the SCs to rotate endlessly around the origin, provided that there is no additional constraint that ties the SC to itself. In order to solve this issue, one can add a term to the mapping that partially \emph{anchors} the codewords to themselves, that is, a one-to-one mapping. This idea was already proposed in the original formulation of \cite{Lazic86} and yields the mapping

\begin{equation} \label{eq:BSC_mapping_damped}
\pmb{F}_{\alpha}\left[C_{\mathrm{S}}\left(m,n\right)\right]=\left\{\underline{\underline{\vec{c}}_{u}+\alpha\underline{\vec{f}}_{u}}\right\}_{u=1}^{n}
\end{equation}
 
\noindent where  $\alpha\in\mathbb{R}^{+}$ is a \emph{damping factor}. The mapping in Eq.~\ref{eq:BSC_mapping_damped} exhibit the same fixed points as that in Eq.~\ref{eq:BSC_mapping}. The iterative process

\begin{equation} \label{eq:BSC_iter}
C_{\mathrm{S}}\left(m,n\right)^{(k+1)}=\pmb{F}_{\alpha}\left(C_{\mathrm{S}}\left(m,n\right)^{(k)}\right), \ k\geq 0
\end{equation}

converges to a fixed point of Eq.~\ref{eq:BSC_mapping_damped}, which is also a fixed point of Eq.~\ref{eq:BSC_mapping}, i.\,e., a local minimum of Eq.~\ref{eq:g_BSCs}. Finding the fixed point for an arbitrarily large $\nu$ implies getting arbitrarily close to the BSC in terms of minimal distance between codewords.

\subsubsection{Best Antipodal Spherical Codes}
\label{fun:SCs:BASC}
A BSC exhibits the lowest possible maximal inner product between its codewords and one is tempted to think that it constitutes a solution to Eq.~\ref{eq:min_coherence_A}. This would be indeed the case if there was no absolute value in the numerator of the argument to be optimized. The existence of the absolute value implies that the BSC cannot be taken directly as the matrix $\pmb{A}$ with lowest worst-case coherence. In order to attain full equivalence between the problem of finding the optimal arrangement of points on the surface of the unit sphere and the packing problem in, e.\,g, $G\left(\mathbb{R}^{m}, 1 \right)$, one has to necessarily consider the \emph{antipodal} of each codeword as a point of the packing \cite{Sloane97}. In the real case, for a codeword $\vec{a}_u\in\mathbb{R}^{m}$, the antipodal would be $-\vec{a}_u$.
BSCs constructed this way were named \emph{Best Antipodal Spherical Codes} (BASC) in \cite{Lazic13}, where the method for generating BSCs by computing the fixed point of a continuous mapping introduced above is conveniently modified to generate BASCs.

The adaptation to the BASC case requires expanding the ERDW equilibrium formulation to include the antipodal codewords, which in the case of real-valued coefficients yields:

\begin{equation} \label{eq:ERDW_antipodal}
\left\{\underline{\vec{c}}_{u}=\underline{\sum_{v\neq u}\left(\frac{\underline{\vec{c}}_{u}-\underline{\vec{c}}_{v}}{\left\|\underline{\vec{c}}_{u}-\underline{\vec{c}}_{v}\right\|_{2}^{\nu}}+\frac{\underline{\vec{c}}_{u}+\underline{\vec{c}}_{v}}{\left\|\underline{\vec{c}}_{u}+\underline{\vec{c}}_{v}\right\|_{2}^{\nu}}\right)}%=\underline{\sum_{v\neq u}\vec{\delta}_{uv}}
\right\}_{u=1}^{n}
\end{equation}

Due to the antipodal codewords, a BASC has a number of codewords that is actually twice the number of columns of the desired optimal measurement matrix, i.\,e., $2n$. Nevertheless, half of the codewords are given by the other half and, additionally, if the starting point is an antipodal SC, the antipodal symmetry of Eq.~\ref{eq:ERDW_antipodal} ensures that antipodality will be preserved through iterations. In other words, the underlying number of degrees of freedom of the problem remains $n$ (still, in an $m$-dimensional space), despite $2n$ points will be effectively packed on the unit sphere. For these reasons and in an effort to be faithful to CS notation, we will denote the \emph{Antipodal Spherical Codes} (ASC) by $C_{\mathrm{AS}}\left(m,n\right)$, despite we know that the total number of codewords is $2n$. The extension to the complex case will support this decision.

For the real-valued case, solving the iterative sequence introduced in Eq.~\ref{eq:BSC_iter} using the right hand of Eq.~\ref{eq:ERDW_antipodal} as set of forces $\left\{\underline{\vec{f}}_{u}\right\}_{u=1}^{n}$ in the mapping of Eq.~\ref{eq:BSC_mapping_damped} yields an ASC that arbitrarily approaches the BASC as $\nu\to\infty$. Then any of the two possible subsets of $n$ non-antipodal codewords in the BASC yields the desired optimal matrix $\pmb{A}$ by columns.

\subsubsection{Best Complex Antipodal Spherical Codes}
\label{fun:SCs:BCASC}
Many real world problems naturally admit a complex formulation, which often simplifies the representation and processing of signals and its understanding. Time-frequency signal processing makes recurrent use of the Fourier basis (be partial or complete, continuous or discrete), which is natively complex. In CS one of the paradigmatic deterministic sensing matrices is the Fourier ensemble, obtained selecting $m$ rows from a DFT matrix, thus a complex-valued matrix. Furthermore, the lower bounds on the coherence given in Section~\ref{fun:SCs:coh} evidence the higher potential of complex-valued codes to minimize the coherence, since the validity ranges of the most attractive (lowest) lower bounds get extended to greater values of $n$. It is thus desirable to extend the concept of BASC to the complex case, so that the composite lower bound in Eq.~\ref{eq:composite_bound} can be tightly approached.

The extension of BASC to the complex case was accomplished in \cite{Zoerlein15}, yielding the advent of \emph{Best Complex Antipodal Spherical Codes} (BCASC). The main question is how the antipodal codewords are generated in the complex-valued case. In the real-valued case the antipodal of each codeword is trivially given by the opposite vector, but in complex domain, which embeds not only one but two dimensions per coefficient, the change of vector sense (sign of the coefficients) translates into a complex rotation of the coefficients. Then a complex codeword $\vec{a}_u\in\mathbb{C}^{m}$ would yield a set of antipodals of the form $\left\{\exp{(\phi i)} \vec{a}_u\right\}_{0\leq\phi<2\pi}$, where $\phi$ is the complex rotation angle and $i$ denotes from now on the complex unit. This means introducing a continuous parameter $\phi\in [0,2\pi)$, which requires a summation over an infinite number of antipodals, i.\,e., an integral, when calculating the forces. Therefore, the two-term sum within the summation in Eq.~\ref{eq:ERDW_antipodal} for the real case becomes an integral in complex phase domain in the complex case, yielding:

\begin{equation} \label{eq:ERDW_complex_antipodal}
\left\{\underline{\vec{f}}_{u}=\underline{\displaystyle\sum_{v\neq u}{\int_{\phi=0}^{2\pi}\frac{\underline{\vec{c}}_{u}-\underline{\vec{c}}_{v}e^{\phi i}}{\left\|\underline{\vec{c}}_{u}-\underline{\vec{c}}_{v}e^{\phi i}\right\|_{2}^{\nu}}d\phi}}%=\underline{\sum_{v\neq u}\vec{\delta}_{uv}}
\right\}_{u=1}^{n}
\end{equation}

As already pointed out in \cite{Zoerlein15}, the integral in Eq.~\ref{eq:ERDW_complex_antipodal} is not easy to solve analytically due to the norm in the denominator of the fraction. Two different options are considered in that work to circumvent this issue: numerical integration using the QAG adaptive integration from the GSL \cite{Galassi09} or casting the integral into a discrete summation for some $K$ discrete values of $\phi$, equally-spaced between $0$ and $2\pi$. The QAG algorithm is an iterative integration procedure in which the integration region is divided into subintervals and at each iteration the subinterval responsible for the largest integration error is bisected, so that the error gets reduced. While this procedure yields accurate approximations, it might become too expensive in computational terms. For this reason we also focus here on the approximate summation case, supposing that $K$ is large enough to ensure the desired accuracy. In this case the integral in Eq.~\ref{eq:ERDW_complex_antipodal} boils down to a discrete summation and yields:

\begin{equation} \label{eq:ERDW_complex_antipodal_approx_sum}
\left\{\underline{\vec{f}}_{u}=\underline{\displaystyle\sum_{v\neq u}{\sum_{k=0}^{K}\frac{\underline{\vec{c}}_{u}-\underline{\vec{c}}_{v}e^{\frac{2\pi k}{K} i}}{\left\|\underline{\vec{c}}_{u}-\underline{\vec{c}}_{v}e^{\frac{2\pi k}{K} i}\right\|_{2}^{\nu}}}}%=\underline{\sum_{v\neq u}\vec{\delta}_{uv}}
\right\}_{u=1}^{n}
\end{equation}

At the light of Eq.~\ref{eq:ERDW_complex_antipodal} or its simplification in Eq.~\ref{eq:ERDW_complex_antipodal_approx_sum} for $K$ large enough, it becomes clear that the following equivalence property between complex rotations of \emph{Complex Antipodal Spherical Code} (CASC) codewords holds:

\begin{equation} \label{eq:CASC_equivalence_prop}
\vec{c}_{u}\equiv\vec{c}_{u}e^{\phi i}, \ 1\leq u<n
\end{equation}

A CASC is denoted by $C_{\mathrm{CAS}}\left(m,n\right)$ when it originates from $n$ $m$-dimensional codewords. Note that in the calculus of the forces a number of $(K-1)$ antipodals is considered for each of the $n$ initial codewords (Eq.~\ref{eq:ERDW_complex_antipodal_approx_sum}) or even an infinite number (Eq.~\ref{eq:ERDW_complex_antipodal}), but in the notation we only write $n$ and not $Kn$ or $\infty$, for the same reason that in the real-valued ASCs we wrote $n$ and not $2n$, despite the antipodals.
Using Eq.~\ref{eq:ERDW_complex_antipodal} (or Eq.~\ref{eq:ERDW_complex_antipodal_approx_sum} for sufficiently large $K$) in the mapping of Eq.~\ref{eq:BSC_mapping_damped} yields a CASC that arbitrarily approaches the BCASC as $\nu\to\infty$.
Considering the complex antipodals enables that the resulting BCASC $C_{\mathrm{BCAS}}\left(m,n\right)$ can be directly considered as the optimal CS measurement matrix $\pmb{A}\in\mathbb{C}^{m\times n}$ in terms of coherence. In general the BCASC will not be attained exactly, but closely approximated, and thus we obtain a \emph{close-to-optimal} solution rather than the absolute optimum.

%\subsubsection{On the Advantage of Best Complex Antipodal Spherical Codes over Random Matrices}
%\label{fun:SCs:BCASCvsRandom}

\subsection{Approximate Nearest Neighbors}
\label{fun:ANN}
Any of the formulas for calculating the forces acting over each codeword of the different SCs introduced in Section~\ref{fun:SCs} exhibit the same general form: a summation whose terms have a distance between two $m$-dimensional vectors to the power of $\nu$ as denominator, with $\nu$ typically large. Regardless of any underlying physical meaning of the calculus, it becomes clear that such a denominator makes summands with low distances between vectors have a much more relevant contribution to the sum than summands with larger distances, whose contribution might be negligible. This observation relates to the core of our approximate method for constructing BCASCs and will be further developed in Section~\ref{met}. For now it is only important to realize the interest of an approach for finding the vectors in some given vector set, say $V : |V|=n, \ V\subset \mathbb{R}^{m}$, that are within a certain neighborhood of some query vector $\vec{v}_{\mathrm{q}}\in\mathbb{R}^{m}$, where $m$ can be relatively large and $n>m$.

The problem of searching the \emph{Nearest Neighbors}, from now on NN, to a given query point in high-dimensional spaces appears often in computer vision problems, most specifically when trying to match visual features between images or between an image and some database of features. Visual features are, in general, points or regions of an image that are considered to be easily recognizable or descriptive of the image content. Feature matching requires computing a vector that describes each feature, known as its \emph{descriptor}, and the problem reduces to finding the best match (or some candidate set of low cardinality) for it. Typical descriptor dimensionalities are $m=128$ (as originally considered for SIFT \cite{Lowe04}) or even larger (as studied for SURF \cite{Bay08}). When matching against feature descriptors from another image, the number of candidates, say $n$, can reach hundreds and easily get over a million when matching against large databases or feature maps.

This minimalistic reference to the feature matching problem is not just intended to highlight the necessity of efficient methods for performing NN searches in high-dimensional spaces, but also to clarify the connection to our case of study. State-of-the-art methods for generating BCASCs \cite{Zoerlein15} cannot cope with large values of $m$ and $n$, due to the square dependency of the complexity on the dimensionality. Similarly, regarding feature matching, it is turns out that there is often no algorithm available that is able to find the nearest neighbor with a computational complexity that is lower than a linear search. Unfortunately, the cost of a linear search often becomes unaffordable for large $n$ and $m$, especially if real-time requirements are to be fulfilled, as it is the case in many computer vision applications. In this context algorithms for performing \emph{approximate} NN (ANN) searches arise as a feasible alternative to performing an exact NN search. When properly tuned these methods are able to speed up the NN searches several orders of magnitude with respect to a linear search. There has been a large amount of research on algorithms for performing ANN searches, a review thereof falls out of the scope of this paper. A common objective is avoiding the linear dependency of the algorithm complexity on $n$ \cite{Har-Peled12}, e.\,g., aiming for a logarithmic dependency. The reader is referred to \cite{Bajramovic06,Ponomarenko14,Li16} for comparisons between different ANN search approaches.

For completeness, finding the $K$ NN of a query point $\vec{v}_{\mathrm{q}}\in\mathbb{R}^{m}$ in the set of vectors $V\subset \mathbb{R}^{m}$ means finding the subset $V_{\mathrm{NN}}\subset V$ with $\left|V_{\mathrm{NN}}\right|=K$ satisfying:

\begin{equation} \label{eq:K-NN}
d\left(\vec{v}_{u}, \vec{v}_{\mathrm{q}}\right)\leq d\left(\vec{v}_{v}, \vec{v}_{\mathrm{q}}\right), \forall \vec{v}_{u}\in V_{\mathrm{NN}}, \vec{v}_{v}\in V \backslash  V_{\mathrm{NN}}%, \vec{v}_{u},\vec{v}_{v}\in V
\end{equation}

\noindent where $d(\cdot)$ denotes some distance function between two points. In this work we restrict our attention to the Euclidean distance, but considering other distances might be a subject of future work. If we are interested in retrieving $K$ ANN, we implicitly allow that the obtained set of neighbors, say $V_{\mathrm{ANN}}$ with $\left|V_{\mathrm{ANN}}\right|=K$, is not exactly equivalent to the set of exact NN $V_{\mathrm{NN}}$. An appropriate measure of the quality of the approximation would be the ratio between the cardinality of the intersection between $V_{\mathrm{ANN}}$ and $V_{\mathrm{NN}}$ and the cardinality of these sets, namely $K$. This can be used to define $V_{\mathrm{ANN}}$ by means of an lower bound on it, which would read:

\begin{equation} \label{eq:K-ANN}
\frac{\left|V_{\mathrm{ANN}} \cap V_{\mathrm{NN}} \right|}{K}\geq\varepsilon
\end{equation}

\noindent where $0<\varepsilon\leq 1$ is a custom parameter that sets the approximation error we can tolerate, typically close to $1$.
Once some approximation parameter is set, the challenge is to design a search structure that allows finding a subset $V_{\mathrm{ANN}}$ satisfying Eq.~\ref{eq:K-ANN} as fast as possible. Provided that the NN are defined in terms of distances to the query point, good search structures will be those clustering the points in $V$ using the same metric $d(\cdot)$. The most widely-used NN search structure is the $k$-d tree \cite{Friedman77}, which is able to obtain search times of $\mathcal{O}(\log{n})$, being the computational cost of the tree construction $\mathcal{O}(mn\log{n})$. The $k$-d tree construction was modified in \cite{Arya98} to allow for ANN searches. If $K=1$ (or, in general, low), i.\,e., if only a single NN is required, the bound in Eq.~\ref{eq:K-ANN} is meaningless and the $\varepsilon$-ANN of $\vec{v}_{\mathrm{q}}$ is given by the point $\vec{v}_{\mathrm{ANN}}$ satisfying:

\begin{equation} \label{eq:ANN}
d\left(\vec{v}_{\mathrm{ANN}}, \vec{v}_{\mathrm{q}}\right)\leq (1+\varepsilon) d\left(\vec{v}_{\mathrm{NN}}, \vec{v}_{\mathrm{q}}\right)
\end{equation}

\noindent where $\vec{v}_{\mathrm{NN}}$ denotes the true NN of $\vec{v}_{\mathrm{q}}$ and $\varepsilon$ should be close to zero. The corresponding ANN search locates first the leaf cell containing $\vec{v}_{\mathrm{q}}$ (in $\mathcal{O}(\log{n})$ time) and then the leaf cells are visited in increasing order of distance from $\vec{v}_{\mathrm{q}}$. The parameter $\varepsilon$ allows interrupting the search when the distance between $\vec{v}_{\mathrm{q}}$ and the current leaf cell exceeds $d\left(\vec{v}_{\mathrm{ANN}}^{\ast}, \vec{v}_{\mathrm{q}}\right)/(1+\varepsilon)$, where $\vec{v}_{\mathrm{ANN}}^{\ast}$ denotes the closest point found so far, and $\vec{v}_{\mathrm{ANN}}=\vec{v}_{\mathrm{ANN}}^{\ast}$ is delivered as solution.
A similar modification of the $k$-d tree for ANN search is described in \cite{Beis97}, where the stopping criterion is given by the maximum number of leaf nodes to examine, which yields better performance than the upper bound on the distance error used in \cite{Arya98}.

When the dimensionality $m$ is large, a $k$-d tree requires an excessive number of points ($n=2^{m}$) in order to have splitting nodes covering every dimension. For lower $n$ the performance of the tree may be far from optimal. As an alternative, multiple randomized $k$-d trees can be created from different rotations of the data \cite{Silpa-Anan04,Silpa-Anan08} or, alternatively, randomly selecting the dimension along which the data is divided from a set of few dimensions in which the data exhibits high variance. This way, each $k$-d tree covers a different set of dimensions in terms of splitting. Additionally, searches on the different trees can be considered as independent \cite{Silpa-Anan08}, thus yielding an exponential decrease of the error with the number of searched nodes.

In \cite{Fukunaga75} the authors propose to construct the search tree by recursive $k$-means clustering of the data, where $k$ denotes the branching factor or number of disjoint clusters to establish at each node of the tree. A similar idea underlies the Geometric Near-neighbor Access Tree (GNAT) proposed in \cite{Brin95}, which avoids the multiple distance computations required to determine the mean point of the clusters taking data points as the cluster centers. The authors also propose adapting the branching factor or degree of each node to the number of points it contains, still keeping a constant average degree. A \emph{hierarchical} $k$-means tree similar to that in \cite{Fukunaga75} is proposed in \cite{Nister06}.

Other search structures and evolutions of the previous exist, but are not included for brevity. Even restricting our attention to the few classical approaches highlighted above, the truth is that there is no structure that can be considered \emph{best} in absolute terms, since their performance largely vary with the nature of the data, its dimensionality $m$, the number of points in the database $n$, and the specific values of the algorithm parameters. The nature of the data relates to the underlying dimensionality of the subspace (or union thereof) where the data lives. If the entire dataset is contained into some $k$-dimensional subspace with $k<m$, the performance will be similar to that obtained using $k$-dimensional data, rather than $m$-dimensional data. Fortunately, a procedure for automatic selection of the optimal algorithm and its parameters is presented in \cite{Muja09}.

The authors consider the algorithm itself one more parameter of a generic NN search routine, whose optimal value for some dataset is to be found. At the light an evaluation of different alternatives for constructing the trees, two methods repeatedly showed the best performance, namely, the \emph{hierarchical $k$-means tree} and the \emph{randomized $k$-d trees}. The automatic algorithm selection and tuning procedure only considers these two alternatives and requires from the user the value of two parameter that control the importance of the tree build time relative to the search time and the importance of the memory
overhead compared to the time overhead, respectively. This algorithm was made available in the nowadays famous Fast Library for Approximate Nearest Neighbors (FLANN) \cite{Muja13_FLANN}, and is the general-purpose ANN search algorithm we implement in our approach\footnote{The authors are aware that the FLANN library has been recently updated to novel search structures showing superior performance in \cite{Muja14}, namely the \emph{randomized $k$-d forest} and the \emph{priority search k-means tree}. Nevertheless, such update is still missing in the latest version in \cite{Muja13_FLANN}, namely, the version 1.8.4, which we use in this work.}.
%Despite one can argue that more recent tree construction methods exist that may outperform a classical $k$-d tree structure, experimental evaluation has shown that such methods, e.\,g., metric trees or cover trees do not perform better than $k$-d trees in general \cite{Kibriya07}.

\section{Methodology}
\label{met} 
The method in \cite{Zoerlein15} was shown to produce BCASCs that closely approach and eventually meet the theoretical lower bounds on coherence. The algorithm is a numerical search whose structure naturally derives from the definition of BCASCs. In spite of approximating the integrals over complex rotations that needs to be evaluated to obtain the forces by discrete summations, the computational load of the method becomes intractable for large numbers of codewords of large dimensionality. This is due to the fact that the complexity of each iteration for a fixed $\nu$ scales asymptotically with $\mathcal{O}(m^2 n^2 K)$, that is, the computational cost grows linearly with the number of discrete complex rotations considered in the approximate integration $K$, but quadratically with the codeword dimensionality $m$ and the number of codewords $n$. The quadratic dependency on $n$ follows directly from the fact that, for each of the $n$ codewords, the difference with respect to each of the $(n-1)K\sim n K$ complex rotations of the other codewords are to be computed. The quadratic dependency on $m$ is due to implementation details and can be circumvented. For a reference on execution times, the reader is referred to Table 6.3 of \cite{Zoerlein15B}, where the method introduced in \cite{Zoerlein15} with approximate integration requires over an hour to construct a BCASC for $n=64$ codewords of dimensionality $m=4$. Without implementing the approximation the original method took nearly $18$ hours for constructing the same BCASC.

Our goal is to achieve the excellent results in \cite{Zoerlein15}, but drastically reducing the execution time that the original algorithm requires for large values of $m$ and $n$, so that the range of applications gets broader. In CS it is often the case that, despite the underlying dimensionality of the problem to solve (say, sparsity or, in general, signal complexity) might be low, the ambient dimension might be very large, e.\,g., in case of a fine discretization of the whole ambient space. In such cases, the method in \cite{Zoerlein15} cannot be directly applied to generate close-to-optimal sensing matrices because of the square dependency on $n$. Furthermore, the required number of measurements $m$ to attain successful reconstruction depends on $\log{n}$ and will be also affect the execution time quadratically.
Additionally, it is a known issue of most methods for generating spherical codes or, in general, best packings, that the performance degrades as the number of points to pack increases, while the computational effort grows \cite{Dhillon08}.
In order to extend the applicability of the BCASCs as CS sensing matrices for real-world problems of large $n$, we need to get rid of the squares in $m$ and $n$, so that the complexity of each iteration reduces to $\mathcal{O}(m n K)$, while still preserving the adequate algorithmic structure presented in \cite{Zoerlein15}.

An analysis of the original program for generating BCASCs with approximate summation whose performance is evaluated in \cite{Zoerlein15,Zoerlein15B} reveals that the square dependency of the complexity on $m$ is implementation-dependent rather than specific to the algorithm. We observe that an optimization of the code easily reduces the square dependency to a linear dependency, thus yielding complexity $\mathcal{O}(m n^2 K)$ per iteration. We would like to stress that this optimized version of the original code does not imply an approximation by any means, since the mathematical operations leading to the construction of the code remain the same and, consequently, the resulting codes are identical up to machine precision. From now on we take this improved implementation as reference and it will be named simply as the BCASC construction algorithm, while any reference to the original implementation will be made explicit.

The square dependency on $n$ is indeed specific to the algorithm, since all other codewords contribute to generate the force that \emph{move} the codeword under consideration from one iteration to the next. Therefore, this dependency cannot be softened without incurring an approximation. More specifically, for each codeword one would need to select few other codewords or, better said, complex rotations thereof, that actually contribute to the calculus of the force that actuates over the codeword under consideration, having the rest no effect in this regard. This is the core idea of this work: how to select few (eventually complex-rotated) codewords that are responsible for the greatest contributions in the force calculus step in such a way that the quality of the final BCASC does not degrade with respect to the reference approach, without this approximation. To this end, we propose using an ANN search to decide which summands in the double summation of Eq.~\ref{eq:ERDW_complex_antipodal_approx_sum} are to be included in the force calculus and which ones can be neglected.

In this work we establish the nearest neighbors to some query codeword $\vec{c}_{\mathrm{q}}\in\mathbb{C}^{m}$ using the Euclidean distance and use FLANN to retrieve them. Note that we deal with complex-valued vectors, while FLANN is implemented for the real-valued case. For simplicity, let $\vec{c}_{u,v,k}\in\mathbb{C}^{m}$ denote the difference vector between the codeword $\vec{c}_{u}$ and the $k^{\mathrm{th}}$ complex rotation of some other codeword $\vec{c}_{v}$. If we denote by $\vec{c}_{u,v,k}^{\Re}=\Re{\left(\vec{c}_{u,v,k}\right)}\in\mathbb{R}^{m}$ the vector of real components and by $\vec{c}_{u,v,k}^{\Im}=\Im{\left(\vec{c}_{u,v,k}\right)}\in\mathbb{R}^{m}$ the vector of real components, we can \emph{pack} the initial complex vector $\vec{c}_{u,v,k}\in\mathbb{C}^{m}$ into a real vector of double dimensionality $\vec{c}_{u,v,k}'=\left[\vec{c}_{u,v,k}^{\Re \top}, \vec{c}_{u,v,k}^{\Im \top}\right]^{\top}\in\mathbb{R}^{2m}$. This is just an operational detail that only affects the way the data is stored, the tree construction and the ANN searches, but not the algorithm itself. Observe that Euclidean distances in this induced $2m$-dimensional real space are equivalent to those in the original $m$-dimensional complex space:

\begin{equation} \label{eq:complex2real}
\begin{split}
\left\|\vec{c}_{u,v,k}\right\|_{2}^{2}&=\vec{c}_{u,v,k}^{\ast}\vec{c}_{u,v,k}=\sum_{w=1}^{m}c_{u,v,k,w}^{\ast}c_{u,v,k,w} \\
&=\sum_{w=1}^{m} \left[\Re{\left(c_{u,v,k,w}\right)}\right]^2 + \left[\Im{\left(c_{u,v,k,w}\right)}\right]^2 \\
&=\left\|\vec{c}_{u,v,k}'\right\|_{2}^{2}
\end{split}
\end{equation}

\noindent where $c_{u,v,k,w}$ denotes the $w^{\mathrm{th}}$ component of $\vec{c}_{u,v,k}$. This domain change implies that in practice we construct the FLANN index with vectors in $\mathbb{R}^{2m}$. The number of points in the index is not $n$, but $Kn$ due to the double summation in Eq.~\ref{eq:ERDW_complex_antipodal_approx_sum}. The set of all complex rotations of all codewords will be denoted by $V$.
From now on $K$ exclusively denotes the number of summands effectively used in the calculus of the forces, that is, $K=\left| V_{\mathrm{ANN}}\right|$ with $V_{\mathrm{ANN}}\subset V$, and no longer denotes the number of discrete rotations considered when approximating the integral on complex rotation domain by a summation, which we will further denote by $n_{\mathrm{rot}}$. Then $\left| V\right|=n n_{\mathrm{rot}}$. Note that in \cite{Zoerlein15} $K=n n_{\mathrm{rot}}$, according to this redefinition, while we will see that $K\ll n n_{\mathrm{rot}}$ in our approach.

For the sake of generality, we consider two possible ways of finding the ANN, namely, by means of a \emph{radius search} and by means of a \emph{KNN search}. In the exact search case, the KNN delivers the subset $V_{\mathrm{NN}}$ of cardinality $K$ satisfying Eq.~\ref{eq:K-NN}. In the approximate case, we obtain a set $V_{\mathrm{ANN}}$ for which the middle term of Eq.~\ref{eq:K-ANN} might be lower than $1$. In such cases one may require that Eq.~\ref{eq:ANN} holds for each of the $K$ neighbors. A radius search delivers a set $V_{\mathrm{NN}}$ such that:

\begin{equation} \label{eq:radius-NN}
\begin{split}
d\left(\vec{v}_{u}, \vec{v}_{\mathrm{q}}\right)&\leq r, \forall \vec{v}_{u}\in V_{\mathrm{NN}} \\
d\left(\vec{v}_{v}, \vec{v}_{\mathrm{q}}\right)&> r, \forall \vec{v}_{v}\in V \backslash V_{\mathrm{NN}}
\end{split}
\end{equation}

\noindent where $r$ denotes the radius, i.\,e., delivers the set of points that are contained in the ball of radius $r$ centered at the query point. This way the resulting $0\leq K<n n_{\mathrm{rot}}$ depends on $0\leq r\leq 2$. Regardless of whether a radius or a KNN search is performed, one ends up with some set of $K$ NN $V_{\mathrm{ANN}}$ and  Eq.~\ref{eq:ERDW_complex_antipodal_approx_sum} boils down to a single $K$-term summation:

\begin{equation} \label{eq:ERDW_complex_antipodal_approx_sum_ANN}
\left\{\underline{\vec{f}}_{u}=\underline{\displaystyle\sum_{\underline{\vec{v}}_{k}\in V_{\mathrm{ANN}}\left(\underline{\vec{c}}_{u}\right)}{\frac{\underline{\vec{c}}_{u}-\underline{\vec{v}}_{k}}{\left\|\underline{\vec{c}}_{u}-\underline{\vec{v}}_{k}\right\|_{2}^{\nu}}}}%=\underline{\sum_{v\neq u}\vec{\delta}_{uv}}
\right\}_{u=1}^{n}
\end{equation}

\noindent where $V_{\mathrm{ANN}}\left(\underline{\vec{c}}_{u}\right)$ simply denotes the set of ANN obtained for the query point $\underline{\vec{c}}_{u}$, whose cardinality $K_{u}$ may vary with $u$ in the radius search case and is fixed to $K_{u}=K, \forall u$ in the KNN search case.
At this point we are prepared to understand the algorithm for ANN-BCASC construction, which is given in Algorithm~\ref{alg:ANN-BCASC}.

%FLANN and implementation details TODO!
\begin{algorithm}
\caption{Approximate Nearest Neighbor - Best Complex Antipodal Spherical Code (ANN-BCASC)}
\label{alg:ANN-BCASC}
\begin{algorithmic}[1]
\INITIALIZE $\alpha=\alpha_{0}, \ \nu=\nu_{0}, \ C_{\mathrm{S}}\left(m,n\right)=\left\{\underline{\vec{c}}_{u}\right\}_{u=1}^{n}\in\mathbb{C}^{m\times n}$	\label{alg:ANN-BCASC:line:0}
\WHILE{$\nu<\nu_{\mathrm{max}}$}	\label{alg:ANN-BCASC:line:1}
	\STATE $FixedPoint_{u}=\FALSE, \ \forall 1\leq u\leq n$	\label{alg:ANN-BCASC:line:2}
	\STATE $\tau=0$	\label{alg:ANN-BCASC:line:3}
	\STATE Initialize damping factors: $\alpha_{u}=\alpha, \ \forall 1\leq u\leq n$	\label{alg:ANN-BCASC:line:4}
	\WHILE{$\left(\tau<\tau_{\mathrm{max}}\right) \AND \left(\left(\overset{n}{\underset{u=1}{\wedge}} FixedPoint_{u}\right)=\FALSE\right)$}	\label{alg:ANN-BCASC:line:5}
		\STATE FLANN index: $\mathfrak{T}\left(V\right), \ V=\left\{\underline{\vec{c}}_{u}e^{\frac{2\pi k}{n_{\mathrm{rot}}} i}\right\}_{\substack{{1\leq u\leq n} \\ {0\leq k \leq n_{\mathrm{rot}}}}}$	\label{alg:ANN-BCASC:line:6}
		\STATE Find ANNs in $\mathfrak{T}\left(V\right)$: $V_{\mathrm{ANN}}\left(\underline{\vec{c}}_{u}\right), \ \forall 1\leq u\leq n$	\label{alg:ANN-BCASC:line:7}
		\FOR{$u=1; u\coloneqq u+1$ \TO $u=n$}	\label{alg:ANN-BCASC:line:8}
			\STATE Calculate forces: $\underline{\vec{f}}_{u}=\underline{\displaystyle\sum_{\underline{\vec{v}}_{k}\in V_{\mathrm{ANN}}\left(\underline{\vec{c}}_{u}\right)}{\frac{\underline{\vec{c}}_{u}-\underline{\vec{v}}_{k}}{\left\|\underline{\vec{c}}_{u}-\underline{\vec{v}}_{k}\right\|_{2}^{\nu}}}}$	\label{alg:ANN-BCASC:line:9}
			\IF{$\left(\left\|\Delta\underline{\vec{f}}_{u}\right\|_{2}<\varepsilon_{\Delta\underline{\vec{f}}}\right) \AND \left(2\alpha_{u}<1 \right)$}	\label{alg:ANN-BCASC:line:10}
				\STATE $\alpha_{u}\coloneqq 2\alpha_{u}$	\label{alg:ANN-BCASC:line:11}
			\ELSE	\label{alg:ANN-BCASC:line:12}
				\STATE $\alpha_{u}\coloneqq\frac{\alpha_{0}}{\nu-1}$	\label{alg:ANN-BCASC:line:13}
			\ENDIF	\label{alg:ANN-BCASC:line:14}
			\STATE Apply forces: $\underline{\vec{c}}_{u}=\underline{\underline{\vec{c}}_{u}+\alpha_{u}\underline{\vec{f}}_{u}}$	\label{alg:ANN-BCASC:line:15}
			\IF{$\left\|\underline{\vec{f}}_{u}-\underline{\vec{c}}_{u}\right\|_{2}<\varepsilon$}	\label{alg:ANN-BCASC:line:16}
				\STATE $FixedPoint_{u}=\TRUE$	\label{alg:ANN-BCASC:line:17}
			\ENDIF	\label{alg:ANN-BCASC:line:18}
		\ENDFOR	\label{alg:ANN-BCASC:line:19}
	\STATE $\tau\coloneqq\tau+1$	\label{alg:ANN-BCASC:line:20}
	\ENDWHILE	\label{alg:ANN-BCASC:line:21}
	\STATE Update $\nu$: $\nu\coloneqq 2\nu$	\label{alg:ANN-BCASC:line:22}
	\STATE Update $\alpha$: $\alpha\coloneqq\frac{\alpha_{0}}{\nu-1}$	\label{alg:ANN-BCASC:line:23}
\ENDWHILE	\label{alg:ANN-BCASC:line:24}
\RETURNBIS $C_{\mathrm{BCAS}}\left(m,n\right)=\left\{\underline{\vec{c}}_{u}\right\}_{u=1}^{n}$	\label{alg:ANN-BCASC:line:25}
\end{algorithmic}
\end{algorithm}

The initial value of $\alpha$ is relatively close to one, e.\,g., $\alpha_{0}=0.9$, so that the codewords are allowed to experience large changes when applying the forces (line~\ref{alg:ANN-BCASC:line:15}). Complementary, the initial value of the exponent $\nu$ has to be low, typically $\nu_{0}=2$, which is the minimum feasible value. At the end of each iteration of the outer \emph{while} loop (line~\ref{alg:ANN-BCASC:line:1}), $\nu$ is doubled (line~\ref{alg:ANN-BCASC:line:22}) and $\alpha$ is conveniently decreased (line~\ref{alg:ANN-BCASC:line:23}), so that $\alpha\to 0$ as $\nu\to\infty$. Before starting the inner \emph{while} loop (line~\ref{alg:ANN-BCASC:line:5}) the damping factor to be used for each codeword is the same for all codewords and equal to $\alpha$. This loop is executed until all codewords are considered a fixed point or some maximum number of iterations $\tau_{\mathrm{max}}$ is reached. Within the loop, first the search structure is created using FLANN. It can be either a randomized $k$-d tree or a $k$-means tree, depending on the data to index $V$, and is denoted by $\mathfrak{T}\left(V\right)$. In line~\ref{alg:ANN-BCASC:line:9} Eq.~\ref{eq:ERDW_complex_antipodal_approx_sum_ANN} is used to compute the approximate force acting on each codeword.
We preserve the idea of \emph{accelerating on straight lines} from \cite{Zoerlein15}: if the direction of $\underline{\vec{f}}_{u}$ did not change between consecutive iterations, the corresponding damping factor $\alpha_{u}$ is increased even within the inner loop (line~\ref{alg:ANN-BCASC:line:10}). The threshold $\varepsilon_{\Delta\underline{\vec{f}}}$ regulates the sensitivity of the direction change detection. A codeword is considered a fixed point if it does not differ from the force acting over it. In practice, a threshold $\varepsilon$ on the $l_2$ norm of the difference is applied (line~\ref{alg:ANN-BCASC:line:16}).
If $\varepsilon$ is sufficiently small and $\tau_{\mathrm{max}}$ not too restrictive, the set of codewords obtained after execution of Algorithm~\ref{alg:ANN-BCASC} can be considered a good approximation of the true BCASC.

%On initialization
In Algorithm~\ref{alg:ANN-BCASC} the initial complex spherical code $C_{\mathrm{S}}\left(m,n\right)$ is randomly generated. Eventual appearance of too close codewords in the initial code can thus occur, yielding an inappropriate seed for the algorithm. One could avoid this situation by implementing some sort of intelligent initialization procedure, e.\,g., generating candidate complex codewords randomly, but only adding them to the initial code if the distances to the previously selected codewords are greater than a threshold, or, equivalently, if the corresponding scalar products are lower than a threshold, similarly to the initialization in \cite{Dhillon08}. %In practice, for fairly large dimensionalties the initialization has no effect on the result and a random construction suffices.
%On freezing the algorithm
Both in \cite{Zoerlein15} and in our improvement the forces acting over the codewords tend to zero as the exponent $\nu$ tends to infinity. It is natural to see here a connection to the \emph{cooling} process in simulated annealing methods \cite{Khachaturyan79,Khachaturyan81}. We would also like to point out a parallelism to the method for optimal frame design in \cite{Rusu13}, which starts with a loose constraint on the variation between consecutive frames, which gets tightened along iterations, progressively \emph{freezing} the process.

Interestingly, after designing and evaluating the proposed approach, we found out that the idea of restricting the calculus of the forces to some \emph{zone of significant influence} around the point under consideration is already contained in \cite{Lazic88}. Differently, in \cite{Lazic88} the codewords lie in a real space of very low dimensionality, while our approach was conceived to deal with codes living in high-dimensional complex spaces. More specifically, in \cite{Lazic88} points are excluded from the force calculus if their influence is weaker than a custom threshold. For a given value of $\nu$ this restriction immediately translates into a radius constraint and is thus equivalent to an NN radius search. Nevertheless, the equivalent search radius depends on $\nu$ and, consequently, in the first stages of the algorithm, where $\nu$ is still very small compared to its final value, the equivalent search radius is very large and most (eventually all) other points are considered in the calculus. This is only motivated by the fact that for small $\nu$ both close and far points yield comparable influences in terms of forces, while, as $\nu$ increases, the influence of far points become negligible compared to that of close ones. Nevertheless, the experience demonstrates that this computational overhead is not really necessary. Fixing a relatively small region of influence from the beginning yields better results than even considering all other points, as our results show. This is due to the fact that the crucial question is not whether the influence is \emph{negligible} or not, but rather whether it is \emph{convenient} or not. Indeed, for low $\nu$ the influence of some points might not be numerically negligible, despite not being close points, but their influence can be neglected because it is not a convenient one. Heuristically, we know that only few neighboring points should influence the displacement of the point under consideration at each iteration, while the aggregate influence of all other points can be regarded as some sort of background clutter that should be ignored.

%On locality
Note that locality, namely, the restriction of the points that will affect a specific codeword to a certain neighborhood of that codeword, is already present in \cite{Xia05}, since the Voronoi cells are defined according to the distance to the previously obtained center of mass. Points not belonging to the new cell will not contribute anymore to updating its center of mass. Unfortunately, integrating probability distributions on high-dimensional complex spaces over Voronoi cells defined by multiple (also highly-dimensional) hyperplanes is not a trivial task. Differently, we do not work with the underlying uniform probability distribution on the complex unit sphere whose optimal quantizer codebook is expected to be an excellent MWBE. We work only with the temporal solutions to such quantizers, that is, with the temporal BCASCs, which (progressively) distill the aforementioned underlying distribution by acquiring an even distribution themselves when regarded as particles. Note that this formulation brings locality back into the game, but without the massive growth of computational cost that the integration of an spherically-supported probabilistic distribution over Voronoi cells would require. Note that tree structures are the natural search structures for efficiently implementing locality in the way we do.

%On degradation of results/performance with number of points (included)
%It is a known issue of most methods for generating spherical codes or, in general, best packings, that the performance degrades as the number of points to pack increases, while the computational effort grows \cite{Dhillon08}.

%Link:
%in the single-user case one might seek to constrain the elements of the codewords to have constant modulus, so as to avoid power imbalances at the transmitter %from \cite{Medra14}

%On complexity
Regarding the complexity of each inner iteration of Algorithm~\ref{alg:ANN-BCASC}, note that an ANN search is to be carried out for each of the $n$ codewords (line~\ref{alg:ANN-BCASC:line:7}). Provided that each search in the tree has a reduced complexity $\mathcal{O}(\log{n n_{\mathrm{rot}}})$, this yields a total search complexity of $\mathcal{O}(n\log{n n_{\mathrm{rot}}})$. Despite this varies from one search structure to another, the complexity of building the tree $\mathfrak{T}$ from the data (line~\ref{alg:ANN-BCASC:line:6}) can be considered to be $\mathcal{O}(m n n_{\mathrm{rot}}\log{(n n_{\mathrm{rot}})})$. The rest of the algorithm has complexity $\mathcal{O}(nK)$ or, more generally, $\mathcal{O}\left(\displaystyle\sum_{u=1}^{n}\left|V_{\mathrm{ANN}}\left(\underline{\vec{c}}_{u}\right)\right|\right)$, due to the force calculus (line~\ref{alg:ANN-BCASC:line:9}). Note that there is no dependency on $m$ because all necessary distances have been already computed in the ANN search. In practice a very small $K\ll n$ is selected and the total complexity of each iteration of the innermost loop of the algorithm can be considered to be $\mathcal{O}(m n n_{\mathrm{rot}}\log{(n n_{\mathrm{rot}})})$ for some specific values of $\nu$ and $K$, regardless of $K$.

\section{Experimental Evaluation}
\label{exp}
In this section we present the results of different experiments destined to evaluate the performance of the ANN-BCASC algorithm. The experiments are divided in three groups. In Section~\ref{exp:ref} we compare both the performance of our approach to that of the reference algorithm it derives from, namely \cite{Zoerlein15}, both in terms of execution time and coherence of the resulting BCASCs. In Section~\ref{exp:tab} the scope is widened in order to offer a general comparison to other approaches for close-to-optimal complex code construction. Finally, Section~\ref{exp:CS} evaluates the performance of our BCASCs for recovering sparse signals from few measurements in a typical CS framework.

\subsection{Comparison to Reference BCASC Algorithm}
\label{exp:ref}
The main advantage of the ANN approximation over the reference BCASC algorithm is the fact that the calculus of the forces acting over each codewords requires some low number of summands, say $K$, rather than $n n_{\mathrm{rot}}$, which will be generally large. A fundamental question is how small can $K$ be while still obtaining complex codes with excellent coherence properties. In the case of performing radius searches for determining the ANN, the same question is transferable to the considered radius $r$. The first series of experiments is intended to provide an empirical answer to these questions and studies the evolution of the coherence and the execution time both with $K$ in the case of KNN search and with $r$ in the case of radius search. We aim to generate $n=64$ codewords of dimensionality $m=8$ considering $n_{\mathrm{rot}}=16$ complex rotations. For this case we use a single $k$-d tree as search structure, without imposing any restriction on the maximum number of leafs to visit. This way the exact NNs can be found. The higher computational cost of exact searches gets widely compensated by the lower computational cost of using a single $k$-d tree in this case.

\begin{figure*}[htpb]
      \centering
	\subfloat[Execution Time]{\includegraphics[width=0.48\textwidth,trim=90 0 100 52, clip]{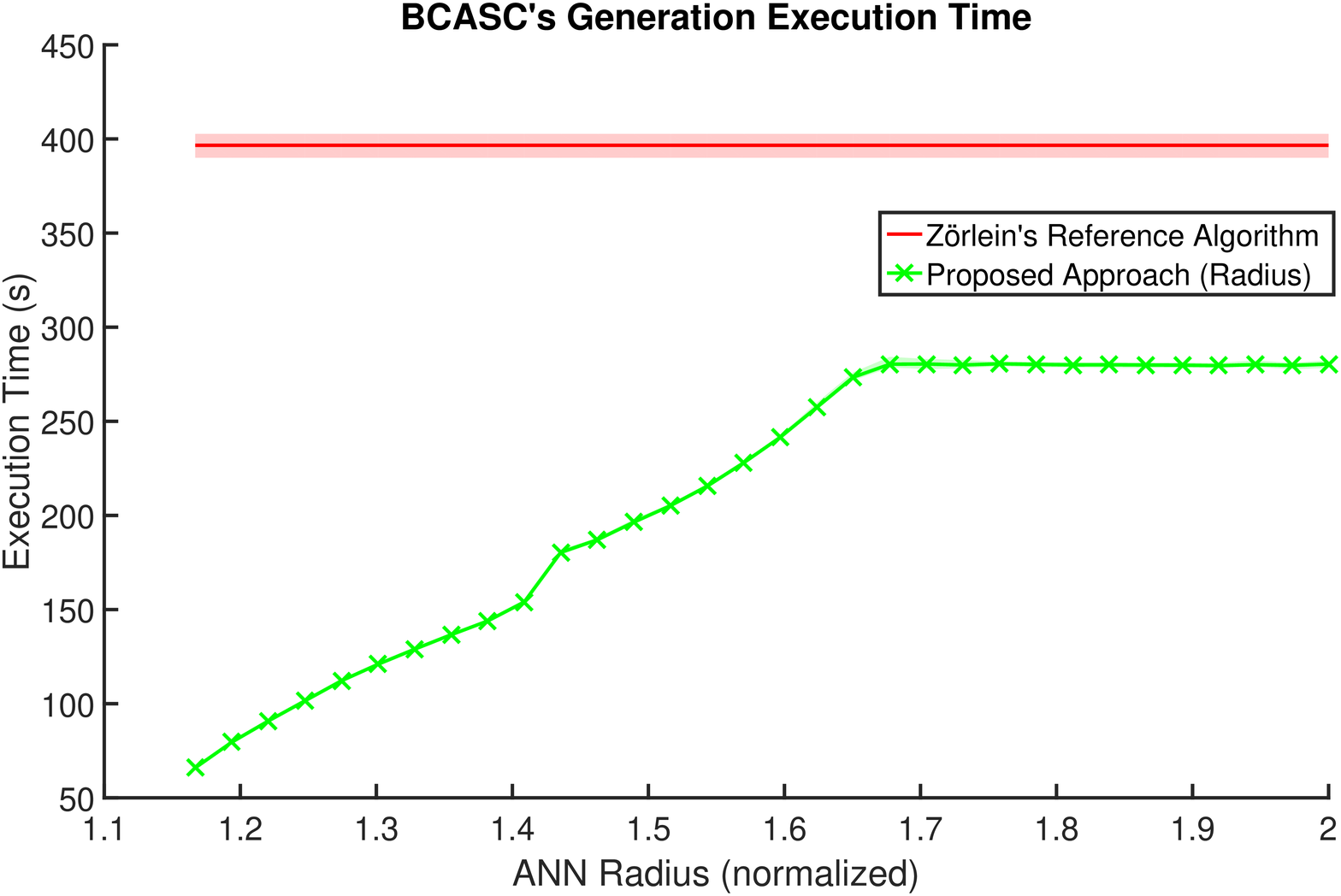}\label{fig:ANN-BCASC_radius:time}} %\hspace{5pt}
	\subfloat[Coherence]{\includegraphics[width=0.48\textwidth,trim=90 0 100 52, clip]{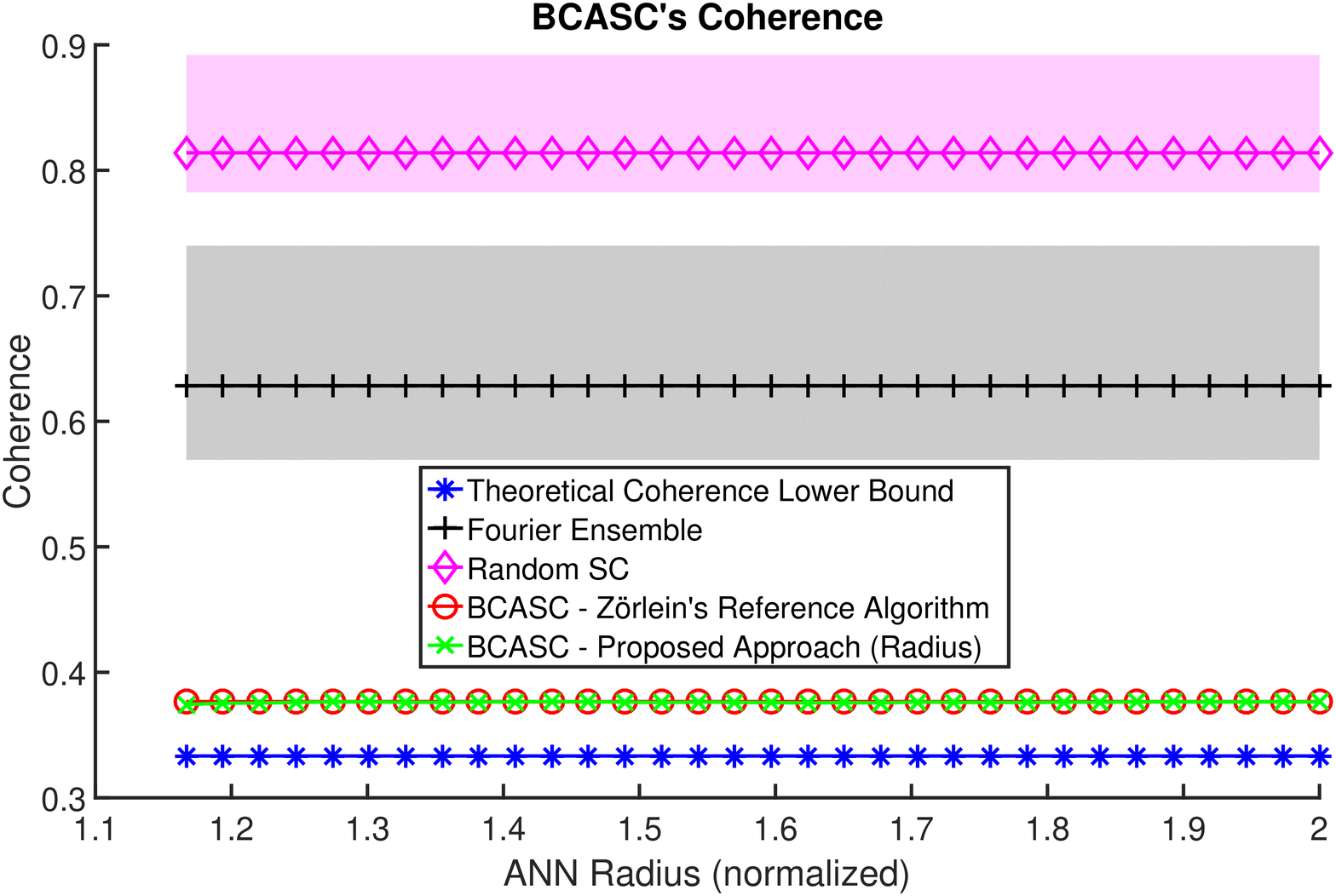}\label{fig:ANN-BCASC_radius:coherence}}\\

	\subfloat[Execution Time]{\includegraphics[width=0.48\textwidth,trim=90 0 100 52, clip]{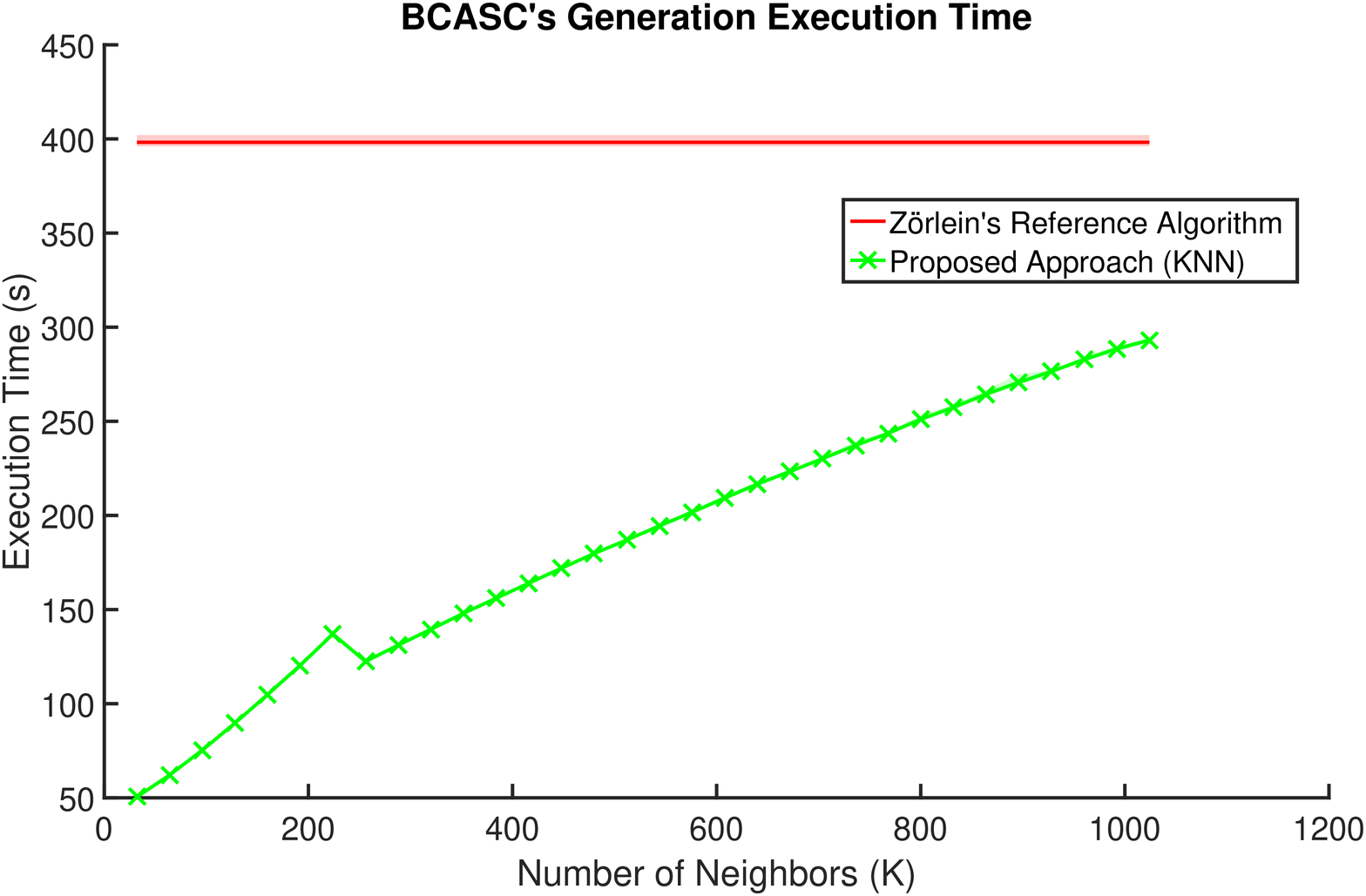}\label{fig:ANN-BCASC_KNN:time}} %\hspace{5pt}
	\subfloat[Coherence]{\includegraphics[width=0.48\textwidth,trim=90 0 100 52, clip]{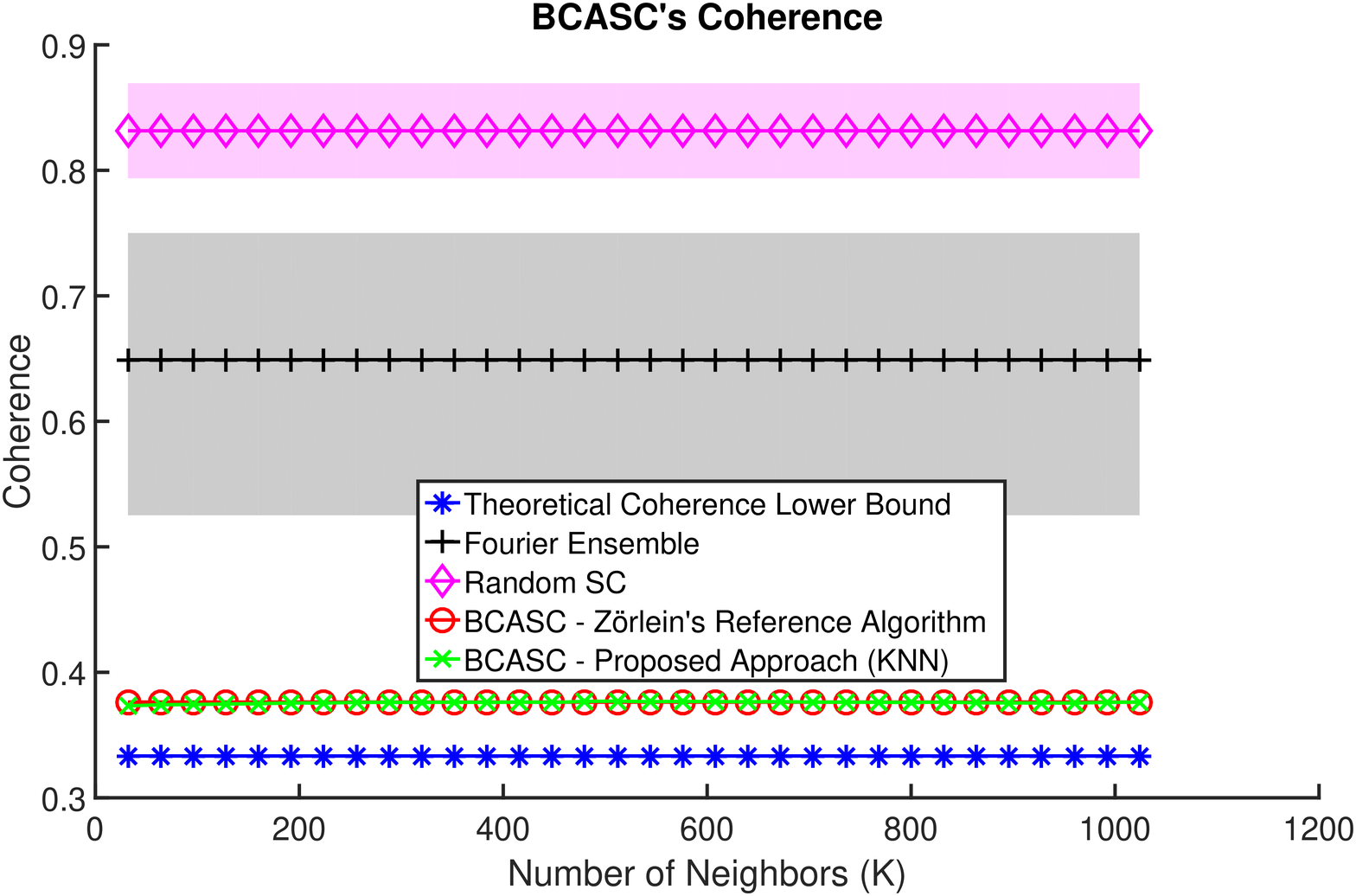}\label{fig:ANN-BCASC_KNN:coherence}}
\caption[Coherence and runtime evaluation of ANN-BCASC with the number of NN and the search radius]{Full range evaluation of the execution time needed to generate the BCASCs by means of our ANN approximation and corresponding coherence of the generated codes, for different numbers of NNs (plots \protect\subref{fig:ANN-BCASC_KNN:time} and \protect\subref{fig:ANN-BCASC_KNN:coherence}) and for different search radii (plots \protect\subref{fig:ANN-BCASC_radius:time} and \protect\subref{fig:ANN-BCASC_radius:coherence}). The execution time is compared to that of the reference algorithm and the coherences of both BCASCs and ANN-BCASCs are compared to the theoretical lower bound and to the coherences of complex random matrices and Fourier ensembles. The solid lines depict the mean of ten experiment runs for each experimental case and the shaded areas are defined between the minimum and maximum values obtained for each case.}
\label{fig:ANN-BCASC_KNN}
\end{figure*}

The results of these experiment series are collected in Fig.~\ref{fig:ANN-BCASC_KNN}. The plots at the left hand side show the evolution of the time required to generate the codes, both with $r$ in the case of radius search (Fig.~\ref{fig:ANN-BCASC_radius:time}) and with $K$ in the case of KNN search (Fig.~\ref{fig:ANN-BCASC_KNN:time}). Similarly, the plots at the right hand side show the evolution of the coherence of the obtained codes, also with $r$ (in Fig.~\ref{fig:ANN-BCASC_radius:coherence}) and with $K$ (in Fig.~\ref{fig:ANN-BCASC_KNN:coherence}). In the radius search experiments, we consider radii $r>1.14$, up to the maximum $r=2$, since our approach failed to generate BCASCs for $r<1.14$. In the KNN experiments we considered all $0<K\leq n n_{\mathrm{rot}}$. In both cases $32$ experimental cases are considered.
We compare the result of our ANN-based approach to the reference algorithm from \cite{Zoerlein15}. Additionally, we include a complex random matrix and a Fourier ensemble in the coherence evaluation. Both the real and complex components of the elements of the random matrix are drawn from a Gaussian distribution of zero mean and unit standard deviation. The Fourier ensemble of size $m\times n$ is constructed by randomly selecting $m$ rows of a DFT matrix on size $n\times n$. Both matrices are then columnwise normalized.
In order to asses the stability of our construction method and obtain some statistically relevant data, we repeat all the experiments ten times. The solid lines in the plots of Fig.~\ref{fig:ANN-BCASC_KNN} are thus average results. The shaded area of the same color around each line depicts the range in which all the results of the individual experiments fall. When the shaded area is not visible, it is because the line width used to plot the corresponding mean values is wider than the width of the range. This is particularly the case for the execution times of the proposed approach, which exhibit negligible variations between executions, and the coherences of the BCASC obtained both via our approach and the reference algorithm.

The first conclusion that can be drawn from Fig.~\ref{fig:ANN-BCASC_KNN} is that, regardless of the search method and the number of NN delivered by the ANN searches, no degradation of the coherence of the generated BCASCs is observed, when comparing to the reference algorithm. Comparing Fig.~\ref{fig:ANN-BCASC_radius:time} to Fig.~\ref{fig:ANN-BCASC_KNN:time} one may think that performing KNN searches is more efficient than radius searches, but the plain region of the curve in Fig.~\ref{fig:ANN-BCASC_radius:time} suggests that for radius in that range the NN are all points in the index, thus further increasing the radius search has no computational cost. Consequently, one should compare only the initial region of the curve in Fig.~\ref{fig:ANN-BCASC_radius:time} (positive slope) to that in Fig.~\ref{fig:ANN-BCASC_KNN:time}, in which case both search options perform equivalently, showing a linear dependency of the execution time on the size of the neighborhood considered.
From Fig.~\ref{fig:ANN-BCASC_radius:coherence} and Fig.~\ref{fig:ANN-BCASC_KNN:coherence} it becomes clear that BCASCs are superior in terms of coherence, not just to randomly-generated SCs, but to SCs derived from a discrete Fourier basis. In fact, it is remarkable how close BCASC coherences are to the theoretical coherence, given in this case by Eq.~\ref{eq:Welch_bound}, regardless of the construction method. The mean coherence of BCASCs generated using our approach is very close to that obtained using the reference algorithm, still always lower than the latter for all experimental cases. In short terms, the ANN-BCASC approximation is able to simultaneously provide a dramatic reduction of the execution time and a slight coherence reduction.

\begin{figure*}[htpb]
      \centering
	\subfloat[Execution Time]{\includegraphics[width=0.48\textwidth,trim=90 0 110 50, clip]{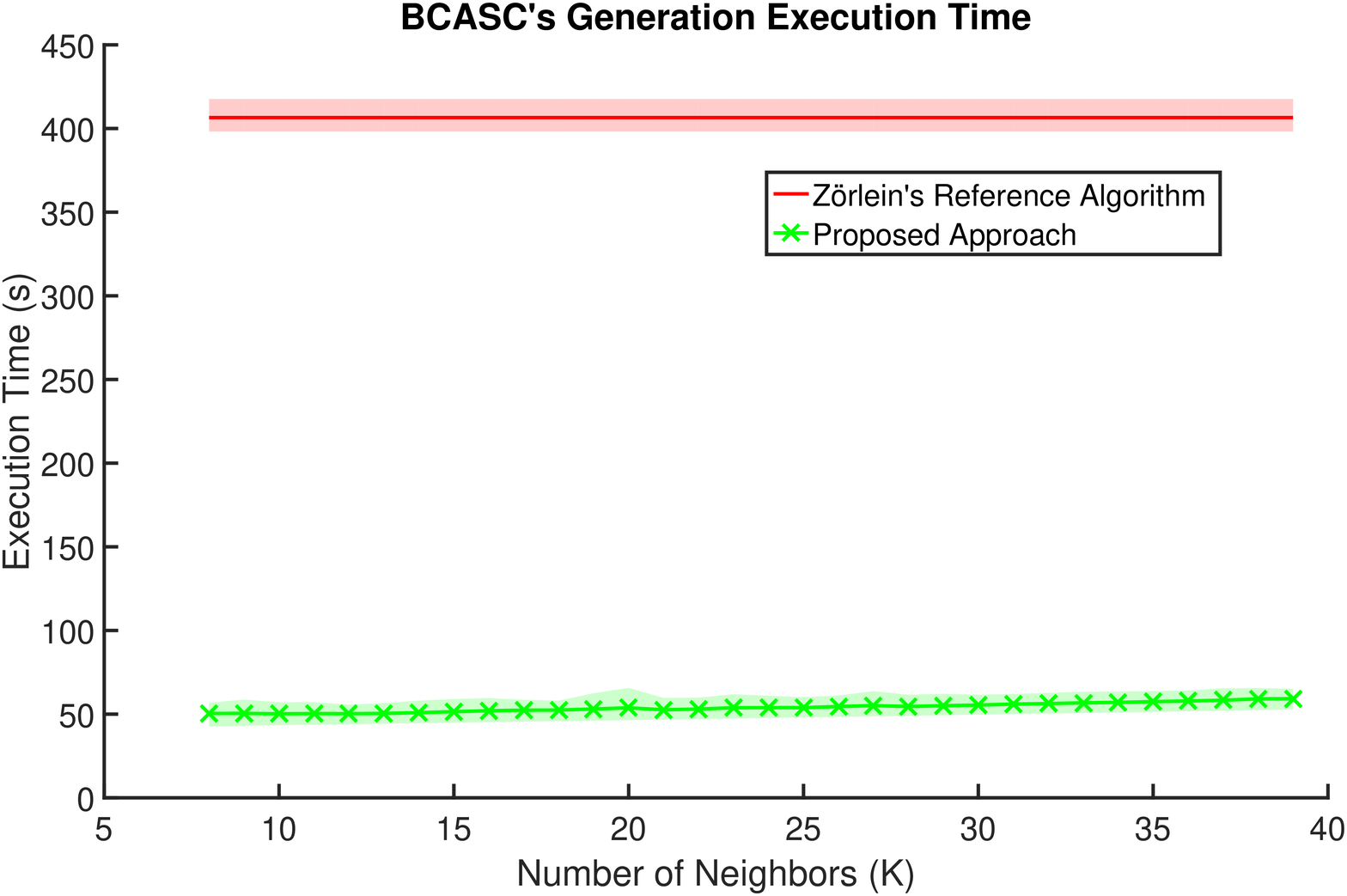}\label{fig:ANN-BCASC_KNN_detail:time}} %\hspace{5pt}
	\subfloat[Coherence]{\includegraphics[width=0.48\textwidth,trim=90 0 110 50, clip]{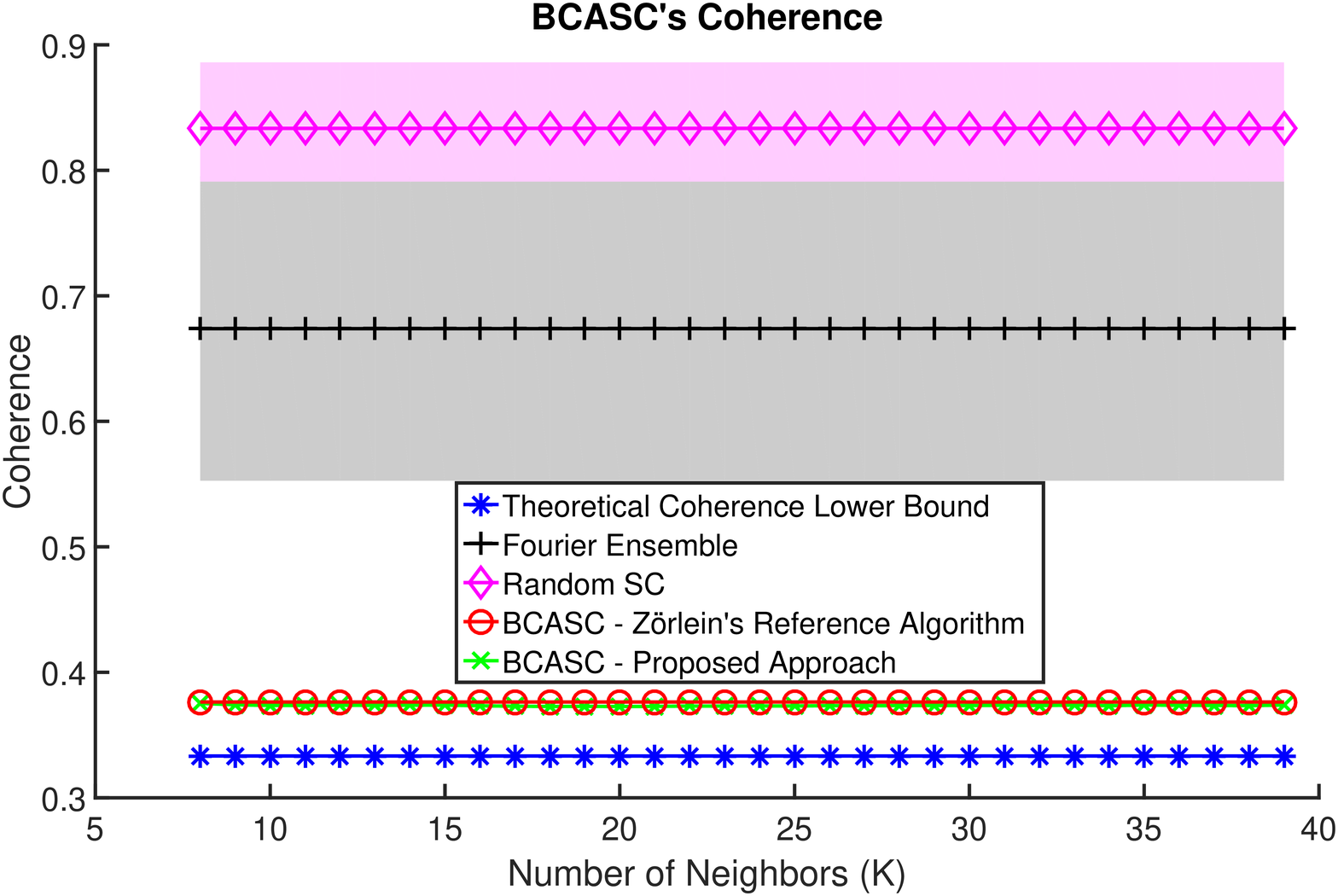}\label{fig:ANN-BCASC_KNN_detail:coherence}}
\caption[Coherence and runtime evaluation of ANN-BCASC with the number of NN (detail)]{Detailed evaluation of the execution time \protect\subref{fig:ANN-BCASC_KNN_detail:time} needed to generate the BCASCs by means of our ANN approximation for different numbers of NNs and corresponding coherence of the generated codes \protect\subref{fig:ANN-BCASC_KNN_detail:coherence}. The execution time is compared to that of the reference algorithm and the coherences of both BCASCs and ANN-BCASCs are compared to the theoretical lower bound and to the coherences of complex random matrices and Fourier ensembles. The solid lines depict the mean of ten experiment runs for each experimental case and the shaded areas are defined between the minimum and maximum values obtained for each case.}
\label{fig:ANN-BCASC_KNN_detail}
\end{figure*}

From now on we adopt the KNN search as default search method for our experiments. A closer observation of the ANN-BCASC coherences in Fig.~\ref{fig:ANN-BCASC_KNN:coherence} reveals that the minimal coherence is obtained for the lowest $K$, while reaching that obtained using the reference algorithm for $K=n n_{\mathrm{rot}}$. This comes at no surprise, since for $K=n n_{\mathrm{rot}}$ our approximation is equivalent to the reference. The remaining question is then how much can we push the ANN approximation so that the execution time and eventually the coherence are further reduced, while still being able to successfully generate the codes. In the search of an answer we carried out detail experiments for very low $K$ and unit increments, which showed that successful ANN-BCASC generation was possible with as few as $K=8$ NN. Fig.~\ref{fig:ANN-BCASC_KNN_detail} shows the execution times and coherences obtained in $32$ different experimental cases, from the lowest possible number of NN, $K=8$, to $K=39$, both included.

The results in Fig.~\ref{fig:ANN-BCASC_KNN_detail} are highly encouraging. As expected, the execution time exhibits a slow linear increase with $K$, while the coherence still remains slightly below that of the reference algorithm for all cases considered. For instance, adopting the minimal $K$, the execution time is approximately $\unit[50]{s}$, instead of the $\unit[>400]{s}$ required by the reference algorithm, i.\,e., the execution time was reduced by a factor of $>8$. In terms of coherence, the minimum average coherence, obtained for $K=20$, was $0.3727$, which means a $1\%$ reduction when compared to that obtained with the reference algorithm ($0.3763$). While this might look negligible at first sight, if differences with respect to the theoretical lower bound are considered, this means a reduction of more than $8\%$.

\begin{figure*}[htpb]
      \centering
	\subfloat[Execution Time]{\includegraphics[width=0.48\textwidth,trim=70 0 110 50, clip]{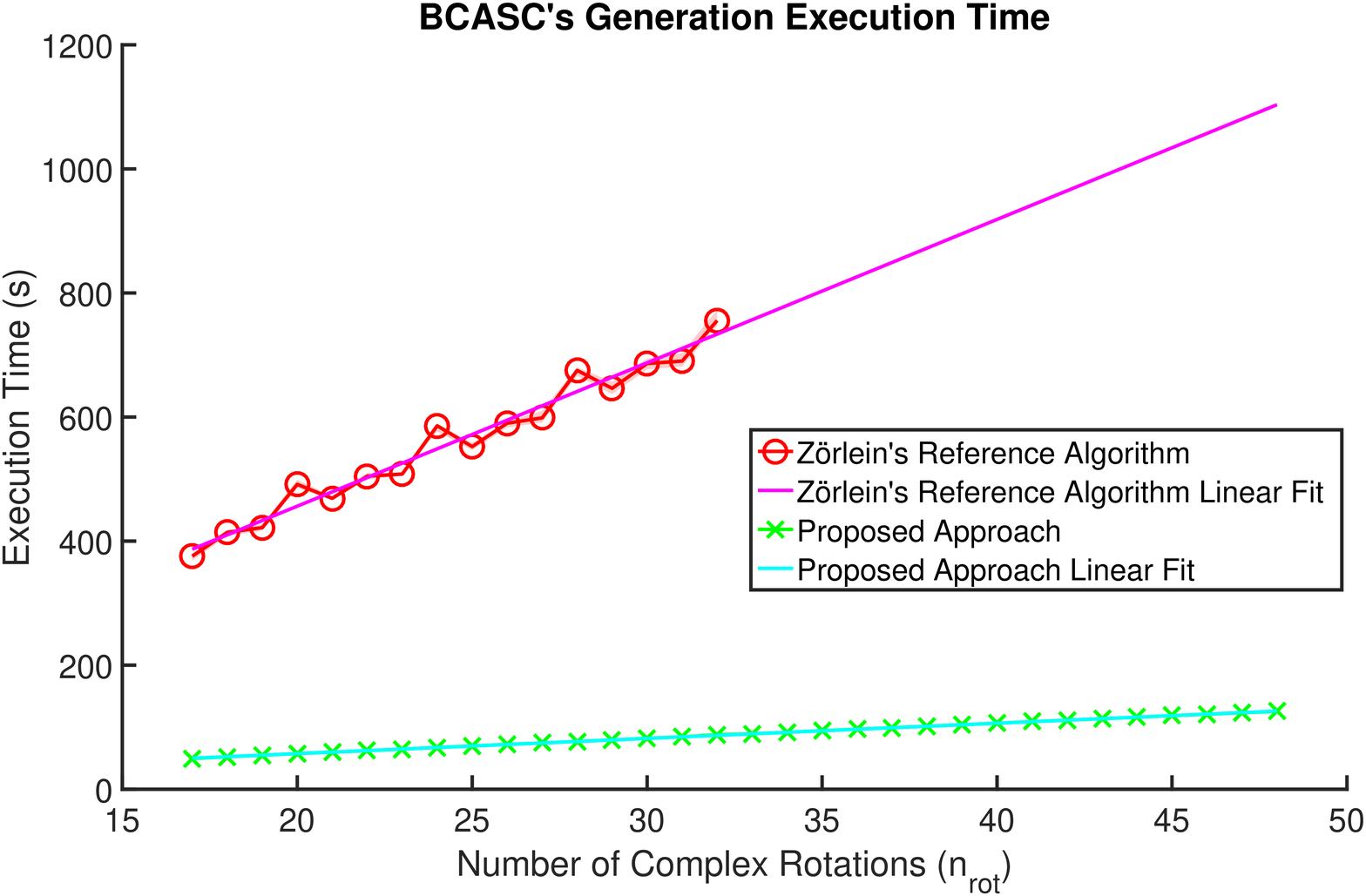}\label{fig:ANN-BCASC_n_rot:time}} %\hspace{5pt}
	\subfloat[Coherence]{\includegraphics[width=0.48\textwidth,trim=70 0 110 50, clip]{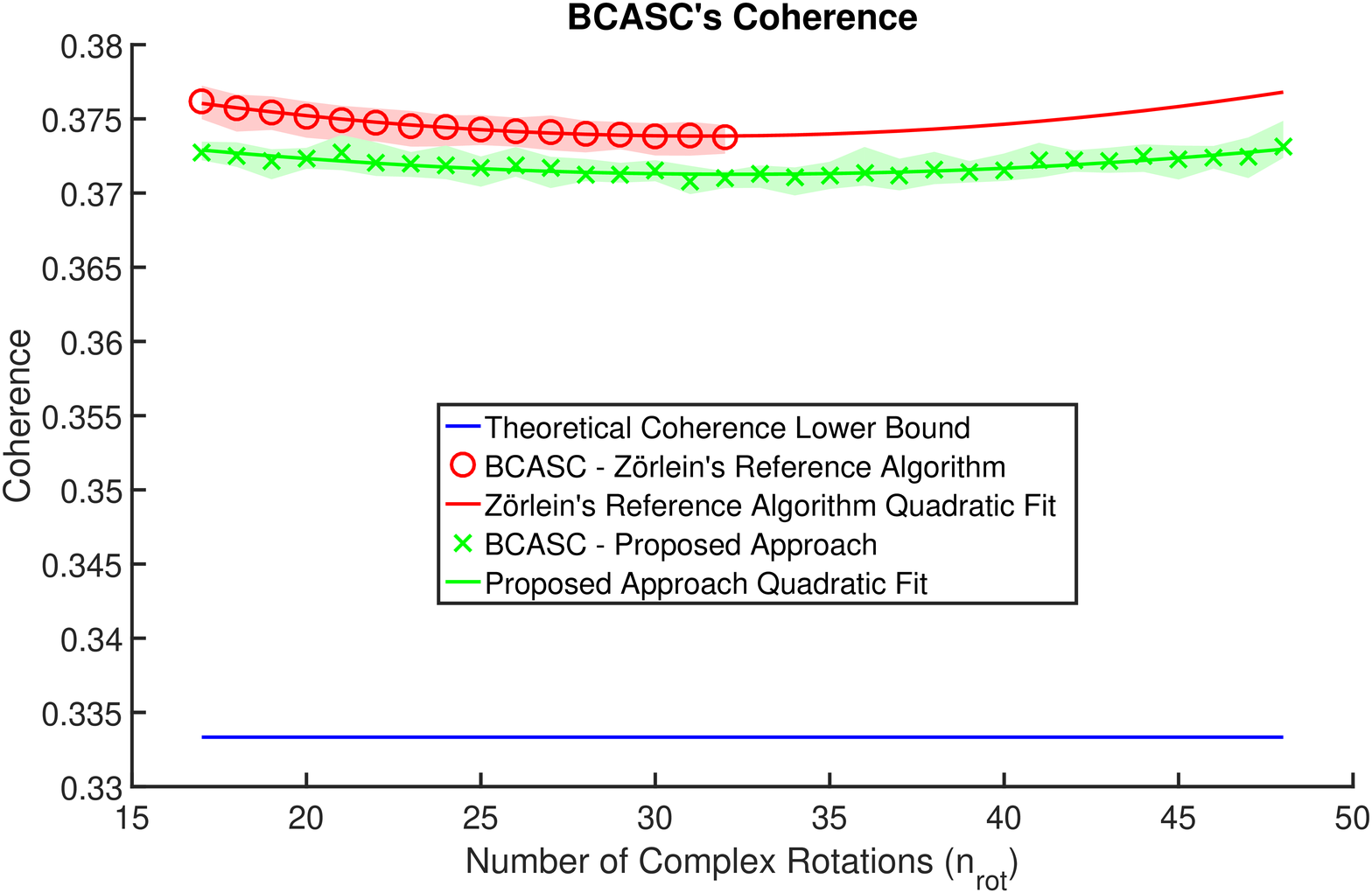}\label{fig:ANN-BCASC_n_rot:coherence}}
\caption[Coherence and runtime evaluation of ANN-BCASC with the number of complex rotations]{Evolution of the execution time \protect\subref{fig:ANN-BCASC_n_rot:time} needed to generate the ANN-BCASCs for different numbers of complex rotations $n_{\mathrm{rot}}$ considered in the algorithm and corresponding coherence of the obtained codes \protect\subref{fig:ANN-BCASC_n_rot:coherence}. The execution time is compared to that of the reference algorithm and the coherences of both BCASCs and ANN-BCASCs are compared to the theoretical lower bound. The solid lines depict the mean of ten experiment runs for each experimental case and the shaded areas are defined between the minimum and maximum values obtained for each case.}
\label{fig:ANN-BCASC_n_rot}
\end{figure*}

Provided the large potential for reducing execution time, one can aim to establish some custom tradeoff between this and an improvement on the quality of the codes, for instance, by considering a larger number of complex rotations in the calculus, i.\,e., refining the approximate integral. Then it is of interest to study the evolution of the execution time with $n_{\mathrm{rot}}$ and the effect of considering larger $n_{\mathrm{rot}}$ on the coherence of the resulting codes. We do so for $16<n_{\mathrm{rot}}\leq 48$ with unit step size, provided that the case $n_{\mathrm{rot}}=16$ was already studied before. The size of the codes remain $8\times 64$ and the number of NNs to search for is set to $K=20$ in this and all the following experiments. Both the execution time and coherence are compared to those obtained with the reference algorithm. As before, all experiments are repeated ten times and the results are given in Fig.~\ref{fig:ANN-BCASC_n_rot}. Due to the large execution times of the reference algorithm, the cases $n_{\mathrm{rot}}>32$ were omitted and estimated by extrapolation of a polynomial fit.

Fig.~\ref{fig:ANN-BCASC_n_rot:time} is truly illuminating: even for the largest number of rotations considered ($n_{\mathrm{rot}}=48$), the execution time required by our approach is three times lower than that required by the reference for $n_{\mathrm{rot}}=16$. Furthermore, the slope of the polynomial fit of the mean values of execution times is much lower for the ANN approximate method, more specifically, $\unit[2.450]{s/rot}$ \emph{versus} $\unit[23.11]{s/rot}$, i.\,e., an order of magnitude lower. Extrapolating the linear fit for our approach one obtains that with the same time budget required by the reference algorithm for only $n_{\mathrm{rot}}=16$, up to $n_{\mathrm{rot}}=150$ complex rotations could be considered in the calculus. Also enlightening is Fig.~\ref{fig:ANN-BCASC_n_rot:coherence}, since it shows that increasing $n_{\mathrm{rot}}$ has in our approach the same effect of further coherence reduction already observed in \cite{Zoerlein15} for the reference algorithm. Furthermore, the shaded areas showing the ranges where the results of each approach live are typically disjoint, that is, in terms of coherence the best code delivered by the reference approach is worse than the worst code delivered by our approach, regardless of the experimental case considered. The mean difference between the quadratic polynomial fits in the considered range is $1.965\times 10^{-3}$, which means over a $4\%$ reduction in terms of mean differences with respect to the theoretical lower bound. Also important is to notice that the rate at which the mean coherence decreases also decreases with $n_{\mathrm{rot}}$, up to the point that for $n_{\mathrm{rot}}>32$ the fitted average coherence starts to degrade and further increase of $n_{\mathrm{rot}}$ will only worsen it. This effect was not observed for the reference algorithm, neither in \cite{Zoerlein15} nor in this work and, in fact, experiments for $n_{\mathrm{rot}}>32$ revealed an asymptotic decrease of the coherence with $n_{\mathrm{rot}}$ for the reference algorithm. The worsening of the coherence for very large $n_{\mathrm{rot}}$ in our approach is due to the relatively low number of NN per search, set to $K=20$ for all experiments to obtain \emph{ceteris paribus} results. Increasing $K$ linearly with $n_{\mathrm{rot}}$ would solve this issue, at the cost of an slightly increased execution time.

\begin{figure*}[htpb]
      \centering
	\subfloat[Execution Time]{\includegraphics[width=0.48\textwidth,trim=70 0 110 50, clip]{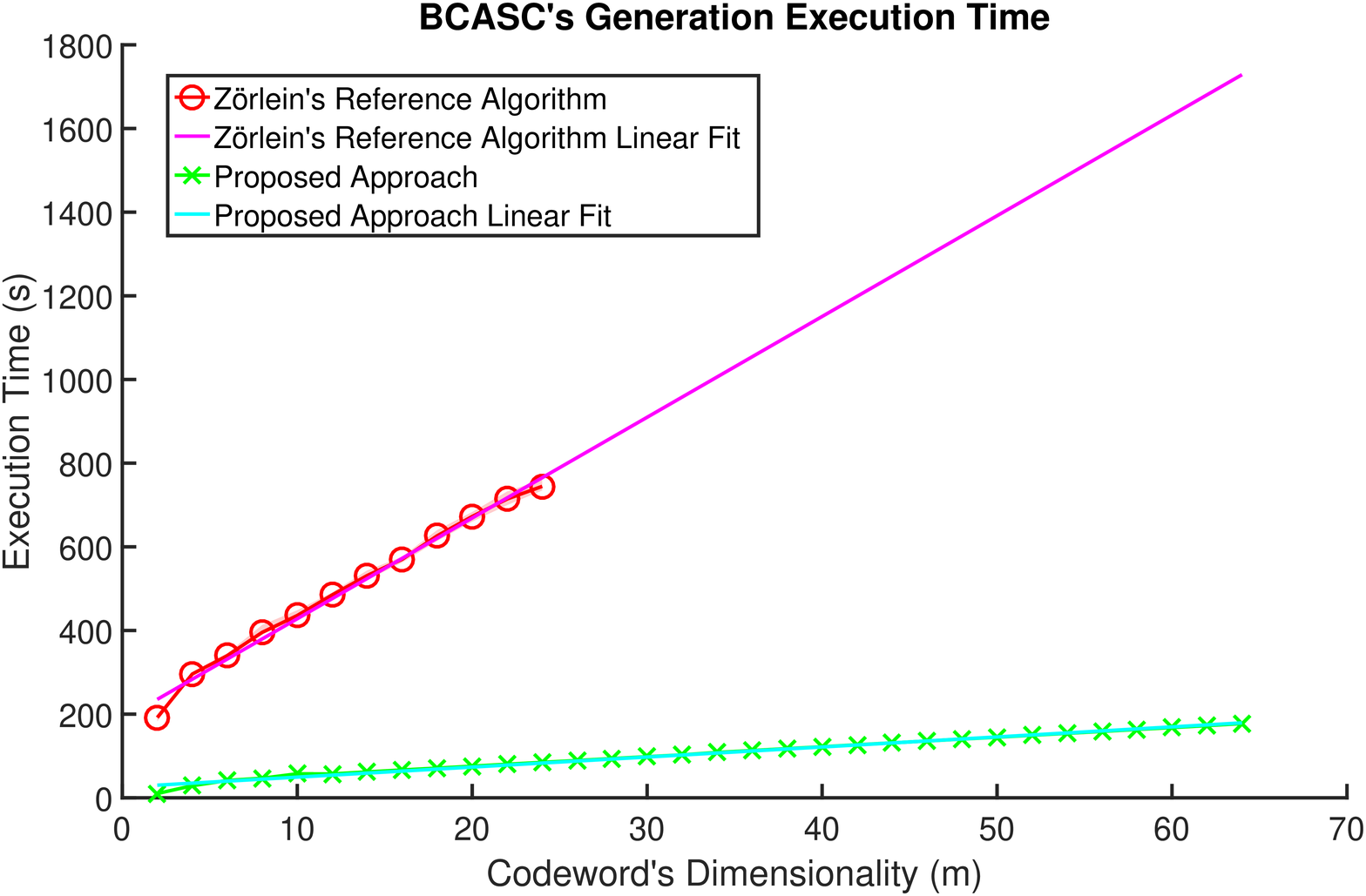}\label{fig:ANN-BCASC_size_sweep:m:time}} %\hspace{5pt}
	\subfloat[Coherence]{\includegraphics[width=0.48\textwidth,trim=70 0 110 50, clip]{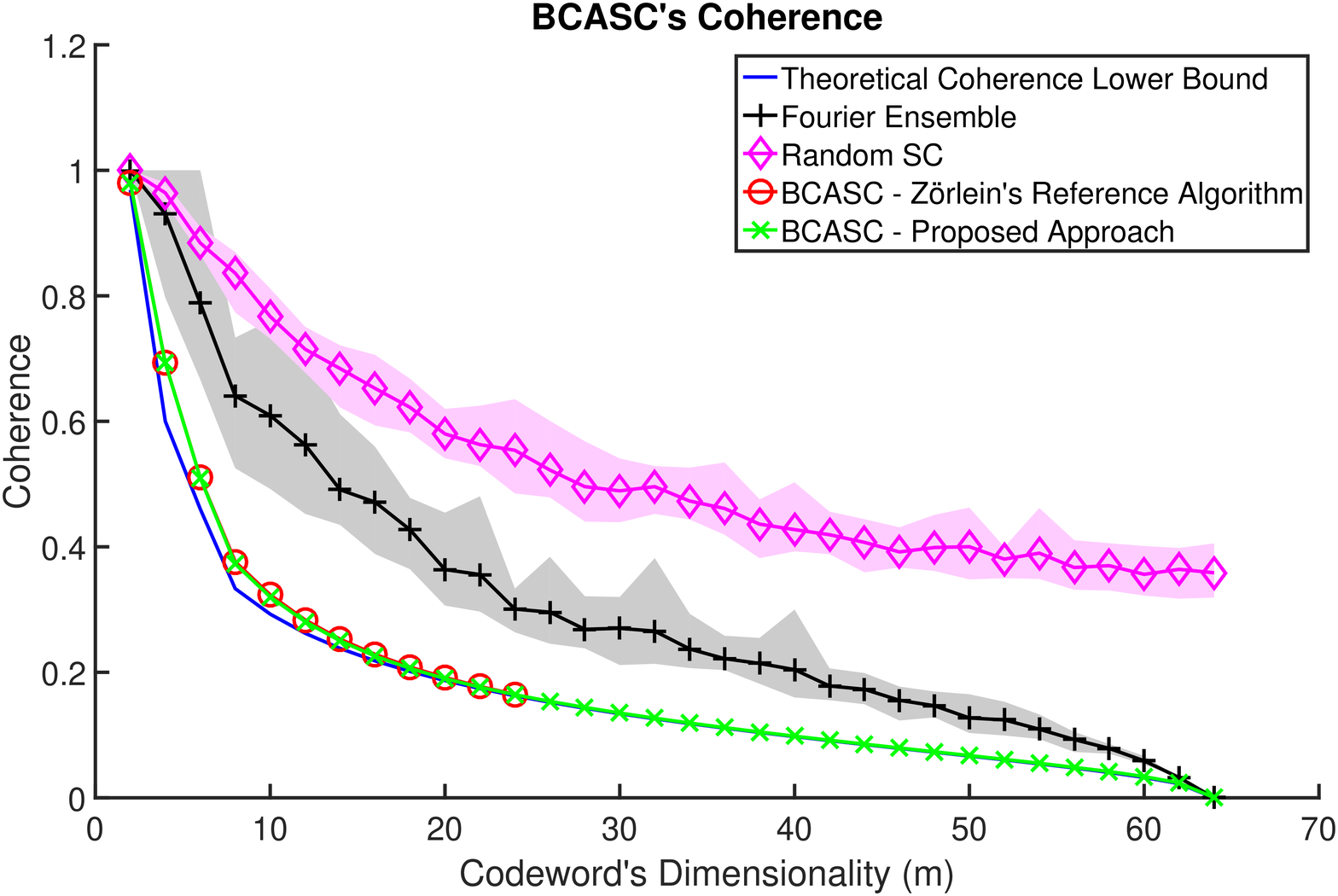}\label{fig:ANN-BCASC_size_sweep:m:coherence}} \\
	\subfloat[Execution Time]{\includegraphics[width=0.48\textwidth,trim=70 0 110 50, clip]{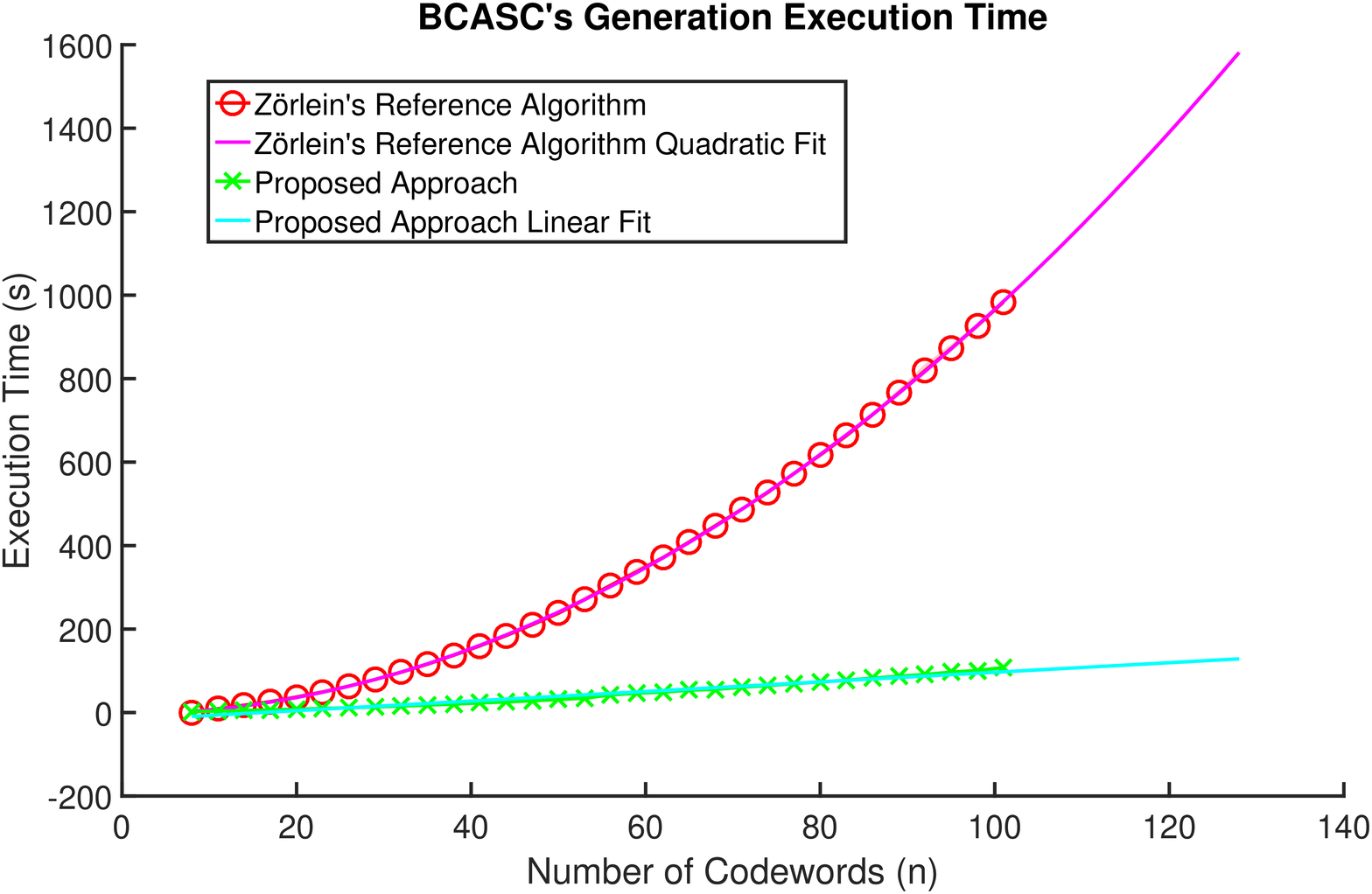}\label{fig:ANN-BCASC_size_sweep:n:time}} %\hspace{5pt}
	\subfloat[Coherence]{\includegraphics[width=0.48\textwidth,trim=70 0 110 50, clip]{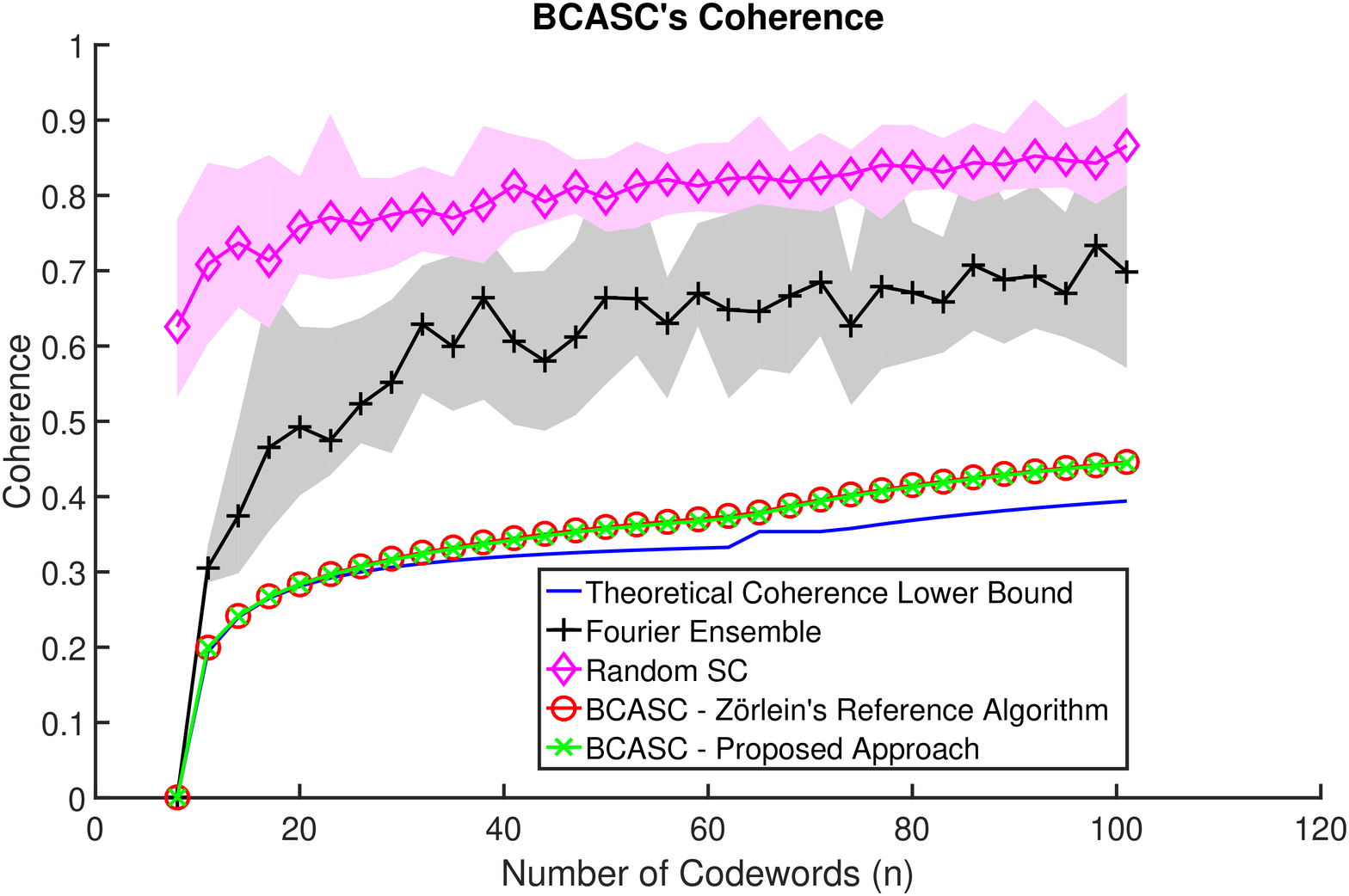}\label{fig:ANN-BCASC_size_sweep:n:coherence}}
\caption[Coherence and runtime evaluation of ANN-BCASC with the size of the codes]{Evaluation of the execution time (left plots) needed to generate the ANN-BCASCs for different code sizes $m\times n$ and corresponding code coherences (right plots). Plots in the first row show the results for varying $m$ and in the second row for varying $n$. The execution time is compared to that of the reference algorithm and the coherences of both BCASCs and ANN-BCASCs are compared to the theoretical lower bound and to the coherences of complex random matrices and Fourier ensembles of the same size. The solid lines depict the mean of ten experiment runs for each experimental case and the shaded areas are defined between the minimum and maximum values obtained for each case.}
\label{fig:ANN-BCASC_size_sweep}
\end{figure*}

Alternatively, for some fixed $n_{\mathrm{rot}}$, considered to be sufficiently high, the time saving that the ANN variant provides can be invested in generating larger codes, i.\,e., of larger dimensionality and with larger number of codewords. This was indeed one of the main motivations of this work, provided that in CS the measurement matrices can be very wide, due to the fact that $n$ often relates to the step used to discretize some continuous domain and not to the underlying dimensionality of the problem. In order to obtain empirical evidence of the expectedly good behavior of the proposed approach with increasing $m$ and $n$, we carry out independent experiments to check the evolution of the execution time and the coherence with these parameters. For the $m$-\emph{sweep} experiments we fix $n=64$ and consider $32$ experimental cases for $2\leq m\leq 64$. Clearly, for $m=n=64$ an orthobasis is obtained and zero coherence is attained. For the $n$-sweep experiments we fix $m=8$ and would like to consider the cases from $n=m=8$ to $n=128$. Unfortunately, due to the quadratic dependency of the execution time on $n$ in the reference algorithm, we only consider $32$ experimental cases for $8\leq n\leq 101$ (step $3$) and obtain the rest cases up to $n=128$ by extrapolation of a polynomial fit. As before, all experiments are repeated ten times to obtain mean, minimum and maximum values. It turned out that we could neither afford the computational cost required for running all the $m$-sweep experiments with the costly reference algorithm. For this reason we were forced to run only the first $12$ experimental cases. We used appropriate polynomial fits to offer an approximation to the results of the omitted cases. The results of these experiments are summarized in Fig.~\ref{fig:ANN-BCASC_size_sweep}. As before, shadowed areas are used to represent the ranges between minimum and maximum values and solid lines are used for mean values.

The first row of plots in Fig.~\ref{fig:ANN-BCASC_size_sweep} correspond to the varying-$m$ experiments. Similarly to Fig.~\ref{fig:ANN-BCASC_n_rot:time}, Fig.~\ref{fig:ANN-BCASC_size_sweep:m:time} also reveals the potential of the ANN approximation for obtaining much more from the same or lower computational effort. Indeed, it is clear than even for the maximum $m$ considered our approach is able to deliver a code in a time that is lower than the time the reference algorithm needs to generate the code of lowest $m$. The linear evolution of the execution time of the reference approach with $m$ is due to the fact we use our own optimized (yet exact) reimplementation, but was originally quadratic. Note the accuracy of the linear fits for both approaches and observe the large ratio between their slopes. The slope of the reference algorithm linear fit is ten times greater than that of our approach. This means that our approach is suitable for generating BCASCs of large $m$, while the reference soon requires unrealistically high execution times.
Further individual experiments revealed that for $m>54$, i.\,e., for $m\to n$ the reference algorithm is able to find fixed points in a reduced amount of time, abruptly breaking the linearity of the regression in Fig.~\ref{fig:ANN-BCASC_size_sweep:m:time}. This is due to the fact that it becomes easier to optimally pack the $n$ codewords in an hypersphere of such a high dimensionality. Unfortunately, the case $m\to n$ is of no interest in CS.
%m sweep: slope BCASC: 24.0841 ANN-BCASC: 2.4017
Regarding the coherence of the codes, Fig.~\ref{fig:ANN-BCASC_size_sweep:m:coherence} shows how tightly BCASCs approach the theoretical lower bound, regardless of the approach used to generate them. A detailed view reveals that the line corresponding to our approach is always below the one corresponding to the reference algorithm, as one could already expect from previous experiments. When $m\ll n$ the coherence globally approaches to one, while tending to zero as $m\to n$. Despite the coherence of the Fourier ensemble also follows this pattern and yields an orthobasis for $m=n$, it is far from the lower bound for relatively low $m$, which is the case of interest in CS. The random construction of complex SCs delivers the worst results and is unable to push the coherence under $0.36$.
Also remarkable is the stability of both our approach and the reference one, both in terms of coherence and execution time.

The second row of plots in Fig.~\ref{fig:ANN-BCASC_size_sweep} summarizes the results of the varying-$n$ experiments. Fig.~\ref{fig:ANN-BCASC_size_sweep:n:time} confirms both the quadratic dependency of the execution time of the reference algorithm on $n$ and the linear dependency of our approach. Compare the relatively fast increase of the slope of the quadratic fit for the reference algorithm with the low constant slope of the linear fit for our approach. As a consequence, the accurate polynomial fits predict that for a moderate $n=128$ our approach only needs around $\unit[2]{min}$ to generate the code, i.\,e., one order of magnitude less than the $\unit[26]{min}$ required by the reference algorithm.
Fig.~\ref{fig:ANN-BCASC_size_sweep:n:coherence} shows that the massive speedup does not degrade the coherence and, in fact, one can observe that the points for the mean coherence obtained with our approach are always below those corresponding to the reference approach. Regardless of the method, BCASC approach the theoretical lower bound more tightly than Fourier ensembles or random complex SCs, whose coherence tends to one as $n\to \infty$. Also in this experiment series the stability of both BCASC generation methods is observed to be very high compared to the other alternatives.

\subsection{General Comparison to Related Work on BCASC Construction}
\label{exp:tab}
In this section we provide a general comparison between the ANN-BCASC approximation and other approaches for generating close-to-optimal complex codes. In order to provide a complete cumulative comparison, the best option for such comparison is to extend the tables in \cite{Zoerlein15}, which are in turn an extension of Table~II of \cite{Xia05} and Table~II of \cite{Medra14}, respectively.
These tables provide the values of coherence obtained from complex SCs generated using different methods and are named Table~\ref{tab:coherence_comp_1} and Table~\ref{tab:coherence_comp_2}, homonimously to the corresponding tables in \cite{Zoerlein15}. For both the BCASC and the ANN-BCASC methods we adopt the best result of ten independent runs. We also measure the time that each algorithm needed to generate the codes. For the other algorithms, we preserve the values in the original tables of \cite{Zoerlein15}. We use the reference BCASC algorithm with approximate summation with $n_{\mathrm{rot}}=32$ summands, in order to have a fair reference for the ANN-BCASC algorithm, for which we adopt the same $n_{\mathrm{rot}}$ and a constant $K=20$ for all cases considered. As in \cite{Zoerlein15}, for both methods $\alpha_{0}=0.9$ and $\tau_{\mathrm{max}}=10^5$ were used.     

\begin{table*}[htbp]
\begin{center}
\begin{tabular}{c c | c c c c c c | c c}
\hline
%\multicolumn{2}{c}{Size}&\multicolumn{6}{c}{Method}\\
\multicolumn{2}{c}{ }&\multicolumn{6}{c}{Coherence}&\multicolumn{2}{c}{Execution Time (s)}\\
\hline
%$m$ & $n$ & Composite Bound& ANN-BCASC & BCASC \cite{Zoerlein15} & Medra \emph{et al.} \cite{Medra14} & Xia \emph{et al.} \cite{Xia05} & Love \cite{Love_spherical_codes}\\
$m$ & $n$ & Composite Bound & ANN-BCASC & BCASC \cite{Zoerlein15} & Medra \emph{et al.} \cite{Medra14} & Xia \emph{et al.} \cite{Xia05} & Love \cite{Love_spherical_codes} & ANN-BCASC & BCASC \cite{Zoerlein15}\\
\hline
2 & 8 & $0.7500$ & $0.7953$ & $0.7954$ & $0.7997$ & $0.8216$ & $0.8415$ & $1.768\times 10^2$ & $4.443\times 10^2$\\
3 & 16 & $0.6202$ & $0.6490$ & $0.6498$ & $0.6590$ & $0.6766$ & $0.8079$ & $5.839\times 10^2$ & $1.336\times 10^3$ \\
4 & 16 & $0.4472$ & $0.4486$ & $0.4480$ & $0.4473$ & $0.4514$ & $0.7525$ & $7.253\times 10^2$ & $5.945\times 10^1$ \\
4 & 64 & $0.6000$ & $0.6878$ & $0.6878$ & $0.7151$ & $0.7447$ & $0.7973$ & $4.859\times 10^3$ & $5.501\times 10^3$ \\
\hline
\end{tabular}
\end{center}
\caption[Coherence comparison of complex codes I]{Comparison of the coherence of close-to-optimal complex codes obtained via different numerical approaches. Based on Table~I of \cite{Zoerlein15}.}
\label{tab:coherence_comp_1}
\end{table*}

\begin{table*}[htbp]
\begin{center}
\begin{tabular}{c c | c c c c c | c c}
\hline
\multicolumn{2}{c}{Size}&\multicolumn{5}{c}{Coherence}&\multicolumn{2}{c}{Execution Time (s)}\\
\hline
$m$ & $n$ & Composite Bound& ANN-BCASC & BCASC \cite{Zoerlein15} & Medra \emph{et al.} \cite{Medra14} & Dhillon \emph{et al.} \cite{Dhillon08} & ANN-BCASC & BCASC \cite{Zoerlein15}\\
\hline
4 & 5 & $0.2500$ & $0.2500$ & $0.2501$ & $0.2502$ & $0.2500$ & $1.642\times 10^2$ & $1.453$\\
4 & 6 & $0.3162$ & $0.3276$ & $0.3278$ & $0.3274$ & $0.3275$ & $2.026\times 10^2$ & $2.387$\\
4 & 7 & $0.3536$ & $0.3540$ & $0.3538$ & $0.3540$ & $0.3536$ & $2.454\times 10^2$ & $4.659$\\
4 & 8 & $0.3780$ & $0.3784$ & $0.3781$ & $0.3787$ & $0.3782$ & $2.844\times 10^2$ & $1.557\times 10^1$\\
4 & 9 & $0.3953$ & $0.4024$ & $0.4023$ & $0.4021$ & $0.4034$ & $3.371\times 10^2$ & $2.018\times 10^1$\\
4 & 10 & $0.4082$ & $0.4114$ & $0.4116$ & $0.4113$ & $0.4114$ & $3.889\times 10^2$ & $2.318\times 10^1$\\
4 & 16 & $0.4472$ & $0.4486$ & $0.4480$ & $0.4473$ & $0.4473$ & $7.294\times 10^2$ & $5.304\times 10^1$\\
4 & 20 & $0.5000$ & $0.5012$ & $0.5008$ & $0.5001$ & $0.5335$ & $9.804\times 10^2$ & $1.136\times 10^2$\\
5 & 6 & $0.2000$ & $0.2000$ & $0.2000$ & $0.2002$ & $0.2001$ & $2.377\times 10^2$ & $0.9564$\\
5 & 7 & $0.2582$ & $0.2668$ & $0.2669$ & $0.2665$ & $0.2669$ & $2.859\times 10^2$ & $6.1576$\\
5 & 8 & $0.2928$ & $0.2957$ & $0.2956$ & $0.2954$ & $0.2955$ & $3.350\times 10^2$ & $1.224\times 10^1$\\
5 & 9 & $0.3162$ & $0.3206$ & $0.3207$ & $0.3203$ & $0.3216$ & $3.898\times 10^2$ & $1.585\times 10^1$\\
5 & 10 & $0.3333$ & $0.3339$ & $0.3334$ & $0.3341$ & $0.3336$ & $4.446\times 10^2$ & $2.174\times 10^1$\\
5 & 16 & $0.3830$ & $0.3892$ & $0.3898$ & $0.3932$ & $0.3959$ & $8.282\times 10^2$ & $1.980\times 10^2$\\
\hline
\end{tabular}
\end{center}
\caption[Coherence comparison of complex codes II]{Comparison of the coherence of close-to-optimal complex codes obtained via different numerical approaches. Based on Table~II of \cite{Zoerlein15}.}
\label{tab:coherence_comp_2}
\end{table*}

In short terms, the proposed ANN approximate approach and the reference algorithm generate BCASCs of equivalent quality in terms of coherence. In other words, Tables~\ref{tab:coherence_comp_1} and~\ref{tab:coherence_comp_2} confirm that the reduction of the algorithm complexity has no significant effect on the quality of the obtained BCASCs. In fact, both alternatives often produce codes with equal coherence, closely approaching and eventually meeting the theoretical lower bound. Cases for which our approach meets the theoretical lower bound are $m=4$, $n=5$ and $m=5$, $n=6$. Furthermore, in some cases the proposed approach outperforms the reference method, which is a remarkable fact, even when the differences are rather negligible in general.
The reader might observe that the coherences obtained for the reference method often do not coincide with those given in \cite{Zoerlein15}. This is due to the fact that we use the variant with approximate discrete summation for calculating the integral over complex rotations, in order to enable a fair comparison to the proposed approach, while the results presented in Tables~I and~II of \cite{Zoerlein15} were obtained using numerical integration.

Regarding the execution time, Table~\ref{tab:coherence_comp_1} suggests that the proposed approach provides a moderate speedup when compared to the reference, but not even in every case. Furthermore, Table~\ref{tab:coherence_comp_2} shows lower execution times for the reference algorithm in all cases considered. At first sight this seems to be in conflict with the reduction of the computational complexity of each algorithm iteration that our approach was claimed to bring. Actually there is no such conflict and the (often largely) reduced execution times obtained for the reference algorithm are due to the fact that fixed points are often found within the first iterations, while the proposed method does not find fixed points so easily due to its locally-restricted operation. Another reason is that we use our own optimized implementation of the reference algorithm, which already makes linear the quadratic dependency of the complexity on $m$.
In practice the early discovery of fixed points and thus abnormally low execution times only happen when $n\approx m$, being both relatively low, which is the case for the experimental cases considered in Tables~\ref{tab:coherence_comp_1} and~\ref{tab:coherence_comp_2}. Such cases are of no relevance in practical CS scenarios, in which $n\gg m$ can be very large. In such cases the reference algorithm is of no use due to the quadratic dependency of its complexity on $n$ (and originally also on $m$) and the proposed method arises as a feasible alternative without coherence degradation. 

\subsection{Compressive Sensing Performance}
\label{exp:CS}
In this last section we come back to the main motivation of the work, namely, the construction of close-to-optimal complex CS measurement matrices. The final measure of the success or failure of the proposed approach is to be measured in terms of the capability of the resulting BCASCs to recover high-dimensional sparse signals from a relatively low number of measurements. We are interested in the cases where $n$ is so large that generating the BCASCs using the reference algorithm becomes unfeasible. In the experiments described in this section the dimensionality of the sparse signal to recover is set to $n\in\{128,256,512\}$. Once $\pmb{A}\in \mathbb{C}^{m\times n}$, namely, the approximate BCASC is generated, the goal is to recover the $s$-sparse $n$-dimensional signal $\vec{x}$ from the $m\ll n$ linear measurements given by $\vec{y}=\pmb{A} \vec{x}$. An estimate of the signal $\hat{\vec{x}}$ is recovered from $\vec{y}$ by $l_1$-minimization in a conventional CS framework using the efficient Chambolle and Pock's algorithm \cite{Chambolle11}. The normalized $l_2$ distance between $\hat{\vec{x}}$ and $\vec{x}$ is adopted as measure of the recovery error. The real and imaginary components of the nonzero entries of $\vec{x}$ are drawn from a Gaussian distribution of zero mean and unit standard deviation and $\vec{x}$ is then $l_2$-normalized. The recovery performance obtained using ANN-BCASC measurement matrices is compared to that obtained using both random complex matrices and Fourier ensembles of the same size. In all cases the columns of the matrices are normalized.

\begin{figure*}[htpb]
	\begin{minipage}{0.91\textwidth}
      \centering
	\subfloat[Gaussian $n=128$]{\includegraphics[width=0.33\linewidth,trim=70 0 110 60, clip]{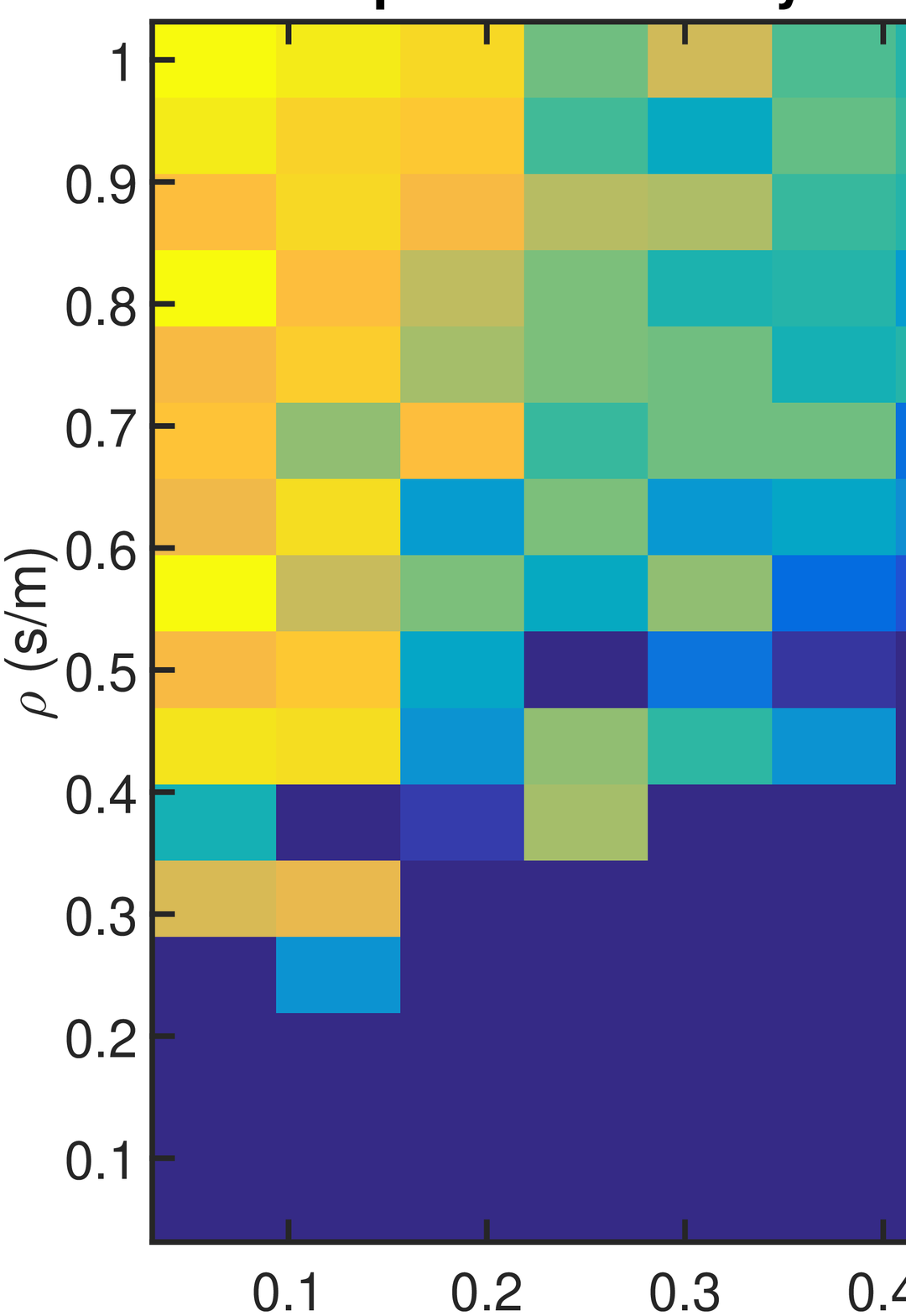}\label{fig:CS_rec:Gauss:128}} %\hspace{5pt}
	\subfloat[Gaussian $n=256$]{\includegraphics[width=0.33\linewidth,trim=70 0 110 60, clip]{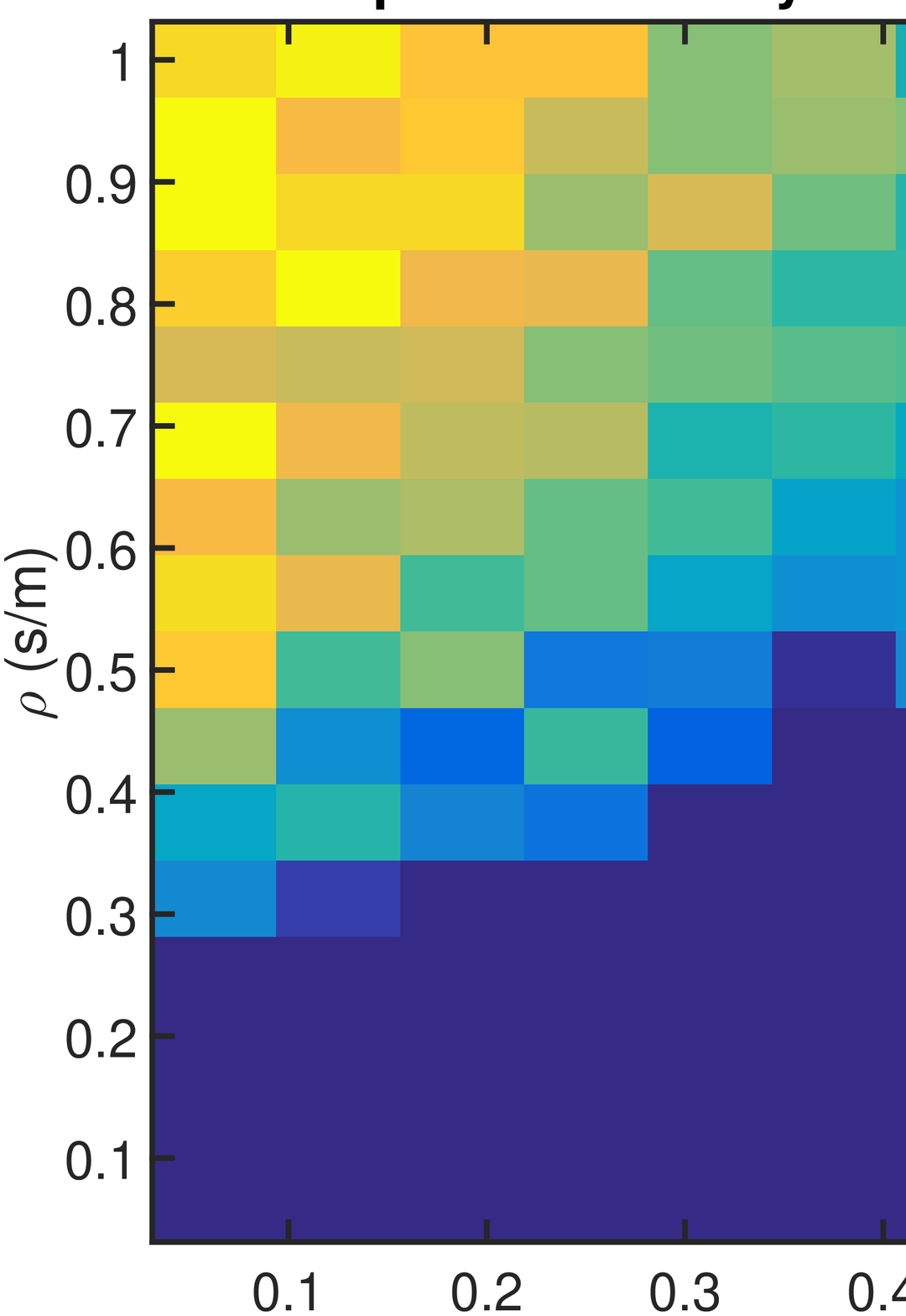}\label{fig:CS_rec:Gauss:256}} %\hspace{5pt}
	\subfloat[Gaussian $n=512$]{\includegraphics[width=0.33\linewidth,trim=70 0 110 60, clip]{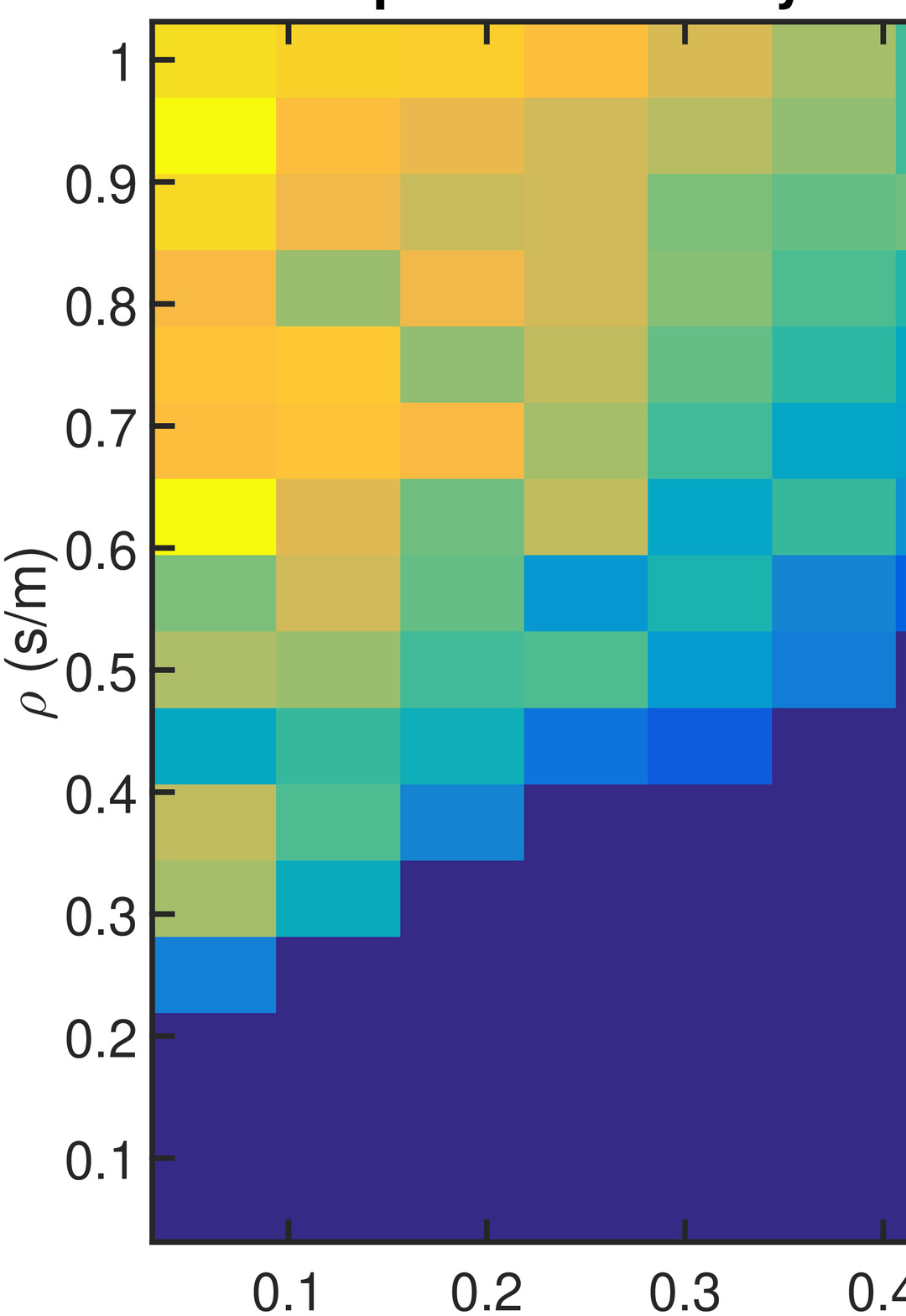}\label{fig:CS_rec:Gauss:512}} \\
	\subfloat[Fourier $n=128$]{\includegraphics[width=0.33\linewidth,trim=70 0 110 60, clip]{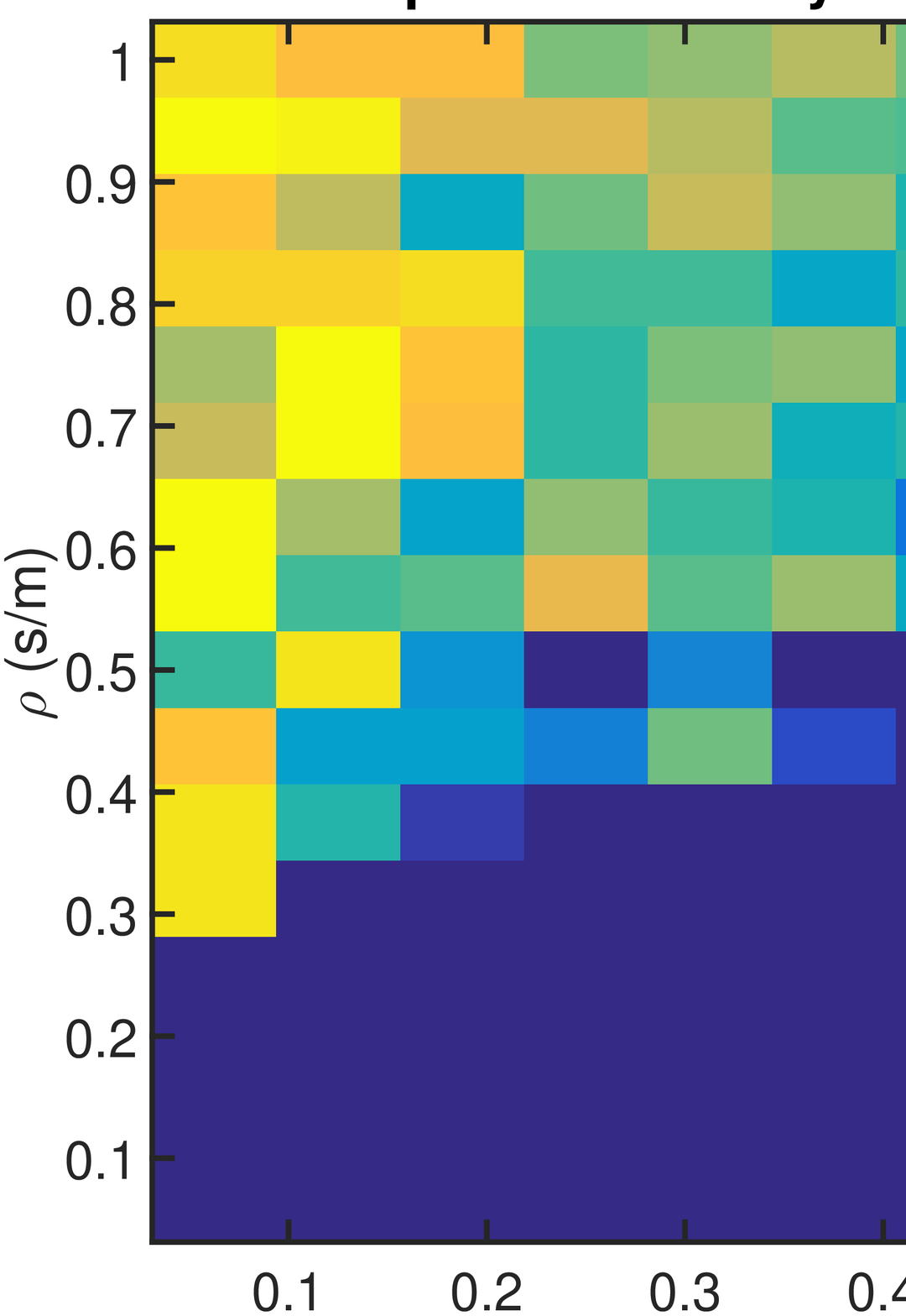}\label{fig:CS_rec:Fourier:128}} %\hspace{5pt}
	\subfloat[Fourier $n=256$]{\includegraphics[width=0.33\linewidth,trim=70 0 110 60, clip]{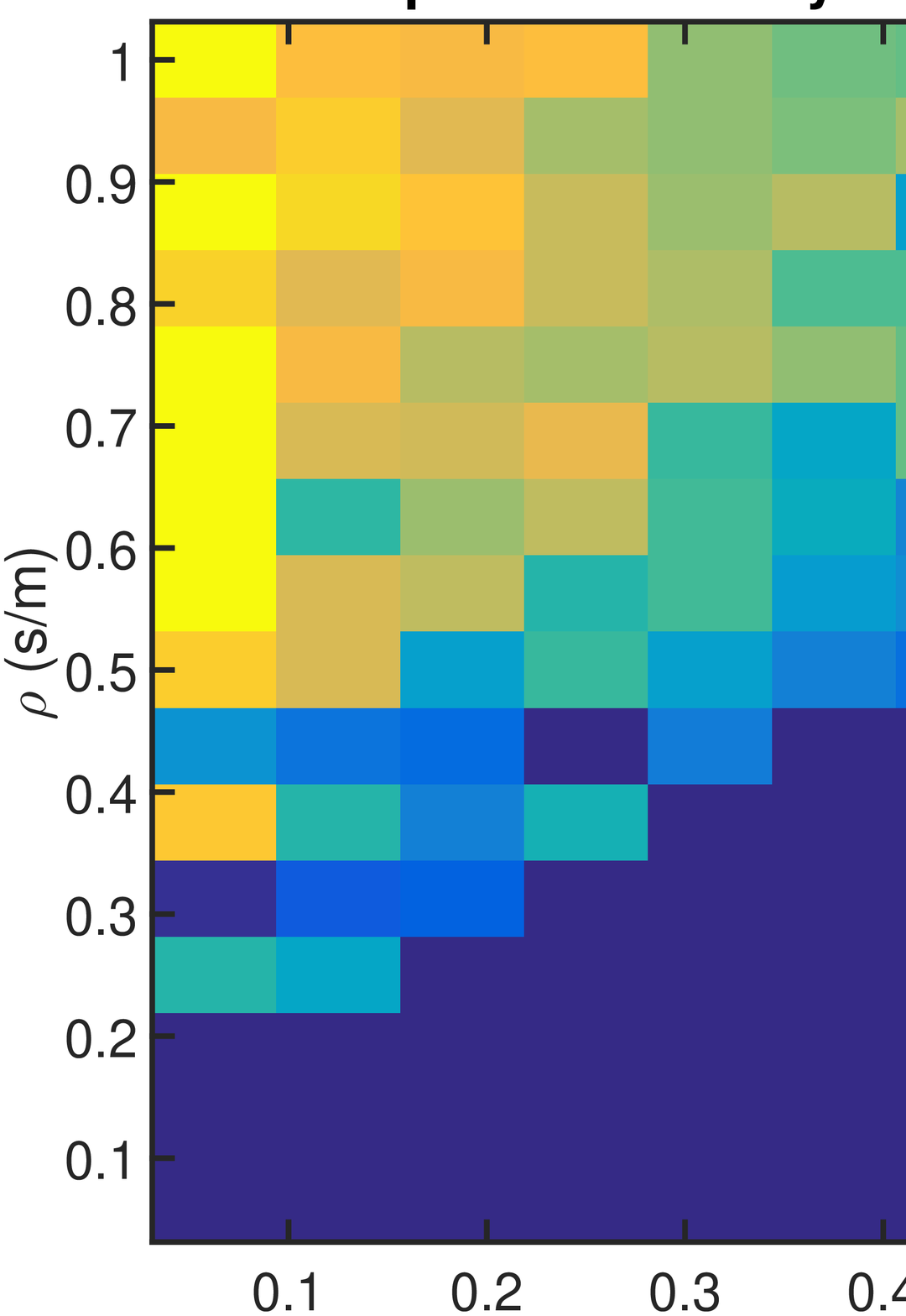}\label{fig:CS_rec:Fourier:256}} %\hspace{5pt}
	\subfloat[Fourier $n=512$]{\includegraphics[width=0.33\linewidth,trim=70 0 110 60, clip]{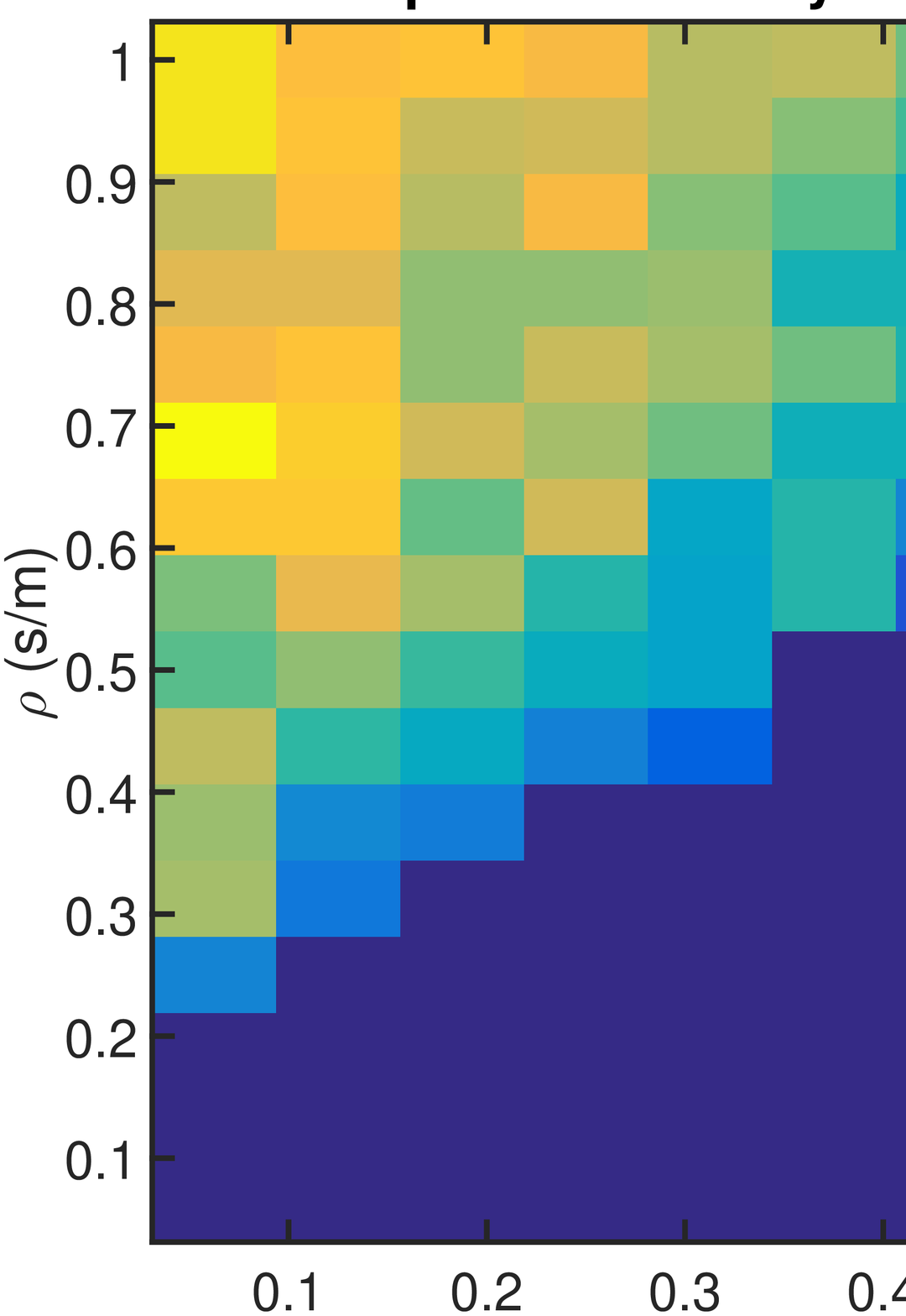}\label{fig:CS_rec:Fourier:512}} \\
	\subfloat[ANN-BCASC $n=128$]{\includegraphics[width=0.33\linewidth,trim=70 0 110 60, clip]{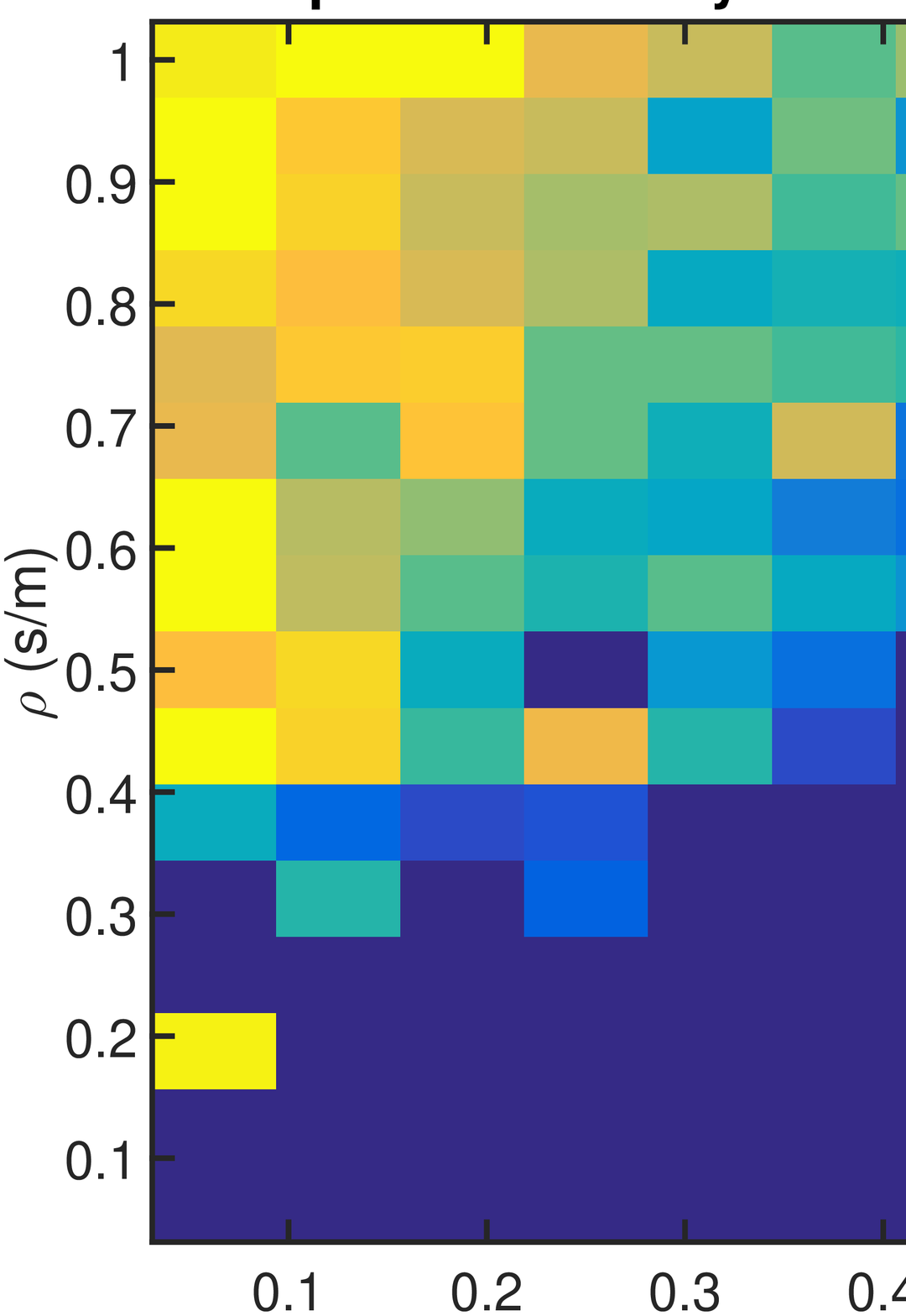}\label{fig:CS_rec:ANN-BCASC:128}} %\hspace{5pt}
	\subfloat[ANN-BCASC $n=256$]{\includegraphics[width=0.33\linewidth,trim=70 0 110 60, clip]{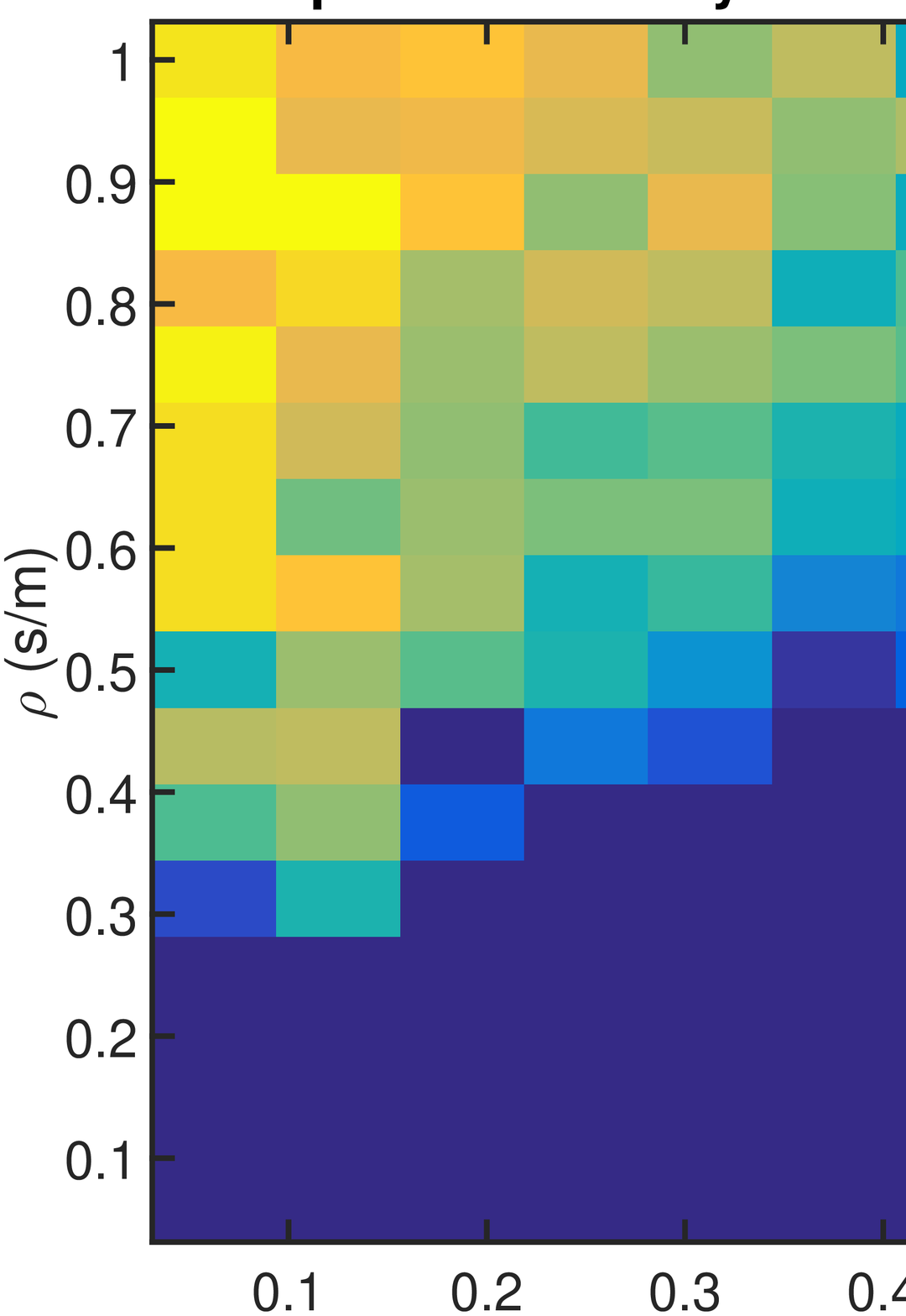}\label{fig:CS_rec:ANN-BCASC:256}} %\hspace{5pt}
	\subfloat[ANN-BCASC $n=512$]{\includegraphics[width=0.33\linewidth,trim=70 0 110 60, clip]{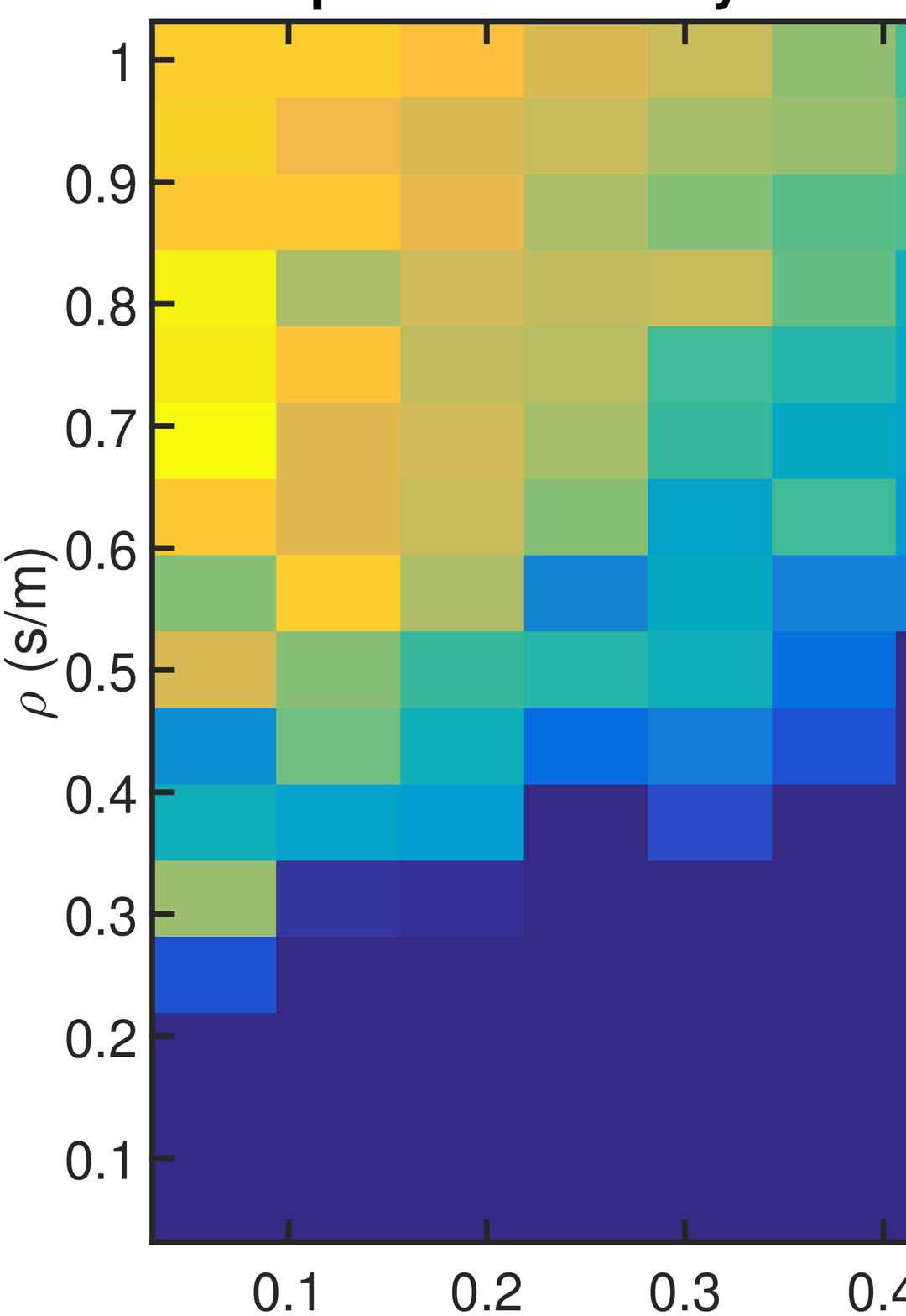}\label{fig:CS_rec:ANN-BCASC:512}} \\
	\end{minipage}
	\begin{minipage}{0.08\textwidth}
      	\centering
		\subfloat{\includegraphics[width=\linewidth,trim=1250 50 100 50, clip]{colorbar_full_screen_line_1_font_10.eps}\label{fig:CS_rec:colorbar}} \\
	\end{minipage}
\caption[Evaluation of the CS recovery error with respect to $\delta$ and $\rho$ for different values of $n$]{Evaluation of the (normalized) CS recovery error with respect to $\delta=m/n$ and $\rho=s/m$ for different large values of $n\in\{128,256,512\}$. Each pixel in the plots corresponds to a single experiment. Each column of each plot was obtained using a single measurement matrix of size $\delta n\times n$. Plots in the first row of the figure provide the results obtained for complex random measurement matrices with Gaussian statistics, plots in the second row are for Fourier ensembles and in the third row for our ANN-BCASCs.}
\label{fig:CS_rec}
\end{figure*}

We consider different experimental cases for different values of the parameters $\delta=m/n$ and $\rho=s/m$, as suggested in \cite{Donoho09}. More specifically, we consider a complete evaluation of the entire $\delta-\rho$ plane, i.\,e., $0\leq \delta \leq1$, $0\leq \rho \leq1$. For each parameter sweep, $16$ equally-spaced discrete steps are considered, yielding $256$ cases per $n$ value. For speed and provided that the size of $\pmb{A}$ only depends on $\delta$, a single matrix is generated for each $\delta$-case and used for all $\rho$. On the contrary, an independent $s$-sparse signal $\vec{x}$ is generated for every case. Fig.~\ref{fig:CS_rec} provides the Donoho-Tanner diagrams ($\delta-\rho$) obtained for each value of $n$ (by columns) for the three different types of matrices considered (by rows). For fairness of comparison the Gaussian random matrices were used as seeds for generating the ANN-BCASCs of equal size.

\begin{figure*}[htpb]
      \centering
	%\setlength{\tabcolsep}{1pt}
%	\subfloat[Histogram $n=128$]{\includegraphics[width=0.33\textwidth,trim=70 0 110 50, clip]{log_CS_recovery_error_histogram_n=128_v2.eps}\label{fig:CS_rec_stats:hist:128}} %\hspace{5pt}
%	\subfloat[Histogram $n=256$]{\includegraphics[width=0.33\textwidth,trim=70 0 110 50, clip]{log_CS_recovery_error_histogram_n=256_v2.eps}\label{fig:CS_rec_stats:hist:256}} %\hspace{5pt}
%	\subfloat[Histogram $n=512$]{\includegraphics[width=0.33\textwidth,trim=70 0 110 50, clip]{log_CS_recovery_error_histogram_n=512_v2.eps}\label{fig:CS_rec_stats:hist:512}} \\
	\subfloat[Histogram $n=128$]{\includegraphics[width=0.32\textwidth,trim=0 0 0 38, clip]{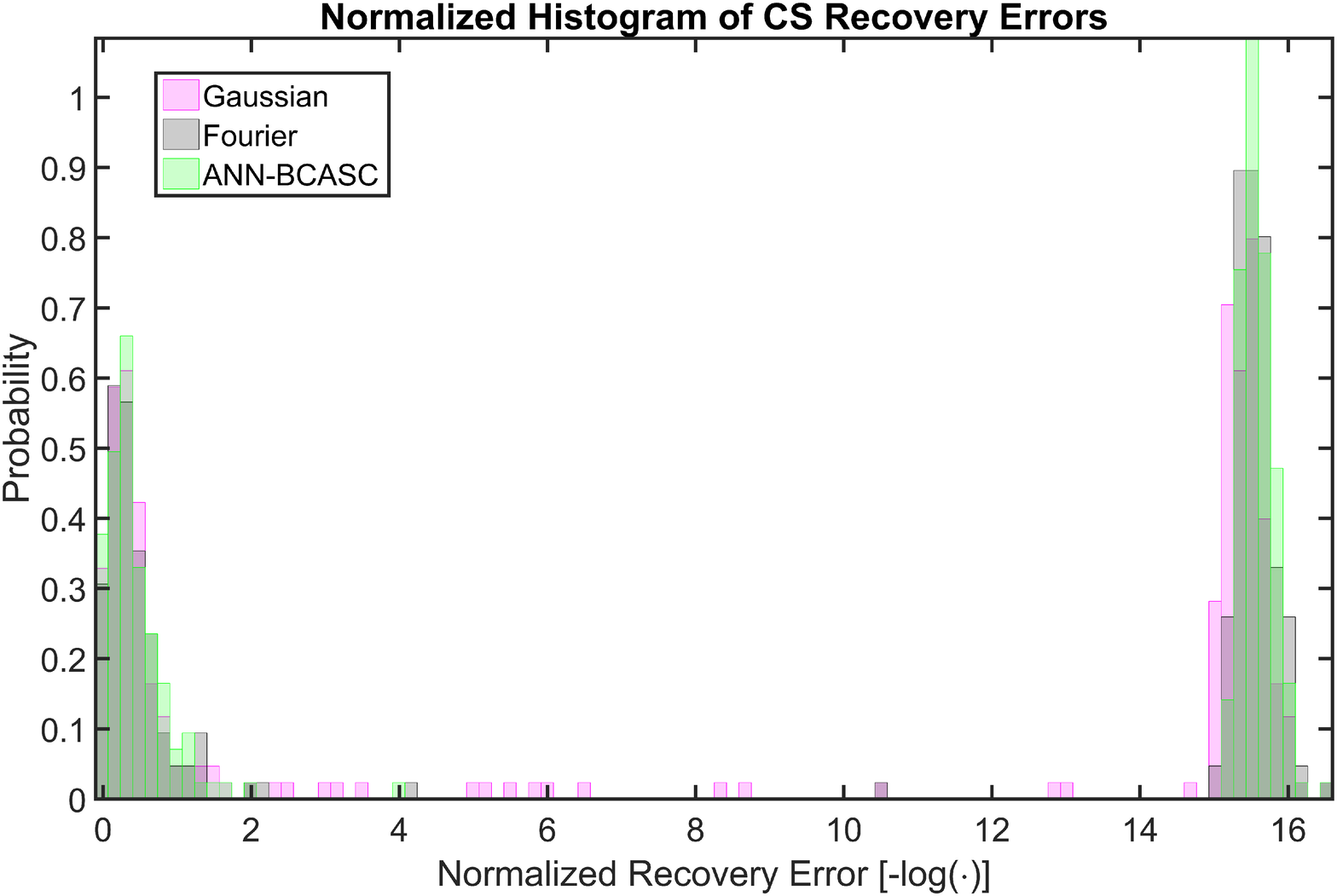}\label{fig:CS_rec_stats:hist:128}} \hspace{4pt}
	\subfloat[Histogram $n=256$]{\includegraphics[width=0.32\textwidth,trim=0 0 0 38, clip]{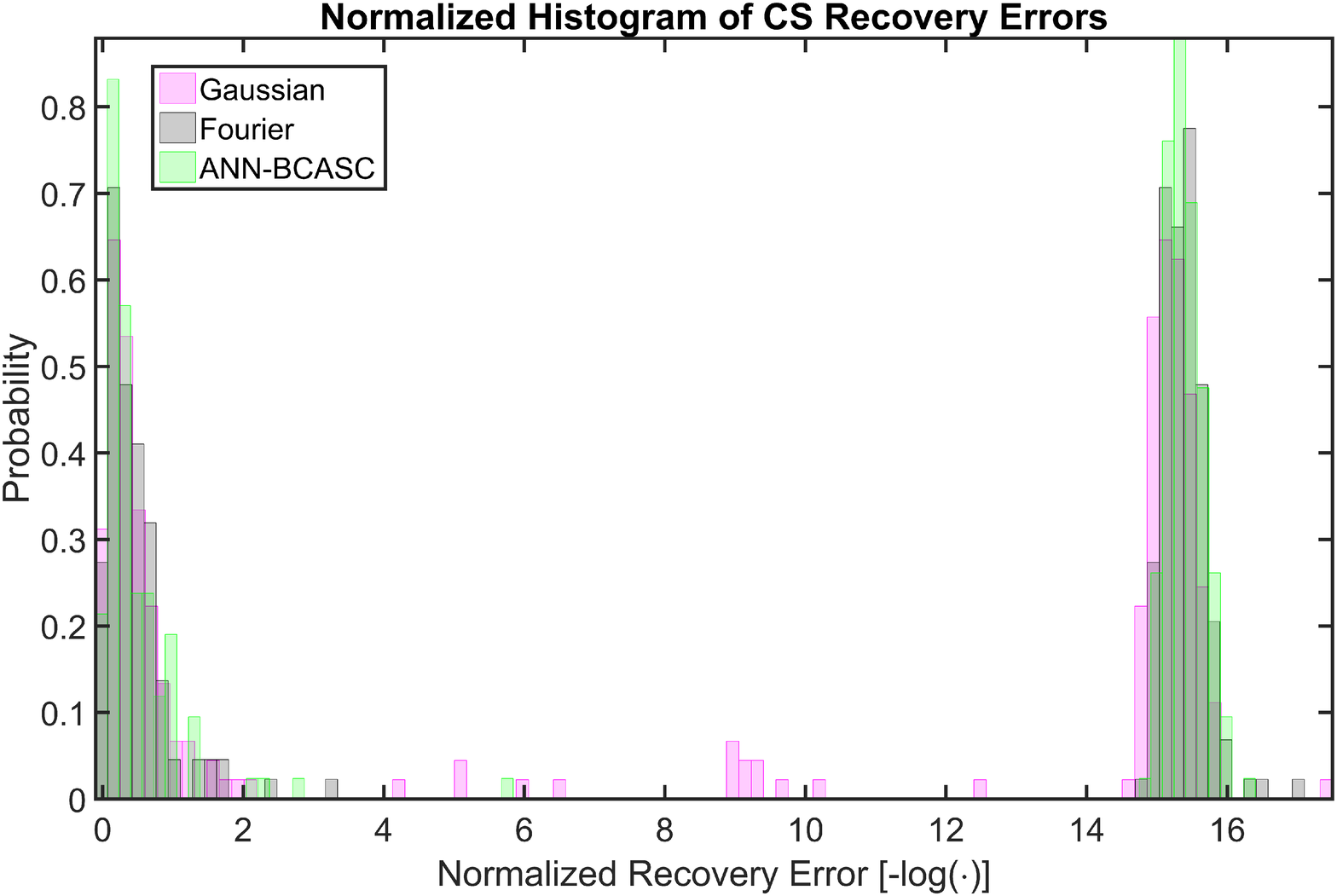}\label{fig:CS_rec_stats:hist:256}} \hspace{4pt}
	\subfloat[Histogram $n=512$]{\includegraphics[width=0.32\textwidth,trim=0 0 0 38, clip]{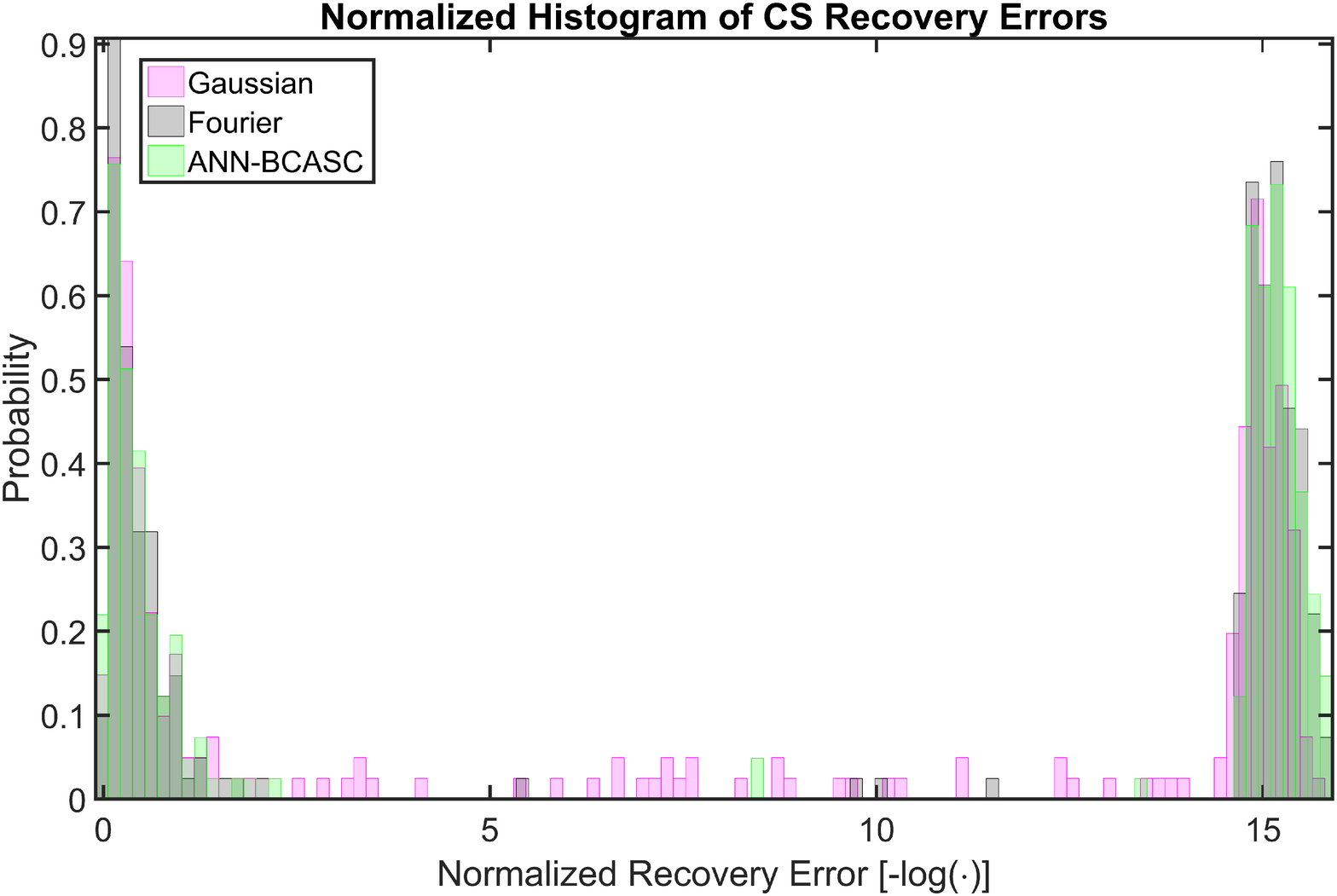}\label{fig:CS_rec_stats:hist:512}} \\
	\subfloat[Survivor $n=128$]{\includegraphics[width=0.33\textwidth,trim=75 0 110 56, clip]{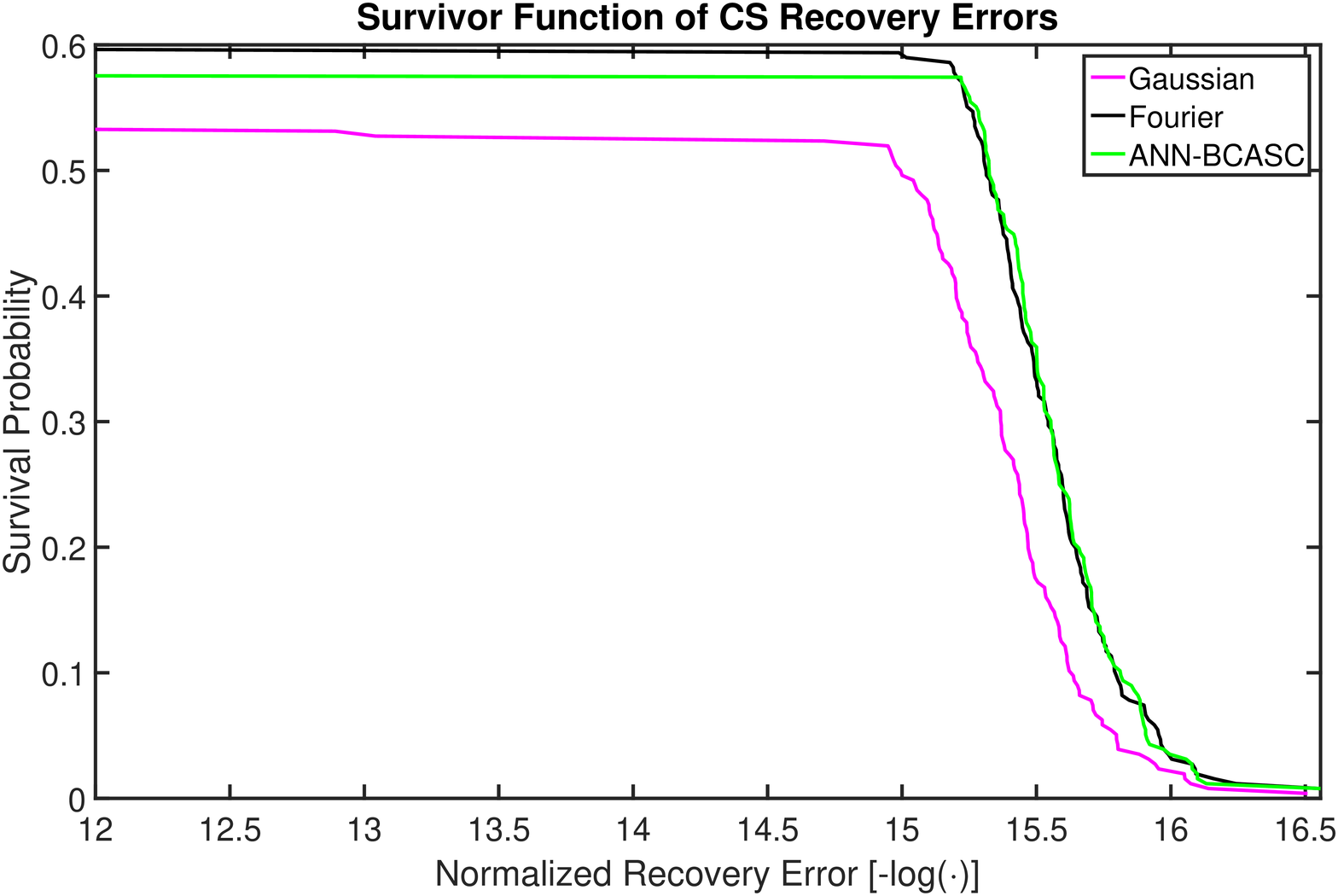}\label{fig:CS_rec_stats:surv:128}} %\hspace{5pt}
	\subfloat[Survivor $n=256$]{\includegraphics[width=0.33\textwidth,trim=75 0 110 56, clip]{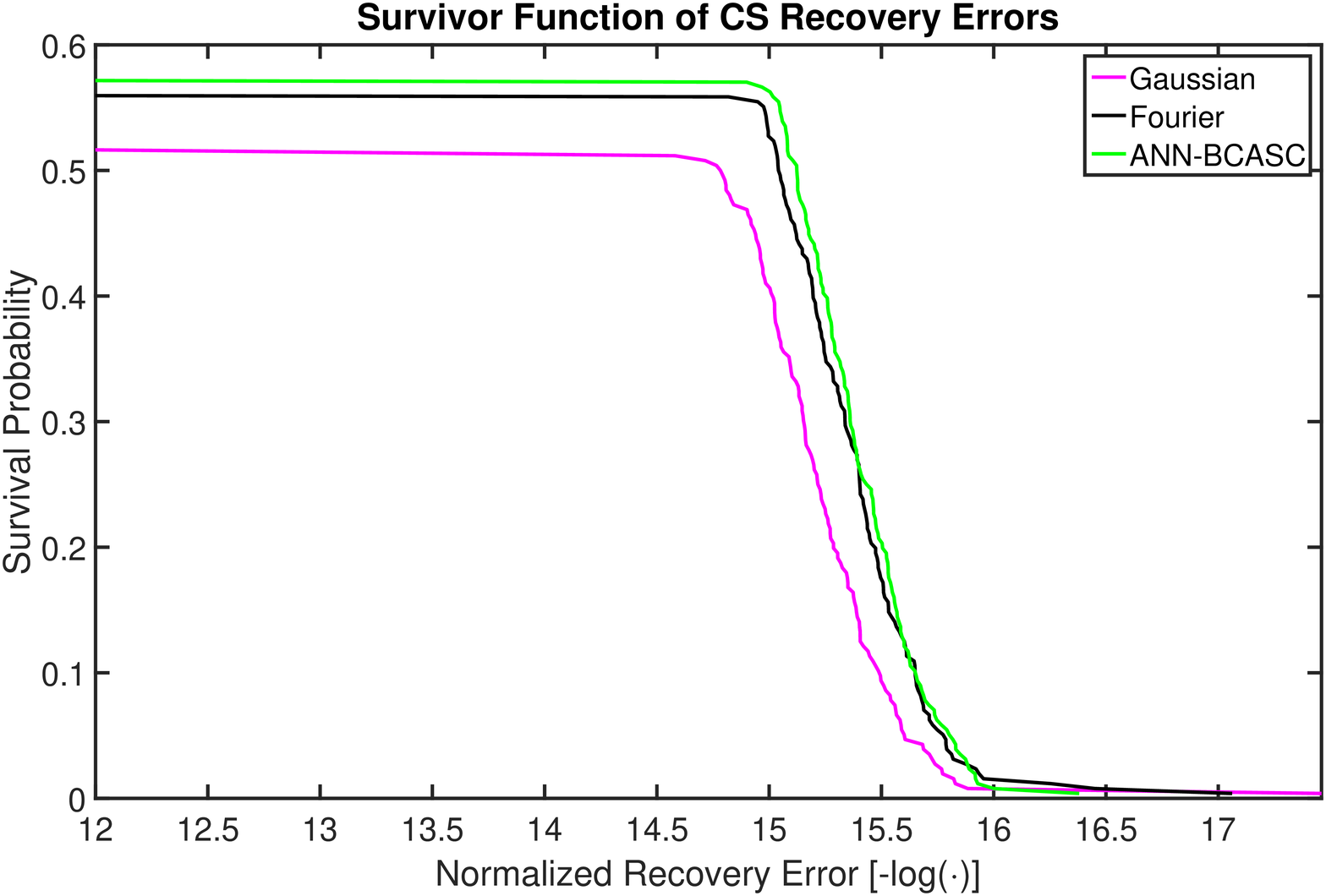}\label{fig:CS_rec_stats:surv:256}} %\hspace{5pt}
	\subfloat[Survivor $n=512$]{\includegraphics[width=0.33\textwidth,trim=75 0 110 56, clip]{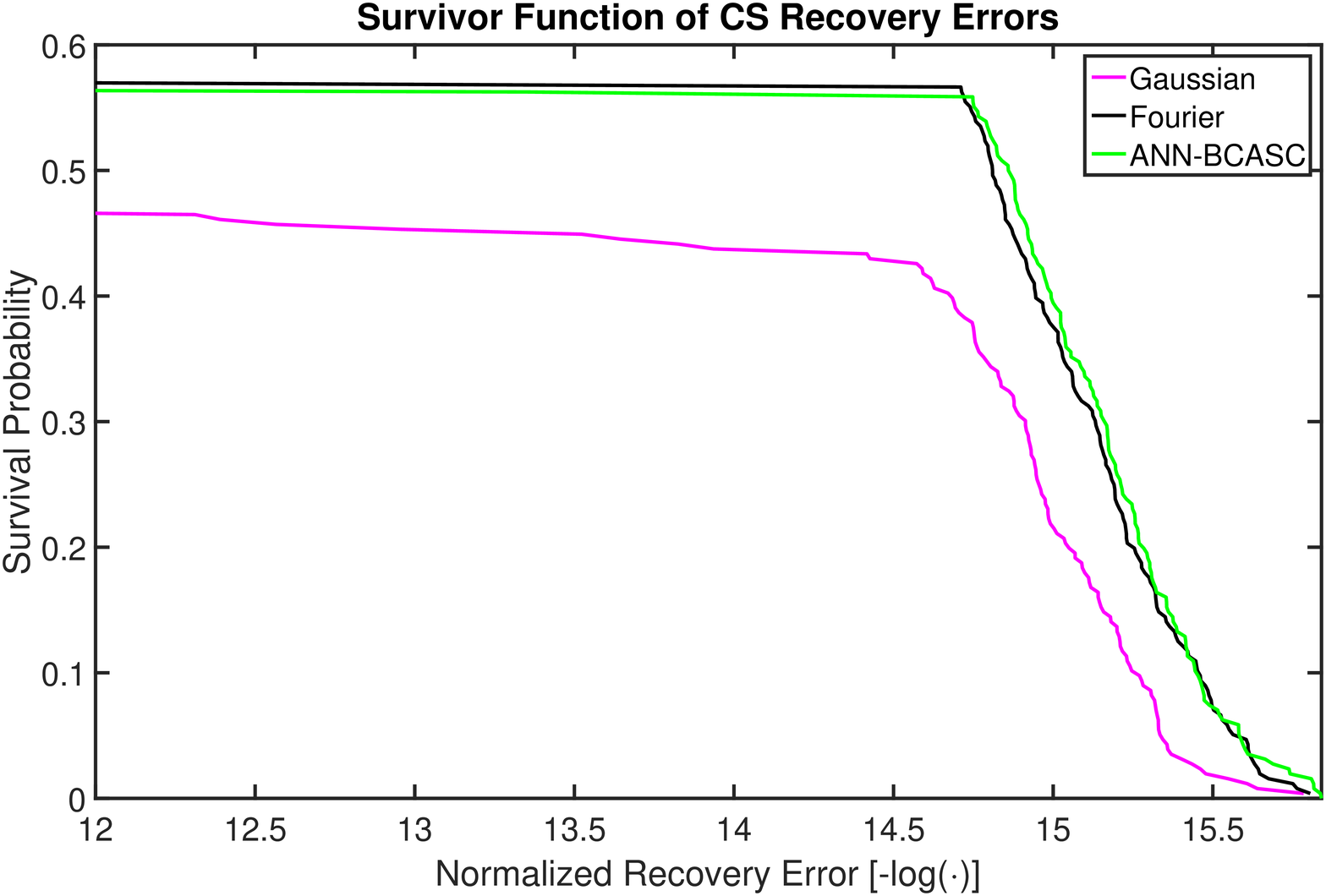}\label{fig:CS_rec_stats:surv:512}} \\
\caption[Statistical evaluation of the CS recovery error]{Statistical evaluation of the recovery error results in Fig.~\ref{fig:CS_rec}. For the sake of visibility, the abscissas of all plots are the negated decimal logarithm of the recovery error instead of the error itself. The first row of plots shows histograms of recovery errors, while the second contains plots of the corresponding survivor functions.}
\label{fig:CS_rec_stats}
\end{figure*}

Pixels in dark blue in the plots of Fig.~\ref{fig:CS_rec} correspond to exact recovery, with normalized error values $<10^{-15}$ in most cases. At first sight it seems that the recovery performance of the three alternatives is equivalent, quite regardless of $n$. It can be confirmed that the area of exact recovery for the ANN-BCASC matrices is always greater or equal than for the other alternatives. Furthermore, errors in that area are lowest for ANN-BCASC matrices. Unfortunately, the advantage of using ANN-BCASCs is slight and cannot be easily observed in the plots of Fig.~\ref{fig:CS_rec}. For this reason, we carry out a minimalist statistical analysis of the data in each of the $\delta-\rho$ graphs. Specifically, we generate both a histogram and a survivor function\footnote{Recall that the survivor function of a random variable $X$ is $S_{X}(\xi)=1-F_{X}(\xi)$, where $F_{X}(\xi)=\displaystyle\int_{-\infty}^{\xi}f_{X}(u)du$ is the cumulative distribution function (CDF) of $X$.} for the empirical distribution of recovery errors. In practice, we use the negated logarithm of the error to ease the visualization. The histograms obtained for each value of $n$ are given in the first row of Fig.~\ref{fig:CS_rec_stats}, while the survivor functions are given in the second row. In principle it is desirable to have two well-concentrated regions in the histogram: a first one corresponding to failure cases, close to zero, and a second one of success cases, which should be centered as far as possible to the right of the previous. Regarding the survivor functions, the superiority of one matrix class over another translates into the latter being over the former, especially in the area of exact reconstruction. 

The histograms in Fig.~\ref{fig:CS_rec_stats} exhibit indeed two well-separated regions, as expected. Furthermore, one can observe that these two regions are most separated for ANN-BCASCs and least for random Gaussian matrices, meaning lower recovery error in the (almost) exact recovery cases. It is also remarkable that almost no cases fall between the two histogram peaks, witnessing a sharp transition between failure or rough recovery and exact recovery. Regarding the survivor functions, the random Gaussian matrices yield curves that are significantly below those for Fourier ensembles and ANN-BCASCs, confirming that significantly poorer coherence also translate into significantly poorer recovery performance. Despite the closeness between the survivor functions of the Fourier ensembles and the ANN-BCASCs, the latter are almost always over the former in the decaying region that corresponds to the last peak in the histogram plots, showing that, despite it might not be of practical relevance, there is indeed a slight improvement with respect to Fourier ensembles too.

In Table~\ref{tab:coherence_CS_rec} we collect the coherence of all the matrices generated for the CS recovery performance evaluation with $m\leq n/2$ (recall that the cases $m\approx n$ are of no interest in CS). In terms of coherence it is clear that ANN-BCASCs widely outperform the other alternative constructions in all cases considered, especially in the interesting cases of $m\ll n$. Furthermore, the coherence of the ANN-BCASCs is relatively close to the theoretical lower bound. To the best of our knowledge, we are the first in generating approximate BCASCs for such large  values of $n$.

\begin{table*}[htbp]
\begin{center}
\setlength{\tabcolsep}{4pt}
\begin{tabular}{c | c c c | c c c | c c c | c c c}
\hline
\multirow{2}{*}{$\delta=m/n$}&\multicolumn{3}{c}{Gaussian}&\multicolumn{3}{c}{Fourier}&\multicolumn{3}{c}{ANN-BCASC}&\multicolumn{3}{c}{Composite Bound}\\
\cline{2-13}
& $n=128$ & $n=256$ & $n=512$ & $n=128$ & $n=256$ & $n=512$ & $n=128$ & $n=256$ & $n=512$ & $n=128$ & $n=256$ & $n=512$\\
\hline
%$0.0625$ & $0.8641$ & $0.7820$ & $0.5471$ & $0.7407$ & $0.5261$ & $0.3873$ & $0.4707$ & $0.3070$ & $0.2169$ & $0.4128$ & $0.2425$ & $0.1713$ \\
%$0.125$ & $0.6468$ & $0.5199$ & $0.4235$ & $0.4826$ & $0.4026$ & $0.2546$ & $0.2649$ & $0.1870$ & $0.1329$ & $0.2348$ & $0.1657$ & $0.1170$ \\
%$0.1875$ & $0.6328$ & $0.4374$ & $0.3380$ & $0.3984$ & $0.3289$ & $0.2149$ & $0.1968$ & $0.1392$ & $9.874\times 10^{-2}$ & $0.1847$ & $0.1304$ & $9.209\times 10^{-2}$ \\
%$0.25$ & $0.5422$ & $0.3997$ & $0.3030$ & $0.2905$ & $0.2730$ & $0.1737$ & $0.1584$ & $0.1120$ & $7.940\times 10^{-2}$ & $0.1537$ & $0.1085$ & $7.662\times 10^{-2}$ \\
%$0.3125$ & $0.4371$ & $0.3595$ & $0.2768$ & $0.2777$ & $0.2260$ & $0.1479$ & $0.1336$ & $9.427\times 10^{-2}$ & $6.667\times 10^{-2}$ & $0.1316$ & $9.288\times 10^{-2}$ & $6.561\times 10^{-2}$ \\
%$0.375$ & $0.4516$ & $0.3368$ & $0.2459$ & $0.2242$ & $0.1583$ & $0.1538$ & $0.1157$ & $8.163\times 10^{-2}$ & $5.769\times 10^{-2}$ & $0.1146$ & $8.085\times 10^{-2}$ & $5.711\times 10^{-2}$ \\
%$0.4375$ & $0.3995$ & $0.3031$ & $0.2453$ & $0.1962$ & $0.1834$ & $0.1250$ & $0.1014$ & $7.155\times 10^{-2}$ & $5.056\times 10^{-2}$ & $0.1006$ & $7.101\times 10^{-2}$ & $5.016\times 10^{-2}$ \\
%$0.5$ & $0.3755$ & $0.3068$ & $0.2117$ & $0.1613$ & $0.1468$ & $0.1236$ & $8.935\times 10^{-2}$ & $6.303\times 10^{-2}$ & $4.453\times 10^{-2}$ & $8.874\times 10^{-2}$ & $6.262\times 10^{-2}$ & $4.424\times 10^{-2}$ \\
$0.0625$ & $0.8641$ & $0.7820$ & $0.5471$ & $0.7407$ & $0.5261$ & $0.3873$ & $0.4707$ & $0.3070$ & $0.2169$ & $0.4128$ & $0.2425$ & $0.1713$ \\
$0.125$ & $0.6468$ & $0.5199$ & $0.4235$ & $0.4826$ & $0.4026$ & $0.2546$ & $0.2649$ & $0.1870$ & $0.1329$ & $0.2348$ & $0.1657$ & $0.1170$ \\
$0.1875$ & $0.6328$ & $0.4374$ & $0.3380$ & $0.3984$ & $0.3289$ & $0.2149$ & $0.1968$ & $0.1392$ & $0.09874$ & $0.1847$ & $0.1304$ & $0.09209$ \\
$0.25$ & $0.5422$ & $0.3997$ & $0.3030$ & $0.2905$ & $0.2730$ & $0.1737$ & $0.1584$ & $0.1120$ & $0.07940$ & $0.1537$ & $0.1085$ & $0.07662$ \\
$0.3125$ & $0.4371$ & $0.3595$ & $0.2768$ & $0.2777$ & $0.2260$ & $0.1479$ & $0.1336$ & $0.09427$ & $0.06667$ & $0.1316$ & $0.09288$ & $0.06561$ \\
$0.375$ & $0.4516$ & $0.3368$ & $0.2459$ & $0.2242$ & $0.1583$ & $0.1538$ & $0.1157$ & $0.08163$ & $0.05769$ & $0.1146$ & $0.08085$ & $0.05711$ \\
$0.4375$ & $0.3995$ & $0.3031$ & $0.2453$ & $0.1962$ & $0.1834$ & $0.1250$ & $0.1014$ & $0.07155$ & $0.05056$ & $0.1006$ & $0.07101$ & $0.05016$ \\
$0.5$ & $0.3755$ & $0.3068$ & $0.2117$ & $0.1613$ & $0.1468$ & $0.1236$ & $0.08935$ & $0.06303$ & $0.04453$ & $0.08874$ & $0.06262$ & $0.04424$ \\
\hline
\end{tabular}
\end{center}
\caption[Coherence comparison of large CS measurement matrices]{Comparison of the coherence of large CS measurement matrices.}
\label{tab:coherence_CS_rec}
\end{table*}

\section{Conclusion}
\label{con}
In this paper an approximate method for constructing BCASCs has been presented and evaluated. The method is a modification of the construction procedure proposed in \cite{Zoerlein15} which uses ANN of each complex codeword to select the most relevant summands to include in an approximate summation that mimics an integral over all the complex rotations of all other codewords during construction of the codes. The main motivation for such an approach was the excessive computational complexity of the reference method in \cite{Zoerlein15}. More specifically, each iteration has complexity $\mathcal{O}(m^2 n^2 K)$, where $n$ is the number of $m$-dimensional codewords in the BCASC and $K$ the number of summands to consider in an approximate integral. The quadratic dependency on both $m$ and $n$ precludes the generation of BCASCs for problems of large dimensionality.
The proposed approach is able to reduce the complexity of each iteration to $\mathcal{O}(m n n_{\mathrm{rot}}\log{(n n_{\mathrm{rot}})})$ (for some fixed number of NN), where $n_{\mathrm{rot}}$ is the number of steps used to discretize the complex rotation domain, which in our case does not equal the number of summands (per codeword) to consider in an approximate integral, as in the original algorithm. In other words, the complexity is no longer quadratic on $m$ and $n$, but linear on both parameters. This yields a large potential for execution time reduction, which easily reaches several orders of magnitude as $m$ and $n$ grow.
Numerical evaluation showed that the method is rather insensitive to the selection of the number of NN to consider and, in fact, a fixed number of NN that is very low compared to $n  n_{\mathrm{rot}}$ suffices to generate BCASCs that reach the coherence of those generated via the reference algorithm. Furthermore, we observed that our approach often results in a reduced coherence when compared to the reference, especially for large $n$.

The capability of constructing BCASCs for large values of $m$ and $n$ is of fundamental interest in the context of Compressive Sensing (CS). CS theory shows that most real signals can be recovered from a reduced number of measurements due to the fact that the minimum number of measurements necessary to retain the signal information is directly related to the amount of information contained in the signal rather than to its ambient dimensionality. In fact, the signal dimensionality $n$ can be arbitrarily large, often arising from a fine discretization of a continuous domain. The proposed approach opens a way of generating close-to-optimal CS measurement matrices in such cases, thanks to its reduced computational complexity.
We have studied the performance of BCASCs generated using the proposed method as CS measurement matrices in the problem of sparse signal recovery. For comparison, common CS measurement matrices, such as (complex) random matrices and Fourier ensembles of the same size have also been evaluated. The BCASCs have been found to outperform the other alternatives both in terms of coherence and recovery error.
%, especially for low ratios $m/n$, which are the cases of interest in CS.

% if have a single appendix:
%\appendix[Proof of the Zonklar Equations]
% or
%\appendix  % for no appendix heading
% do not use \section anymore after \appendix, only \section*
% is possibly needed

% use appendices with more than one appendix
% then use \section to start each appendix
% you must declare a \section before using any
% \subsection or using \label (\appendices by itself
% starts a section numbered zero.)
%

%\appendices
%\section{Proof of the First Zonklar Equation}
%Appendix one text goes here.

% you can choose not to have a title for an appendix
% if you want by leaving the argument blank
%\section{}
%Appendix two text goes here.

% use section* for acknowledgment
\section*{Acknowledgment}
The authors would like to thank Dr. H. Z\"orlein for providing the code of his algorithms for generating BASCs and BCASCs, his readiness to help, and his interest in this work.

% Can use something like this to put references on a page
% by themselves when using endfloat and the captionsoff option.
\ifCLASSOPTIONcaptionsoff
  \newpage
\fi

% trigger a \newpage just before the given reference
% number - used to balance the columns on the last page
% adjust value as needed - may need to be readjusted if
% the document is modified later
%\IEEEtriggeratref{8}
% The "triggered" command can be changed if desired:
%\IEEEtriggercmd{\enlargethispage{-5in}}

% references section

% can use a bibliography generated by BibTeX as a .bbl file
% BibTeX documentation can be easily obtained at:
% http://mirror.ctan.org/biblio/bibtex/contrib/doc/
% The IEEEtran BibTeX style support page is at:
% http://www.michaelshell.org/tex/ieeetran/bibtex/
%\bibliographystyle{IEEEtran}
% argument is your BibTeX string definitions and bibliography database(s)
%\bibliography{IEEEabrv,../bib/paper}
%
% <OR> manually copy in the resultant .bbl file
% set second argument of \begin to the number of references
% (used to reserve space for the reference number labels box)
%\begin{thebibliography}{1}
%
%\bibitem{IEEEhowto:kopka}
%H.~Kopka and P.~W. Daly, \emph{A Guide to \LaTeX}, 3rd~ed.\hskip 1em plus
%  0.5em minus 0.4em\relax Harlow, England: Addison-Wesley, 1999.
%
%\end{thebibliography}

%\bibliographystyle{IEEEtran_no_urls}
\bibliographystyle{IEEEtran}
\bibliography{references}

% biography section
% 
% If you have an EPS/PDF photo (graphicx package needed) extra braces are
% needed around the contents of the optional argument to biography to prevent
% the LaTeX parser from getting confused when it sees the complicated
% \includegraphics command within an optional argument. (You could create
% your own custom macro containing the \includegraphics command to make things
% simpler here.)
%\begin{IEEEbiography}[{\includegraphics[width=1in,height=1.25in,clip,keepaspectratio]{mshell}}]{Michael Shell}
% or if you just want to reserve a space for a photo:

\begin{IEEEbiography}[{\includegraphics[width=1in,height=1.25in,clip,keepaspectratio]{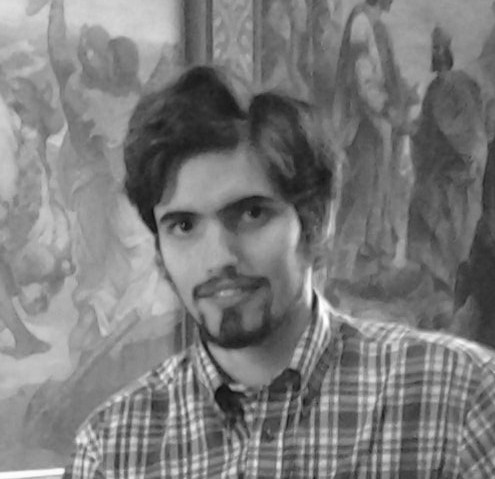}}]{Miguel Heredia Conde}
received the Diploma degree in industrial engineering, specialized in automation and electronics, from the University of Vigo, Vigo, Spain, in 2012. He wrote his Diploma thesis on ``Fast and Robust Localization for RGD-D Sensors'' at the Autonomous Intelligent Systems group of the University of Freiburg, Freiburg, Germany, within an exchange program with the University of Vigo. After graduation he started to work as researcher in the research group on Photogrammetry and Close Range Teledetection of the University of Vigo.

In 2013, he joined the Center for Sensorsystems (ZESS) and the Research Training Group GRK 1564 "Imaging New Modalities", University of Siegen, Siegen, Germany, where he pursued his Ph.D. degree with main focus on applying Compressive Sensing techniques to improve the quality of Time-of-Flight (ToF) Imaging systems based on the Photonic Mixer Device (PMD) until its conclusion in 2016. Since then he leads the research group on Compressive Sensing for the Photonic Mixer Device at ZESS.

Miguel Heredia was one of the recipients of the 2006 Academic Excellence Prices, awarded by the Government of Galicia, Spain, in acknowledgment of outstanding academic performance. %His Diploma thesis was awarded with a Mention of Honor by the University of Vigo and his doctoral thesis with a \emph{summa cum laude} by the University of Siegen.

His current research interests include computer vision, depth sensing, with focus on ToF imaging sensors, PMD technology, Compressive Sensing, coding strategies, and different aspects of signal processing. He is a member of the IEEE and ITG/VDE professional societies.
\end{IEEEbiography}

%\begin{IEEEbiography}[{\includegraphics[width=1in,height=1.25in,clip,keepaspectratio]{photos/Hartmann_photo.jpg}}]{Klaus Hartmann}
%received the Diploma degree in electrical engineering from the Technical University of Siegen, Siegen, Germany. Up to 1985 he was with the company Krohne, Duisburg, as a development engineer. He receives the Dr. Eng. degree in the field of Real-Time Embedded Computing and Sensorics from the University of Siegen, in 1988.
%
%In 1989, he became a Member of the Center for Sensorsystems (ZESS), which is a central scientific-research establishment at the University of Siegen, where since 1990; he has been the General Manager. Since 2000 he is further Leader of the research group ``Embedded Systems and Multi Sensors''.
%
%In 1993 he was cofounder of the company aicoss GmbH (Inspection Systems). In 1997 he was cofounder of the company S-TEC GmbH (3D-Imager, after an investment the today's PMDTec GmbH).
%
%Dr Hartmann is holder of several patents in the field of Sensors and Embedded Systems. His current research interests include Real-Time Computing in Sensor Networks, multisensor-data fusion, Image processing and optical sensors.
%
%\end{IEEEbiography}

\begin{IEEEbiography}[{\includegraphics[width=1in,height=1.25in,clip,keepaspectratio]{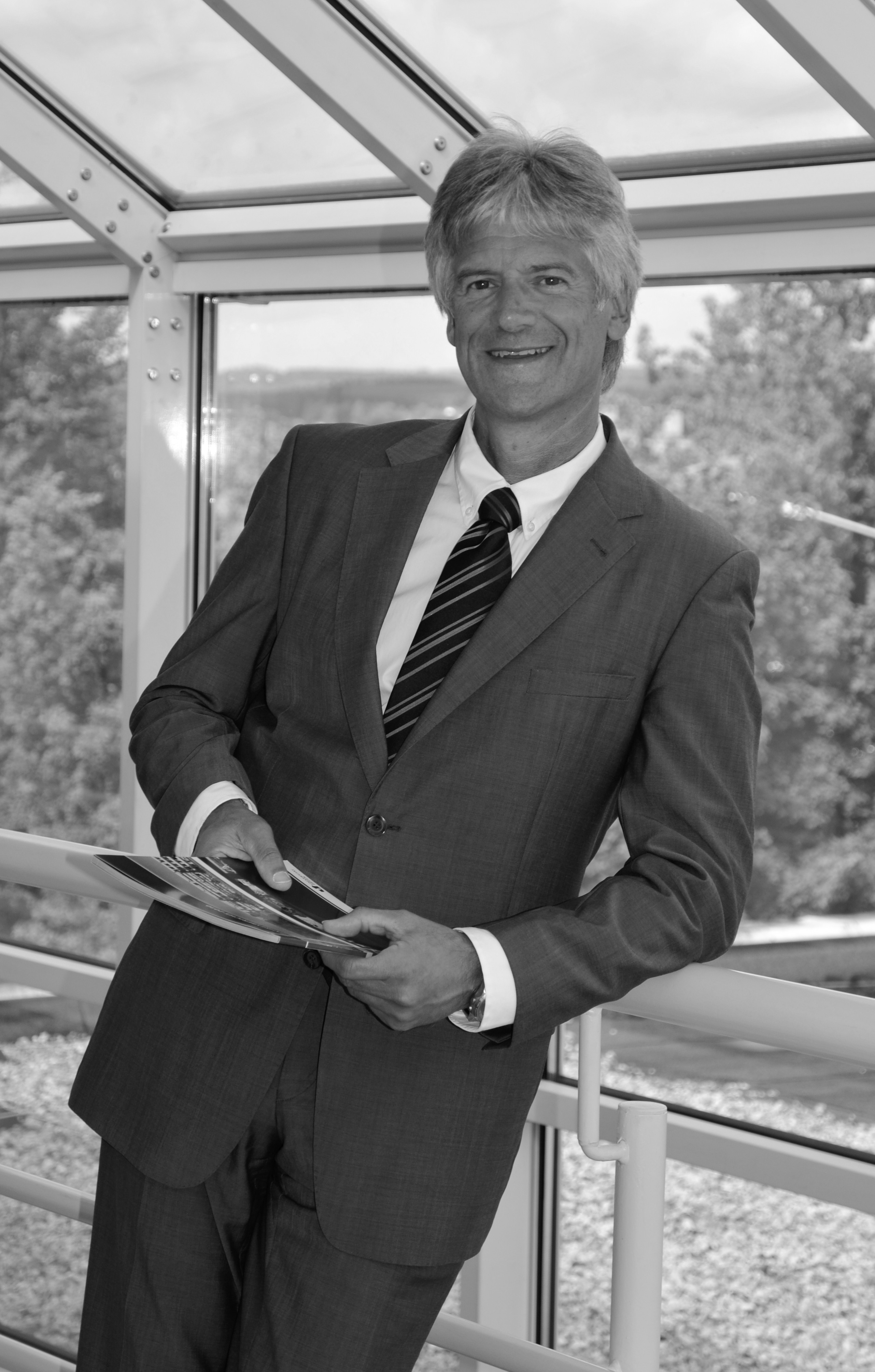}}]{Otmar Loffeld}
(M'05--SM'06) received the Diploma degree in electrical engineering from the Technical University of Aachen in 1982, the Dr. Eng. degree and the “Habilitation” in the field of digital signal processing and estimation theory in 1986 and 1989, respectively, both from the University of Siegen. In 1991, he became professor for digital signal processing and estimation theory at the University of Siegen. He lectures on General Communication Theory, Digital Signal Processing, Stochastic Models and Estimation Theory and Synthetic Aperture Radar, and is author of two textbooks on estimation theory. In 1995, he joined the Center for Sensorsystems (ZESS) at the University of Siegen and became the chair in 2005.

In 1999, Prof. Dr. Loffeld became Principal Investigator (PI) on Baseline Estimation for the X-Band part of the Shuttle Radar Topography Mission (SRTM), to which ZESS contributed to DLR’s baseline calibration algorithms. He is a PI for interferometric techniques in the German TerraSAR-X mission, anda PI for a bistatic spaceborne airborne experiment, where TerraSAR-X serves as the bistatic illuminator while FGAN’s PAMIR system mounted on a Transall airplane is used as a bistatic receiver.

In 2002, he founded the International Postgraduate Program (IPP) “Multi Sensorics,” and in 2008 established the “NRW Research School on Multi Modal Sensor Systems for Environmental Exploration and Safety (MOSES)” at the University of Siegen. 

His current research interests include multisensor-data fusion, Kalman filtering techniques for data fusion, optimal filtering and process identification, SAR processing and simulation, SAR interferometry, phase unwrapping, baseline estimation and, recently, bistatic SAR processing. He is a member of the ITG/VDE and Senior Member of the IEEE/GRSS.
\end{IEEEbiography}

% insert where needed to balance the two columns on the last page with
% biographies
%\newpage

% You can push biographies down or up by placing
% a \vfill before or after them. The appropriate
% use of \vfill depends on what kind of text is
% on the last page and whether or not the columns
% are being equalized.

%\vfill

% Can be used to pull up biographies so that the bottom of the last one
% is flush with the other column.
%\enlargethispage{-5in}

% that's all folks
\end{document}